\documentclass[acmlarge]{acmart}

\usepackage{xspace}
\usepackage{balance}
\usepackage{multirow}
\usepackage{graphicx}
\usepackage{epstopdf}
\usepackage{microtype}
\usepackage{booktabs}
\usepackage{caption}
\usepackage{subcaption}
\usepackage{amsmath}

\usepackage{enumitem}
\usepackage{pifont}
\newcommand{\amark}{\ding{42}\xspace}%

\newcommand{\fakepar}[1]{\smallbreak{}}
\newcommand{\boldpar}[1]{\smallbreak\noindent\textbf{#1.}}

\iffalse
\newcommand{\highlight}[1]{\color{red}#1\color{black}\xspace}
\newcommand{\newtext}[1]{\color{blue}#1\color{black}\xspace}
\else
\newcommand{\highlight}[1]{#1}
\newcommand{\newtext}[1]{#1}
\fi

\usepackage{soul}
\soulregister\cite7
\soulregister\ref7
\iffalse
\newcommand{\response}[1]{{\color{blue}#1}}
\else
\newcommand{\response}[1]{#1}
\fi

\iffalse
\newcommand{\remove}[1]{{\color{blue}\st{#1}}}
\else
\newcommand{\remove}[1]{}
\fi

\newcommand{\ieee}{\mbox{IEEE~802.15.4}\xspace}
\newcommand{\wifi}{\mbox{Wi-Fi}\xspace}
\newcommand{\blefive}{\mbox{BLE\,5}\xspace}
\newcommand{\dcubee}{\mbox{D-Cube}\xspace}
\newcommand{\jamlabng}{\mbox{JamLab-NG}\xspace}

\usepackage{tabularx}
\usepackage{booktabs}
\newcommand{\PreserveBackslash}[1]{\let\temp=\\#1\let\\=\temp}
\newcolumntype{M}[1]{>{\PreserveBackslash\centering}m{#1}}
\newcolumntype{C}[1]{>{\PreserveBackslash\centering}p{#1}}
\newcolumntype{R}[1]{>{\PreserveBackslash\raggedleft}p{#1}}
\newcolumntype{L}[1]{>{\PreserveBackslash\raggedright}p{#1}}

\usepackage{manyfoot}
\DeclareNewFootnote{A}
\DeclareNewFootnote{B}

\let\footnoteR\footnoteB
\let\footnote\footnoteA
\iftrue
\newcommand{\michael}[1]{\footnoteR{{\color{red}\bf MB: #1}\color{red}}}
\newcommand{\carlo}[1]{\footnoteR{{\color{red}\bf CB: #1}\color{red}}}
\newcommand{\markus}[1]{\footnoteR{{\color{red}\bf MS: #1}\color{red}}}
\else
\newcommand{\michael}[1]{}
\newcommand{\carlo}[1]{}
\newcommand{\markus}[1]{}
\fi

\AtBeginDocument{%
  \providecommand\BibTeX{{%
    \normalfont B\kern-0.5em{\scshape i\kern-0.25em b}\kern-0.8em\TeX}}}

\acmConference{Under submission}
\acmDOI{}
\renewcommand\footnotetextcopyrightpermission[1]{} 
\begin{document}



\title{\textbf{Understanding Concurrent Transmissions}}
\subtitle{\textbf{The Impact of Carrier Frequency Offset and RF Interference on Physical Layer Performance}}

\renewcommand{\shorttitle}{The Impact of Carrier Frequency Offset and RF Interference on the Physical Layer}

\author{Michael Baddeley}
\orcid{0000-0002-9202-8582}
\affiliation{%
  \institution{Secure Systems Research Center, Technology Innovation Institute}
  \streetaddress{Mazdar City}
  \city{Abu Dhabi}
  \country{UAE}
  \postcode{P.O. Box 9639}
}
\email{michael.baddeley@tii.ae}

\author{Carlo Alberto Boano}
\orcid{0000-0001-7647-3734}
\affiliation{%
  \institution{Institute of Technical Informatics, Graz University of Technology}
  \streetaddress{Inffeldgasse 16}
  \city{Graz}
  \state{Steiermark}
  \country{Austria}
  \postcode{8010}
  }
\email{cboano@tugraz.at}

\author{Antonio Escobar-Molero}
\orcid{0000-0003-3541-2956}
\affiliation{%
  \institution{RedNodeLabs UG}
  \streetaddress{Ickstattstr. 18}
  \city{Munich}
  \country{Germany}
  \postcode{80469}
  }
\email{antonio@rednodelabs.com}

\author{Ye Liu}
\orcid{0000-0001-9156-9515}
\affiliation{%
  \institution{College of Artificial Intelligence, Nanjing Agricultural University}
  \streetaddress{No. 40, Dian Jiang Tai Road}
  \city{Nanjing}
  \country{China}
  \postcode{210031}
  }
\email{yeliu@njau.edu.cn}

\author{Xiaoyuan Ma}
\orcid{0000-0001-6655-3001}
\affiliation{%
  \institution{SKF Group}
  \streetaddress{No. 1189, Yuan Qi Road}
  \city{Shanghai}
  \country{China}
  \postcode{201814}
  }
\email{ma.xiaoyuan.mail@gmail.com}

\author{Victor Marot}
\orcid{0000-0001-7832-9023}
\affiliation{%
  \institution{University of Bristol}
  \streetaddress{Queens Road}
  \city{Bristol}
  \country{United Kingdom}
  \postcode{BS81QU}
  }
\email{hk19236@bristol.ac.uk}

\author{Usman Raza}
\orcid{0000-0002-8957-2384}
\affiliation{%
	\institution{Waymap Ltd.}
	\streetaddress{Dowgate Hill House}
	\city{London}
	\country{United Kingdom}
	\postcode{EC4R 2SU}
}
\email{usman.raza@waymapnav.org}

\author{Kay R\"{o}mer}
\orcid{0000-0002-4248-4424}
\affiliation{%
  \institution{Institute of Technical Informatics, Graz University of Technology}
  \streetaddress{Inffeldgasse 16}
  \city{Graz}
  \state{Steiermark}
  \country{Austria}
  \postcode{8010}
  }
\email{roemer@tugraz.at}

\author{Markus Schu{\ss}}
\affiliation{%
  \institution{Institute of Technical Informatics, Graz University of Technology}
  \streetaddress{Inffeldgasse 16}
  \city{Graz}
  \state{Steiermark}
  \country{Austria}
  \postcode{8010}
  }
\email{markus.schuss@tugraz.at}

\author{Aleksandar Stanoev}
\affiliation{%
  \institution{Bristol Research and Innovation Lab, Toshiba Europe Ltd.}
  \streetaddress{32 Queen Square}
  \city{Bristol}
  \country{United Kingdom}
  \postcode{BS1 4ND}
  }
\email{Aleksandar.Stanoev@toshiba-bril.com}

\renewcommand{\shortauthors}{M. Baddeley et al.}



\begin{abstract}
The popularity of concurrent transmissions (CT) has soared after recent studies have shown their feasibility on the four physical layers specified by \blefive, hence providing an alternative to the use of \ieee for the design of reliable and efficient low-power wireless protocols. 
However, to date, the extent to which physical layer properties affect the performance of CT has not yet been investigated in detail. 
This paper fills this gap and provides an extensive study on the impact of the physical layer on CT-based \mbox{solutions using \ieee and \blefive}. 
We first highlight through simulation how the impact of errors induced \newtext{by relative carrier frequency offsets} on the performance of CT highly depends on the choice of the underlying physical layer. 
We then confirm these observations experimentally on real hardware \newtext{and with varying environmental conditions } through an analysis of the bit error distribution across received packets, unveiling possible techniques to effectively handle these errors. 
We further study the performance of CT-based \newtext{data collection and} dissemination protocols in the presence of RF interference on a large-scale testbed, deriving insights on how the employed physical layer affects their dependability. 
\end{abstract}



\maketitle

\thispagestyle{empty}


\section{Introduction} \label{sec:introduction}

\newtext{
Throughout the last decade, the development of communication protocols based on Concurrent Transmissions (CT) has driven a breakthrough in highly-performant and dependable low-power wireless networking protocols\footnote{Different terminologies exist in the literature to refer to this concept, including \textit{synchronous} transmissions~\cite{zimmerling20synchronous}, \textit{simultaneous} transmissions~\cite{debardashini20lcn}, and \textit{concurrent} transmissions~\cite{wilhelm14concurrent}. We will stick to the latter in the remainder of this paper.}. 
}
CT-based protocols intentionally let multiple relaying nodes forward packets by simultaneously broadcasting them on the same carrier frequency.
Thanks to the capture effect~\cite{leentvaar76capture} and to non-destructive interference~\cite{liao2016revisiting}, nodes overhearing these CT have a high probability to receive at least one transmission correctly, enabling the creation of reliable and efficient cyber-physical systems and Internet of Things (IoT) applications~\cite{zimmerling20synchronous}.  

The key benefit of CT is the ability to exploit sender diversity to realize simple flooding and synchronization services across large-scale multi-hop wireless networks~\cite{ferrari2011efficient}.
Relaying nodes in a mesh network utilizing CT-based protocols do not need to explicitly avoid collisions using conventional techniques such as carrier sensing, 
and can hence avoid the overhead of routing and link-based communication~\cite{zimmerling20synchronous}. 

A large body of work has proposed CT-based data collection~\cite{ferrari2012low, suzuki13choco, istomin2016data} and dissemination~\cite{doddavenkatappa2013splash, du17pando, debardashini20mass} protocols that can achieve unprecedented gains in terms of reliability, end-to-end latency, and energy efficiency. 
These protocols can outperform existing solutions even in the presence of harsh radio interference~\cite{lim2017competition, istomin18crystal, ma20harmony, escobar2019competition}, as shown by four editions of the EWSN dependability competition~\cite{boano17competition}.

However, the vast majority of these protocols have only been implemented and verified experimentally using off-the-shelf platforms based on \ieee radios operating in the 2.4\,GHz ISM band (e.g., the popular, but rather outdated TelosB mote~\cite{polastre05telos}).  
These solutions employ a physical layer (PHY) based on Orthogonal Quadrature Phase Shift Keying (OQPSK) and Direct Sequence Spread Spectrum (DSSS), as specified by the \ieee standard~\cite{802154_ieee}, where DSSS provides the coding robustness needed by CT to be sufficiently reliable~\cite{wilhelm14concurrent}. 

\boldpar{The rise of multi-PHY IoT platforms and the role of the physical layer} 
An experimental study by Al Nahas et al.~\cite{alnahas2019concurrentBLE5} has shown the feasibility of CT also when using Bluetooth Low Energy~(BLE).
Their preliminary results show that reliable and efficient \mbox{CT-based} flooding is possible on BLE-based mesh networks, but highlight that the performance \emph{largely depends on the employed PHY}, as confirmed by the measurements reported in~\cite{schaper2019truth, jakob20experimental}. 
Indeed, the most recent version of Bluetooth Low Energy~(\blefive) supports four PHYs that largely differ in terms of data rate and robustness~\cite{spoerk19ble5phy}: 2M (2\,Mbps), which doubles the nominal throughput of the original 1M\,PHY (1\,Mbps), and two coded PHYs with coding rates of 1/2~and~1/8 respectively \mbox{(i.e., the 500K and 125K\,PHYs}). 

These preliminary observations are important in light of the increasing number of commodity IoT platforms that embed low-power radios supporting multiple wireless standards; hinting that developers need to carefully select the physical layer used for CT-based communication.
\newtext{ 
Examples of such single-chip, multi-PHY platforms are the Texas Instruments'~\texttt{CC2652R}~\cite{cc2652_productsheet}, the Silicon Labs'~\texttt{EFR32MG}~\cite{efr32mg_productsheet}, the 
STMicroelectronics'~\texttt{STM32WB}~\cite{stm32wb_productsheet}, the NXP Semiconductors'~\texttt{K32W061}~\cite{nxp_k32w061_productsheet}, as well as the Nordic Semiconductors'~\texttt{nRF52840}~\cite{nrf52840_productsheet}~and~\texttt{nRF5340}~\cite{nrf5340_productsheet}, which all support 2.4\,GHz \ieee OQPSK-DSSS alongside the four \blefive PHYs}%
\footnote{Especially the \texttt{nRF52840} has become popular in the research community in recent years, as a number of researchers started to use it to experimentally validate their low-power wireless solutions~\cite{stanoev20demo, jakob20experimental, geissdoerfer19shepherd, alnahas2019concurrentBLE5, baddeley216pp, spoerk21endtoend, spoerk20tiot, langiu19upkit, giovanelli18rssi, biri19totternary}.}. 

However, to date, the extent to which physical layer properties affect CT-based solutions employing \ieee and \blefive has not yet been investigated in detail. 
Firstly, there is no experimental study systematically analyzing how physical layer effects such as \textit{beating}
(a pulsating interference pattern between two or more signals at slightly different frequencies, introduced in Sect.\,\ref{sec:background})
induced by relative carrier frequency offset~\cite{liao2016revisiting}, and de-synchronization due to clock drift~\cite{escobar19imprel} affect the reliability of the received signal in the presence of multiple concurrent transmitters. \response{Importantly, as a function of underlying carrier frequency offsets, the use of non temperature controlled oscillators in low-power radios ensures that beating and clock drift will respond in-line with environmental (both indoor and outdoor) temperature fluctuations - and therefore the impact of these effects may change over time.} Furthermore, there is no experimental work studying whether the performance of CT in harsh RF environments (i.e., in the presence of RF interference) differs depending on the underlying PHY. 

Shedding light on these aspects is important to (i)~provide a better understanding about the role of the physical layer on the reliability and efficiency of CT, as well as to (ii)~empower developers to use the physical layer as a means to fine-tune the performance of CT-based protocols at runtime.

\boldpar{Our contributions}
This paper represents the first in-depth experimental study on the impact of the PHY on CT-based solutions employing \ieee and \blefive, providing valuable observations intended to inform and direct the engineering of future CT-based network protocols. 

Firstly, there exists no other work extensively examining and empirically proving
the significance of the beating effect on CT performance, while the highly novel technique we introduce to achieve this can potentially allow observation in real-time, without a Software Defined Radio~(SDR) \newtext{and low level access to the received baseband signal}. 

Secondly, while recent selected works have demonstrated the feasibility of CT over BLE in single-hop experiments ~\cite{alnahas2019concurrentBLE5, schaper2019truth, jakob20experimental,alnahas2020blueflood}, this paper is the first to systematically examine the impact of different PHYs in a multi-hop scenario, and under benchmarkable interference. With modern chips introducing real-time PHY switching capabilities with \emph{no additional radio overhead}
~\cite{cc2652_productsheet,nrf52840_productsheet}, these insights can provide a blueprint for the design of CT-based networking protocols capable of switching PHY at runtime in response to changing network conditions towards a more dependable performance.

We start by simulating the performance of CT for the \ieee PHY and for the different \blefive PHYs, highlighting the role of beating under different interference scenarios. 
We then set up an extensive experimental campaign to confirm these results and to systematically study CT performance across all \ieee and \blefive PHYs supported by the nRF52840 platform. 
To this end, we use the \dcubee public testbed~\cite{schuss17competition, schuss18benchmark}, recently enhanced with $50$ nRF52840-DK devices \newtext{and with the ability of controlling the temperature of the nodes~\cite{schuss20dcube}}, to observe both beating frequencies and de-synchronization effects on real hardware through an analysis of the error distribution across received packets. 
Our experiments demonstrate that the impact of errors induced by de-synchronization and beating on CT performance is highly dependent on the choice of the underlying PHY, on the relative carrier frequency offset between transmitting devices, \newtext{on environmental temperature fluctuations}, as well as on the number of concurrent transmitters. 
Specifically, we observe that: (i)~high data rate PHYs experience wider beating and we can mitigate its impact through \newtext{retransmissions}; (ii)~if the power delta between signals is insufficient, then \blefive convolutional coding does not sustain reliable CT; (iii)~the pattern mapper used in \blefive125K allows it to effectively handle narrow beating.
\newtext{
Among others, our experimental evaluation shows that: (iv)~observed beating frequencies at the receiver can be quantified by applying a Fast Fourier Transform (FFT) over a histogram of received bit errors; (v)~beating frequencies are highly sensitive to the temperature at transmitting nodes, in line with the temperature response of the RF oscillator; (vi)~with steady temperature at the transmitters, the beating frequency will remain consistent over time. 
}


We further use \dcubee to perform the first experimental study on the performance of different CT-based \newtext{data collection} and dissemination protocols as a function of the underlying PHY in the presence of RF interference on a large scale. 
To this end, we use \dcubee's \jamlabng functionality~\cite{schuss19jamlabng} to generate artificial \wifi interference and stress-test the performance of CT-based data dissemination protocols such as Glossy~\cite{ferrari2011efficient} and robust flooding (RoF)~\cite{lim2017competition} under \emph{no}, \emph{mild}, and \emph{strong} interference. 
\newtext{
We then repeat the same study for data collection protocols such as Crystal~\cite{istomin2016data} and its multi-channel and interference-resilient extension~\cite{istomin18crystal}.} 
Our results allow us to derive important insights on which PHYs are effective to help CT-based protocols in mitigating the impact of RF interference. 
Such insights \newtext{apply to both data collection and dissemination protocols, and} include: (i)~the superiority of \ieee and \blefive500K\,PHY under strong interference, (ii)~the fact that the BLE 125K\,PHY should not be used in conjunction with long payload lengths under interference, as well as (iii)~the need to dynamically change PHY at runtime to provide the best trade-off between reliability, latency, and energy efficiency.

\fakepar{}
After providing some background knowledge on CT and on the beating effect in Sect.~\ref{sec:background}, this paper makes the following specific contributions:
\begin{itemize}
    \item We simulate the performance of CT for \newtext{the \ieee PHY} and all \blefive PHYs, highlighting the strong impact of beating on the resulting packet error rate (Sect.~\ref{sec:simulation}).
    \item We are the first to experimentally observe beating frequencies and de-synchronization effects on real hardware and wireless channel for different PHYs through \newtext{a novel technique whereby} analysis of the bit error distribution across received packets \newtext{exposes the underlying beating frequency} (Sect.~\ref{sec:beating}). 
    \item \newtext{We demonstrate that the beating frequency can be quantified by applying a Fast Fourier Transform~(FFT) across the received bit error distribution, potentially allowing continuous and real-time monitoring of beating in future CT-based systems (Sect.~\ref{sec:beating}).}
	\item \newtext{We show that the sensitivity of radio oscillators to temperature variations impacts the relative carrier frequency offset between devices and consequently the CT-induced beating frequency at the receiver~(Sect.~\ref{sec:beating}).}
    \item We experimentally evaluate the performance of CT-based \newtext{data collection and} dissemination protocols in the presence of RF interference, and provide insights on how the employed PHYs affect their reliability, end-to-end latency, as well as energy efficiency (Sect.~\ref{sec:interference}). 
\end{itemize}
We then describe related work in Sect.~\ref{sec:related_work} to highlight how our insights align with existing literature, and conclude our paper in Sect.~\ref{sec:conclusions}, along with a discussion on future work.


\section{Demystifying Concurrent Transmissions in Low-Power Wireless Networks} \label{sec:background}

\begin{figure}[!t]
	\centering
	\includegraphics[width=0.70\columnwidth]{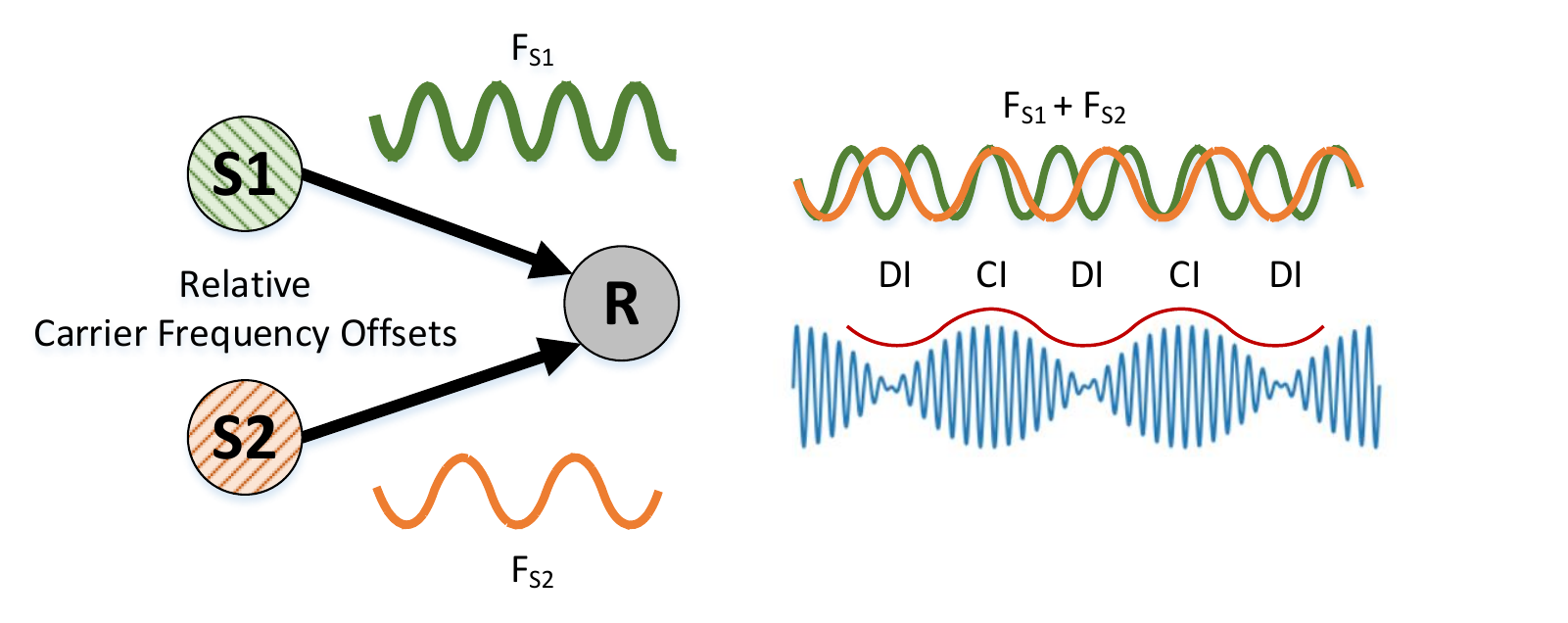}
	\vspace{-5.50mm}
	\caption{\textbf{Illustration of the beating effect.}  Concurrent transmitters (S1 and S2) experience a mismatch between the frequencies of their carrier signals due to imperfect crystal oscillators. As a consequence, the concurrent signals combine to produce periods of \emph{constructive} and \emph{destructive} interference (denoted as CI~and~DI, respectively) at the receiver (R).}
	\label{fig:ct-beating-background}
\end{figure}

\newtext{
Humans, as well as wireless devices, typically communicate with each other \emph{politely}, meaning that basic principles of conversational manners and wireless design favored strategies that prevent multiple speakers/transmitters to talk over each other. People's tendency to listen before talking and not interrupting others did inspire wireless MAC designs such as carrier sense multiple access with collision avoidance (CSMA/CA) when more intrusive approaches like ALOHA had proven to achieve less effective communication throughput. Time Division Multiple Access (TDMA), the other dominant communication paradigm, emulates the turn-taking in human conversation. 

CT-based communication is, however, a more disruptive approach and is beyond the polite conversational conventions mentioned above. CT encourages multiple wireless devices to transmit data \emph{exactly at the same time}. Finding analogies of CT in a normal human situation is not difficult though. Think about listening, understanding, and potentially remembering an unheard song sung by multiple people at the same time in a pub. The ability to successfully decode data transmitted by multiple concurrent transmitters was pioneered by Glossy~\cite{ferrari2011efficient} a decade ago for low-power wireless networks. In Glossy, all the transmitters send the same data just like the same song is sung by all the pub attendees. Now, imagine listening to an argument between a group of people who are talking over each other, trying to get their \emph{different} point of views across. Yet, the loudest voice and often its contents can be understood even in this seemly chaotic situation. Low power wireless radios, not very different than human listeners, can understand the loudest of many signals due to the \emph{capture effect} of radios~\cite{leentvaar76capture, whitehouse05capture}\footnote{The capture effect allows a receiver to correctly receive a packet despite multiple (and with potentially \emph{different} data) CT when the incoming RF signals satisfy certain power and timing constraints. The latter relate to the power and time with which the stronger signal is received (compared to all other signals), and are linked to the concepts of capture threshold and window~\cite{zimmerling20synchronous}. With CT, the (power) capture threshold is greatly extended compared to the classical capture effect by synchronously sending \emph{the same} data.~\cite{wilhelm14concurrent}.}. 
This phenomenon enabled new flexible designs that permit multiple concurrent transmitters to send different data packets simultaneously. Chaos~\cite{landsiedel13chaos} and Crystal~\cite{istomin2016data} are two well-known examples of CT-based protocols building on top of Glossy, but the literature on CT-based protocols is constantly growing~\cite{zimmerling20synchronous}. 
}

Low power CT-based protocols are tested with several wireless technologies, \newtext{including different variations of IEEE 802.15.4 (narrowband/ultra-wideband, 2.4\,GHz/sub-Ghz), BLE, LoRa, etc. \response{In fact, diverse physical models which are based on SINR have been proposed and taken as an important factor when designing CT protocols~\cite{chen2016nopsm}.} Most experimental studies, however, are based on 2.4 GHz-based short range technologies, which include the narrowband IEEE 802.15.4  and BLE physical layers. The narrowband IEEE 802.15.4 specifications use OQPSK modulation with half-sine pulse shaping in the 2.4 GHz band. Many commercial off-the-shelf transceivers realize this functionality using Minimum-Shift Keying (MSK) with frequency modulation~\cite{pasupathy1979minimum}. This puts such transceivers in the same boat as most BLE transceivers that similarly use  binary frequency-shift keying (BFSK)\footnote{For the same reason, it is very common today to see both standards to coexist in a single RF integrated circuit.}. Thus, it would be accurate to say that CT-based protocols are dominantly tested for BFSK based wireless technologies, something that also defines the scope of this paper. For a comprehensive overview of BFSK-based CT protocols, please refer to~\cite{802154_ieee, spoerk19ble5phy, ble5specs}.}


Early works such as Glossy~\cite{ferrari2011efficient} showed that, when using frequency-based modulations, if nodes are sufficiently synchronized, then transmissions of the \emph{same data} will sufficiently align and the packet will be correctly received with cooperative gain. Meanwhile, later works \newtext{such as Chaos and Crystal} have shown that transmissions of \emph{different data} greatly benefit from the capture effect due to energy diversity between transmitters. CT hence constitute a robust technique to deploy simple and latency-optimal mesh networks. Nevertheless, recent literature has shown that CT introduce two types of errors that degrade communication performance.  
\begin{enumerate}[leftmargin=+5.75mm]
	\item \emph{Time synchronization errors}. Concurrent transmitters are not perfectly synchronized. This introduces \newtext{additional inter-symbol interference (ISI) than what would be caused by multipath fading alone.}
	ISI happens when different symbols overlap on the air on the receiver side. To minimize this effect, packet transmissions must be triggered within a time interval ideally lower than half the symbol period~\cite{escobar19imprel}. \newtext{Even when transmitters are perfectly synchronized, ISI would still be present due to different time-of-flight from each transmitter to a receiver. Nevertheless, this typically adds only a small fraction of the symbol period for short-range wireless personal area network technologies such as Bluetooth or IEEE 802.15.4, and is thus not a dominant factor.}  

\item \newtext{\emph{Frequency synchronization errors}. Transmitters do not only send data over the air at slightly different times, but they also experience a mismatch between the frequencies of their carrier signals due to their imperfect crystal oscillators. When these carrier signals overlap on the air at the receiver side,} the resulting waveform has alternating periods of high and low amplitude signals, giving rise to constructive and destructive interference\,\,(beating). This is called the \textit{beating effect}. As shown in Fig.~\ref{fig:ct-beating-background}, with \emph{two} concurrent transmitters, the waveform's envelope has a sinusoidal shape with periods of constructive and destructive interference, whereas it features more complex shapes when \emph{more than two} transmissions overlap~\cite{alnahas2020blueflood}. While potentially introducing a certain degree of energy gain during peaks, beating increases the chance of bit errors during low-energy periods (valleys). 
\newtext{The beating effect is a major focus of this paper, as it has not been explored in sufficient depth in earlier literature. }
\end{enumerate}

\begin{figure*}[!t]
	\centering
	\begin{subfigure}[t]{0.54\textwidth}
		\centering
		\includegraphics[width=1\columnwidth]{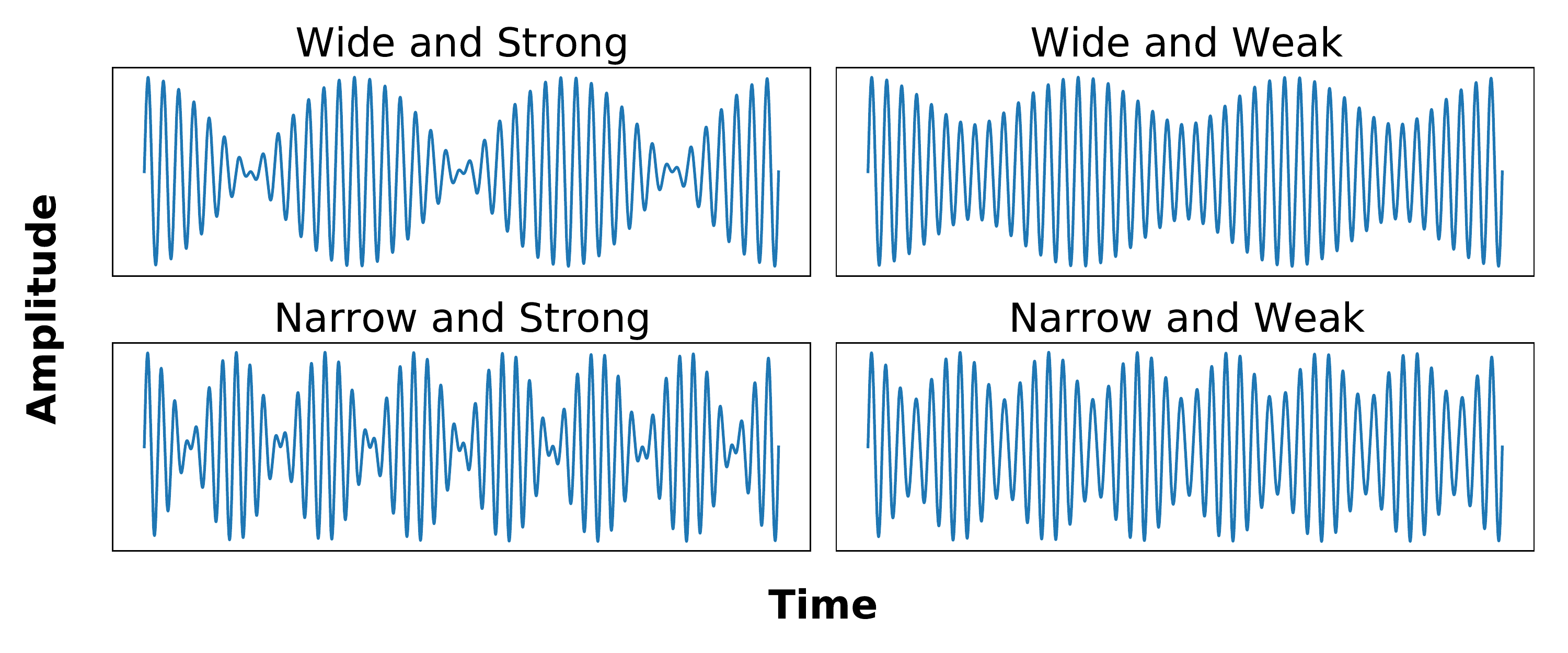}
		\vspace{-7.50mm}
		\caption{\textit{Simple} beating when there are only 2 transmitters.}
		\label{fig:ct-beating-simple}
	\end{subfigure}%
	\vskip\baselineskip
	\vspace{-3.50mm}
	\begin{subfigure}[t]{0.495\textwidth}
		\centering
		\includegraphics[width=1\columnwidth]{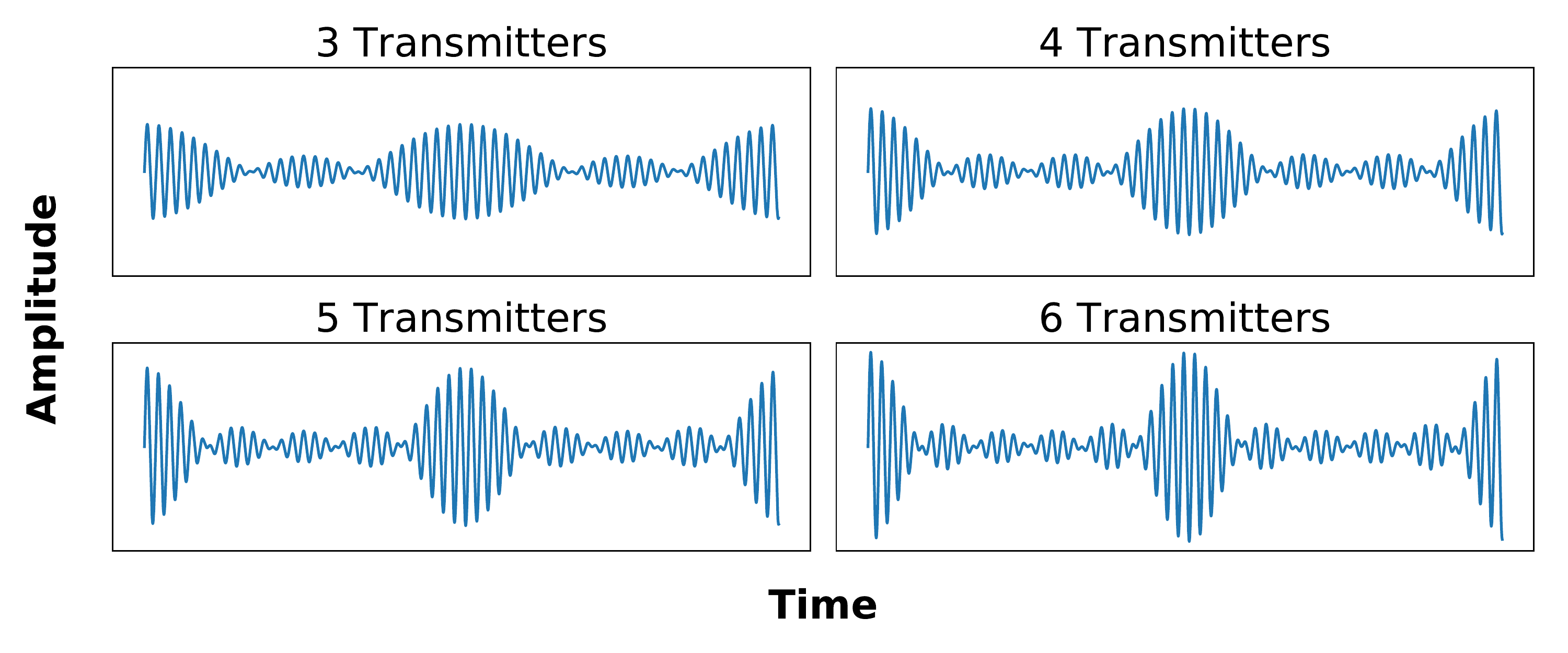}
		\vspace{-7.50mm}
		\caption{Wide and strong \emph{complex} beating.}
		\label{fig:ct-beating-complex-wide-strong}
	\end{subfigure}%
	\hfill
	\begin{subfigure}[t]{0.495\textwidth}
		\centering
		\includegraphics[width=1\columnwidth]{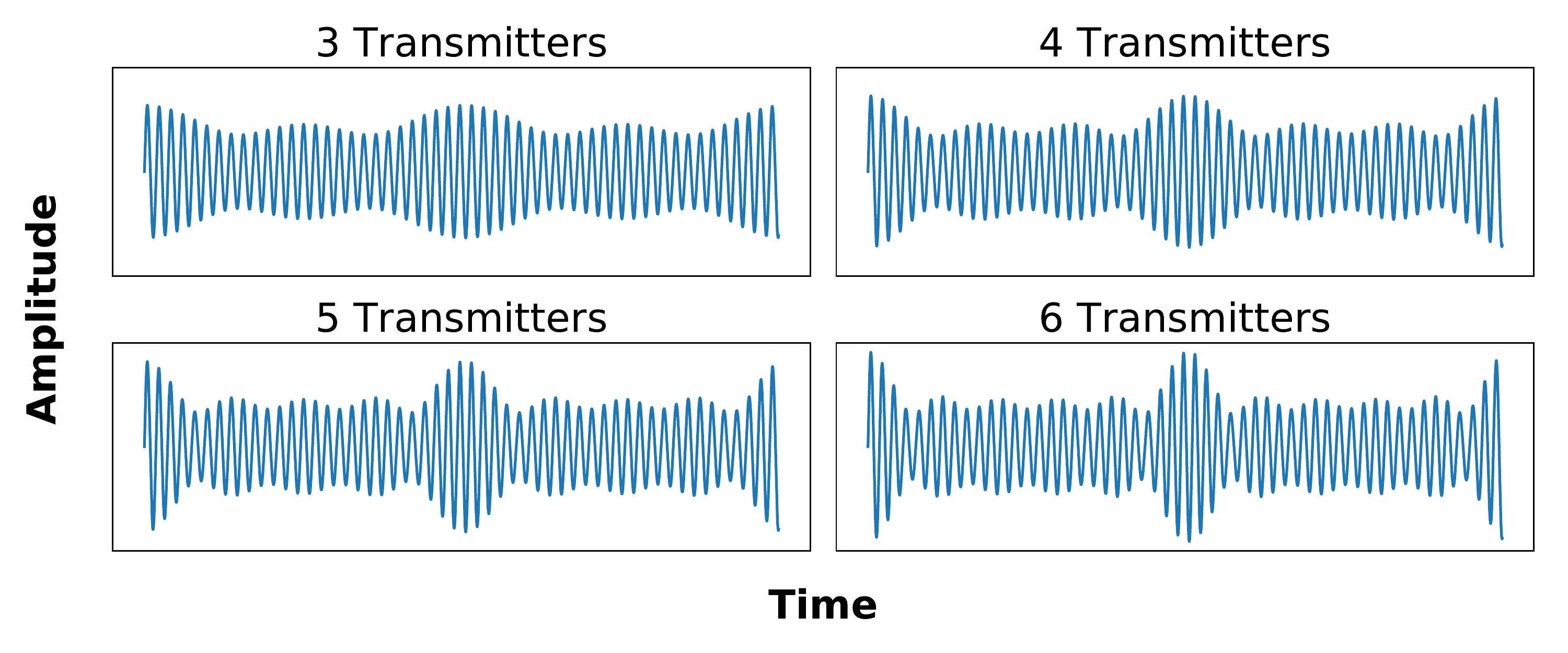}
		\vspace{-7.50mm}
		\caption{Wide and weak \emph{complex} beating.}
		\label{fig:ct-beating-complex-wide-weak}
	\end{subfigure}%
	\vskip\baselineskip
	\vspace{-3.50mm}
	\begin{subfigure}[t]{0.495\textwidth}
		\centering
		\includegraphics[width=1\columnwidth]{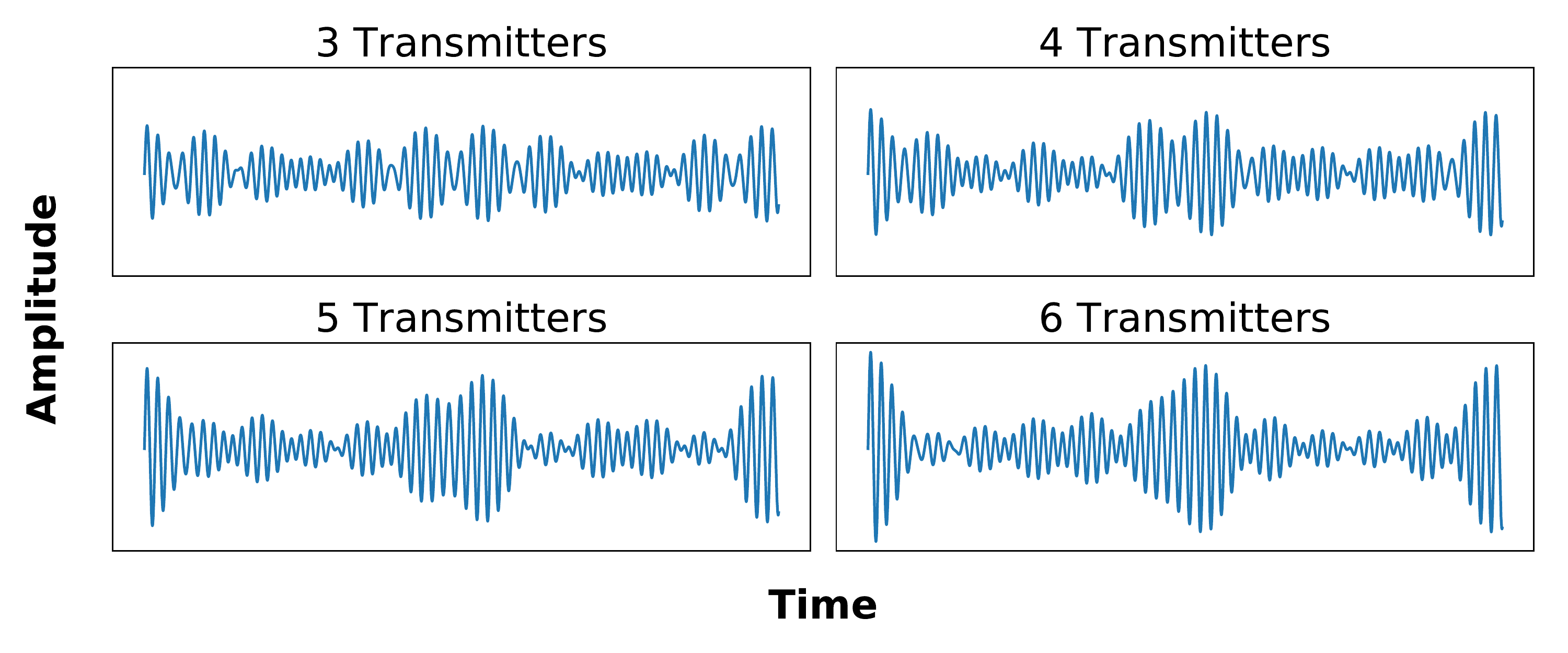}
		\vspace{-7.50mm}
		\caption{Narrow and strong \emph{complex} beating.}
		\label{fig:ct-beating-complex-narrow-strong}
	\end{subfigure}%
	\hfill
	\begin{subfigure}[t]{0.495\textwidth}
		\centering
		\includegraphics[width=1\columnwidth]{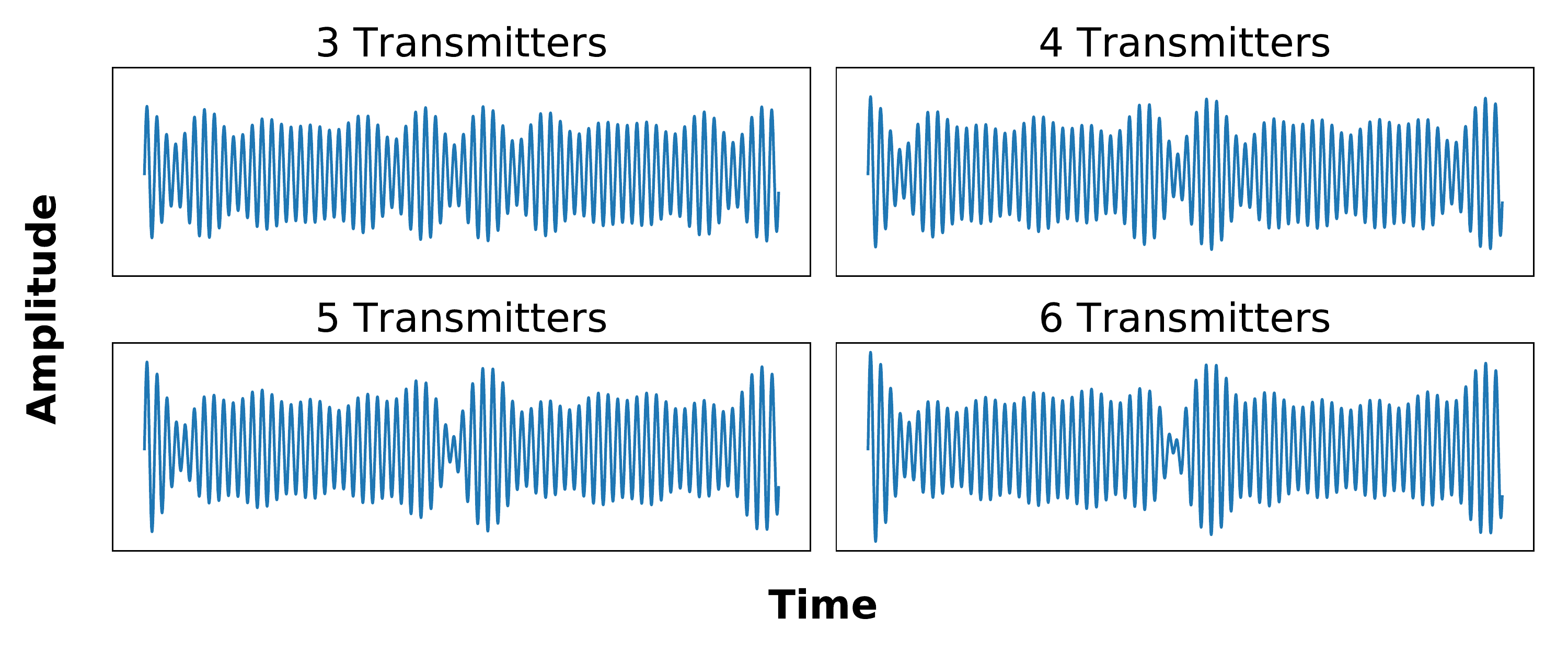}
		\vspace{-7.50mm}
		\caption{Narrow and weak \emph{complex} beating.}
		\label{fig:ct-beating-complex-narrow-weak}
	\end{subfigure}
	\vspace{-3.50mm}
	\caption{\textbf{Beating manifests in many forms depending on the number of concurrent devices, their relative frequencies, and their channel gains.} \newtext{These simplistic simulations are intended to demonstrate the occurrence of beating when multiple signals are superimposed. Sine waves are generated at 15\,Hz steps, between 400-500\,Hz, and are superimposed. If there is no dominant signal (i.e., the signal powers are equal), then \emph{strong} beating attenuates a signal toward zero in a beating valley. Conversely, the presence of dominant signal (simulated as 5x the amplitude of the other signals) \emph{weakens} the beating. As the number of transmissions increases, the resultant amplitude modulation gets more complex, with a tendency to narrower periods of varying amplitude.}}
	\label{fig:ct-beating-types}
	\vspace{-2.00mm}	
\end{figure*}

\newtext{Thus, intuitively, the beating effect under concurrent transmissions has the potential to cause significant packet losses. The extent of this partly depends on two factors:} 

\begin{itemize}
    \item \newtext{\emph{How destructive is the interference?} When concurrently transmitted signals superimpose out of phase at the receiver, these tend to cancel each other out. If the envelope of the signal drops below the sensitivity threshold of the receiver, albeit for a short duration, the receiver sometimes cannot decode symbols correctly. We refer to this as \emph{strong beating}, in contrast to \emph{weak beating} in which case the envelope stays above the sensitivity threshold.} Whether beating will manifest as strong or weak depends on the relative difference in received signal energy. \newtext{When signals with similar power superimpose out of phase, strong beating is most likely to happen. A high power difference would typically result in a weak beating or capture effect.} 

    \item \newtext{\emph{How wide is the destructive interference?} When concurrently transmitted signals superimpose out of phase for a duration that is long in relation to the packet on-air time, we refer to \emph{wide beating} (as opposed to \emph{narrow beating}).} Beating will be randomly narrow or wide depending on the relative carrier frequency offset (CFO) between transmitters. \newtext{Smaller offsets result in wide beating; larger offsets, instead, result in narrow beating.}
\end{itemize}

\newtext{That being said, it is unlikely that a radio will experience continuous (i.e., long periods of) destructive interference. Therefore, it should be noted that if the transmission lasts long enough, something will eventually be received. Following this logic, if you cannot accurately synchronize, or do not wish to synchronize, then beating potentially avoids this interference when the carrier frequencies of senders are (almost) equal and the phase offset is $\pi$. Early work on CT focused on OOK modulated physical layers~\cite{ringwald2005bitmac}. In that case, the bit time is longer than the beating period, so one would eventually see a beating `valley' during one bit. However, the move to \ieee and \blefive allows the beating to avoid continuous destructive interference, as the bit time is (usually) shorter than the beating period.}


In Fig.~\ref{fig:ct-beating-simple} we categorize beating as \emph{wide and strong}, \emph{wide and weak}, \emph{narrow and strong}, or \emph{narrow and weak}. \newtext{However, based on this classification, it is not straightforward to provide qualitative statements on packet error rates. It is useful here to highlight several other factors that dictate the reliability of CT-based protocols in presence of beating:}

\begin{itemize}
\item \newtext{\emph{Error correction.} This refers to the robustness of the coding scheme employed by the underlying PHY, i.e., to the ability of such coding scheme to cope with the beating pattern.}

\item \newtext{\emph{Packet length.} Reliability generally reduces as packet length increases. But an unpredictable beating length} affects the error rate for narrow and wide beating differently. If the beating period is very narrow then several peaks and valleys can appear within a packet transmission and if it is very wide then a packet can be completely shadowed within a valley.

\item \newtext{\emph{Number of concurrent transmitters.}} Receivers listening from more than two concurrent transmitters experience more complex beating patterns as shown in (Fig.~\ref{fig:ct-beating-complex-wide-strong} to Fig.~\ref{fig:ct-beating-complex-narrow-weak}). 
The period between valleys decreases, and peaks and valleys of different amplitudes appear. Remarkably, stronger peaks and deeper valleys may appear, potentially disrupting the reception even when there is a dominant transmitter. If the combined strength of all the other, non-dominant, transmissions is comparable to the strength of the dominant one, which is typical in dense topologies, the reception may still be heavily affected.

\end{itemize}

\newtext{With background into CT-based protocols and factors affecting their performance, it is clear that reliability of a \emph{single CT link} depends on not one but multiple factors. 
We next provide insights into \emph{link-level} performance through simulation in Sect.~\ref{sec:simulation} and then use empirical experiments in Sect.~\ref{sec:beating}. 
Moving from \emph{link} to a \emph{network wide} performance, we resort to large-scale testbed experiments using known CT protocols in Sect.~\ref{sec:interference}. 
}




\section{Impact of Physical Layer on CT Performance: Analysis and Simulation} \label{sec:simulation}

While synchronization errors can be greatly reduced by a proper low-level design of the CT network protocol and its retransmission strategy, \textit{beating cannot be avoided}. Beating always appears when signals from non-coherent transmitters overlap in the air, due to their different carrier frequency offset (CFO) \response{ -- a random deviation from the nominal transmission frequency that occurs due to the tolerance of the quartz crystals used in the radio oscillator, usually expressed in parts per million (ppm). For comparison, in classical MIMO communication systems CT come from the same local oscillator, so multipath effects dominate instead of beating, which is a unique property of systems using independent transmitters}.

To better analyze the impact that different PHYs have on the performance of CT under beating, we simulate different low-power communication systems supported by the nRF52840 platform using MATLAB to obtain the average Packet Error Rate (PER) vs. Signal-to-Noise Ratio (SNR) for \response{Concurrent Transmissions from two nodes (a scenario later referred to as CT\_2 in this paper)} recovered with a non-coherent BFSK receiver, as in~\cite{escobar20phd}. 
We assume additive white Gaussian noise (AWGN) and no synchronization errors. 
We repeat this for different uncoded (\blefive1M and 2M) and coded (\blefive500K, \blefive125K, and \ieee) PHYs, for different oscillator inaccuracies (which result in either wide or narrow beating), and for power deltas. Both \blefive coded PHYs are based on the 1M PHY, adding a convolutional code of rate 1/2, and are received with a hard-decision Viterbi decoder~\cite{viterbi67code}. In addition, the 125K PHY adds a Manchester pattern mapper of four elements per coded bit\footnote{When using \blefive125K PHY's Manchester Pattern Mapper, a `0' is translated into `0011', whereas a `1' is translated to `1100'~\cite{ble5specs, ble5distance}.}. The \ieee coding for devices operating in the 2.4\,GHz ISM band is based on DSSS. 

We define the Relative Frequency Offset (RFO) -- which determines the beating frequency -- as the difference between the CFO of each individual transmitter. 
We further define the power delta $\Delta$P as the power ratio with which both CT are received. The SNR is defined relative to the strongest transmission (assuming $P_{R1}>P_{R2}$) and $N$ being the noise power: \vspace{-3.00mm} \\
\begin{equation}
RFO(Hz) = |CFO_1 - CFO_2| = 1 / T_{Beating},
\end{equation}
\vspace{-4.00mm}
\begin{equation}
\Delta P(dB) = 10\log_ {10}(P_{R1}/P_{R2}),
\end{equation}
\vspace{-4.00mm}
\begin{equation}
SNR(dB) = 10\log_ {10}(P_{R1}/N).
\end{equation}

The BLE standard requires the CFO to be within $\pm$150\,kHz~\cite{ble5specs}, which results in RFOs lower than 300\,kHz. Therefore, the RFO is always lower than the symbol frequency (i.e.,  2\,MHz in \blefive2M and 1\,MHz for the other three \blefive PHYs). \newtext{Similarly, the \ieee standard limits the CFO to $\pm$100\,kHz~\cite{802154_ieee}.}
	
\newtext{\boldpar{Experimental setup} We build a simulator using MATLAB and release its code open-source\footnote{\url{https://github.com/ADEscobar/ct-simulator.git}}. Specifically, we follow these steps:}

\begin{enumerate}

\item \newtext{We start with the baseband model, typically available in MATLAB
for common communication systems, and use it to generate a random packet stream. Since we want to accurately characterize the beating effect within a symbol period, we obtain our baseband-equivalent stream, $B(t)$, with 8 samples per symbol.
}

\item \newtext{Then, we add to the baseband signal the effect of the amplitude distortion that occurs when two concurrent transmissions overlap in the air sending the same bitstream. When they are both received with the same energy ($\Delta P=0\,dB$), this is equivalent to introduce a periodical amplitude modulation of period equal to twice the beating period \cite{escobar20phd}:
\begin{equation}
	B_{CT\_2}(\Delta P=0\,dB)(t)=2B(t)\left\lvert\cos{(\pi t/T_{Beating})}\right\rvert
\end{equation}
If we consider energy differences between the two CT, the baseband-equivalent signal can be modeled as \cite{escobar20phd}:
\begin{equation}
	B_{CT\_2}(t)=B(t)\left(1 - A_2/A_1 + 2(A_2/A_1)\left\lvert\cos{(\pi t/T_{Beating})}\right\rvert\right)
\end{equation}
Where $A_1$ and $A_2$ denote the amplitude of the received signals, and $A_1 > A_2$:
\begin{equation}
P_{R1}/P_{R2} = A_1^2/A_2^2
\end{equation}
For more than two CT, the model of the amplitude distortion of the baseband envelope is much more complex, as sketched in Fig.~\ref{fig:ct-beating-complex-wide-strong} to Fig.~\ref{fig:ct-beating-complex-narrow-weak}. For simplicity, we will stick to two CT throughout the simulations, which gives illustrative-enough results for typical scenarios that can be modelled as having a dominant transmission.
}

\item \newtext{Finally, we recover the original bitstream by demodulating the distorted baseband signal, $B_{CT\_2}(t)$ with a suitable receiver. For the particular case of a simple non-coherent BFSK receiver, the bit error rate (BER) for $\Delta P=0\,dB$ can be obtained analytically \cite{escobar20phd}:
	\begin{equation}
		BER_{CT\_2\:BFSK}(\Delta P=0\,dB)  =\frac{1}{2}\exp(-E_{b0}/N_{0})\;\;\mathrm{I_{0}}\left(-E_{b0}/N_{0}\right)
	\end{equation}
where $\mathrm{I_{n}}(z)$ is the modified Bessel function of the
first kind. Analytical derivations of the BER when introducing energy imbalances, or for more complex communication systems such as the coded modes of BLE 5 and IEEE 802.15.4, are very complex. Obtaining the analytical packet error rate (PER) is also difficult even for the most simple case (BFSK), since bit errors are not randomly distributed and tend to appear during the energy valleys. Similarly, the shape of the envelope becomes too complex to be analytically tractable for more than two CT.}

\end{enumerate}

\begin{figure*}[!t]
	\vspace{-1.75mm}
	\begin{subfigure}[t]{0.485\columnwidth}
		\centering
		\includegraphics[width=0.875\columnwidth]{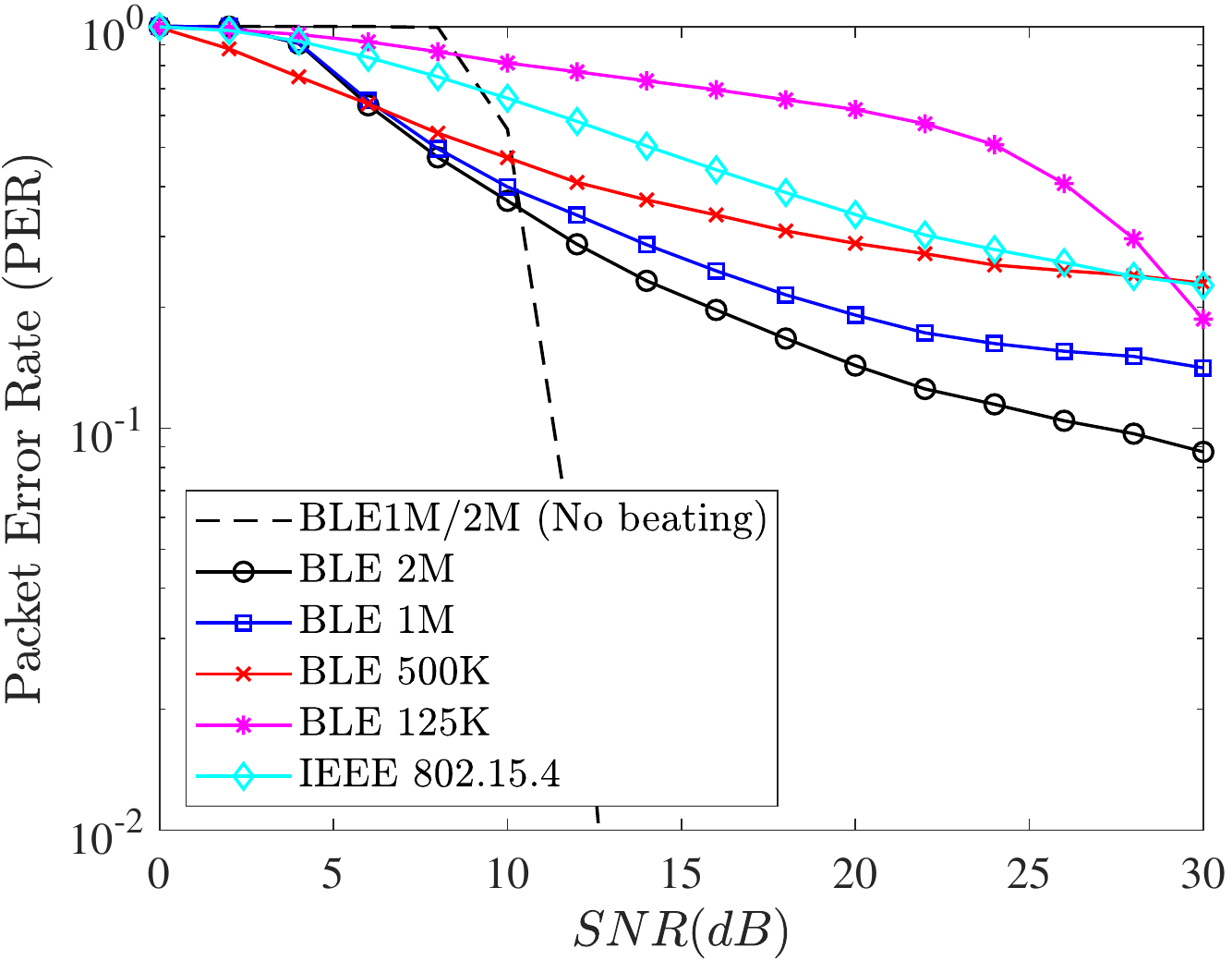}
		\vspace{-1.85mm}
		\caption{Wide \& strong beating (CT\_2, RFO=500Hz, $\Delta$P=0dB)}
		\label{fig:rednode_sim1}
	\end{subfigure}%
	\hfill
	\begin{subfigure}[t]{0.485\columnwidth}
		\centering
		\includegraphics[width=0.875\columnwidth]{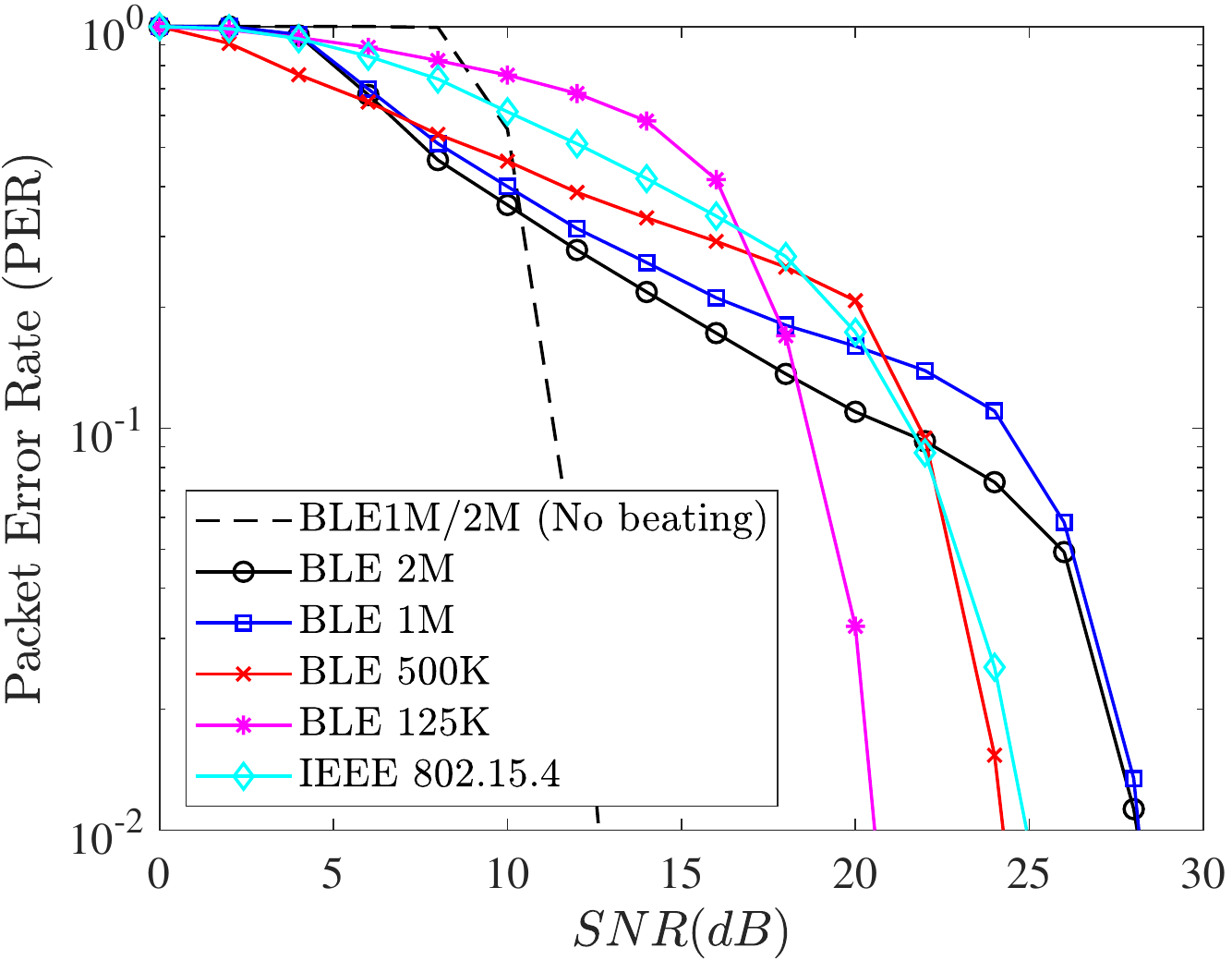}
		\vspace{-1.85mm}
		\caption{Wide \& weak beating (CT\_2, RFO=500Hz, $\Delta$P=1dB)}
		\label{fig:rednode_sim2}		
	\end{subfigure}%
    \vskip\baselineskip
    \vspace{-2.00mm}
	\begin{subfigure}[t]{0.485\columnwidth}
		\centering
		\includegraphics[width=0.875\columnwidth]{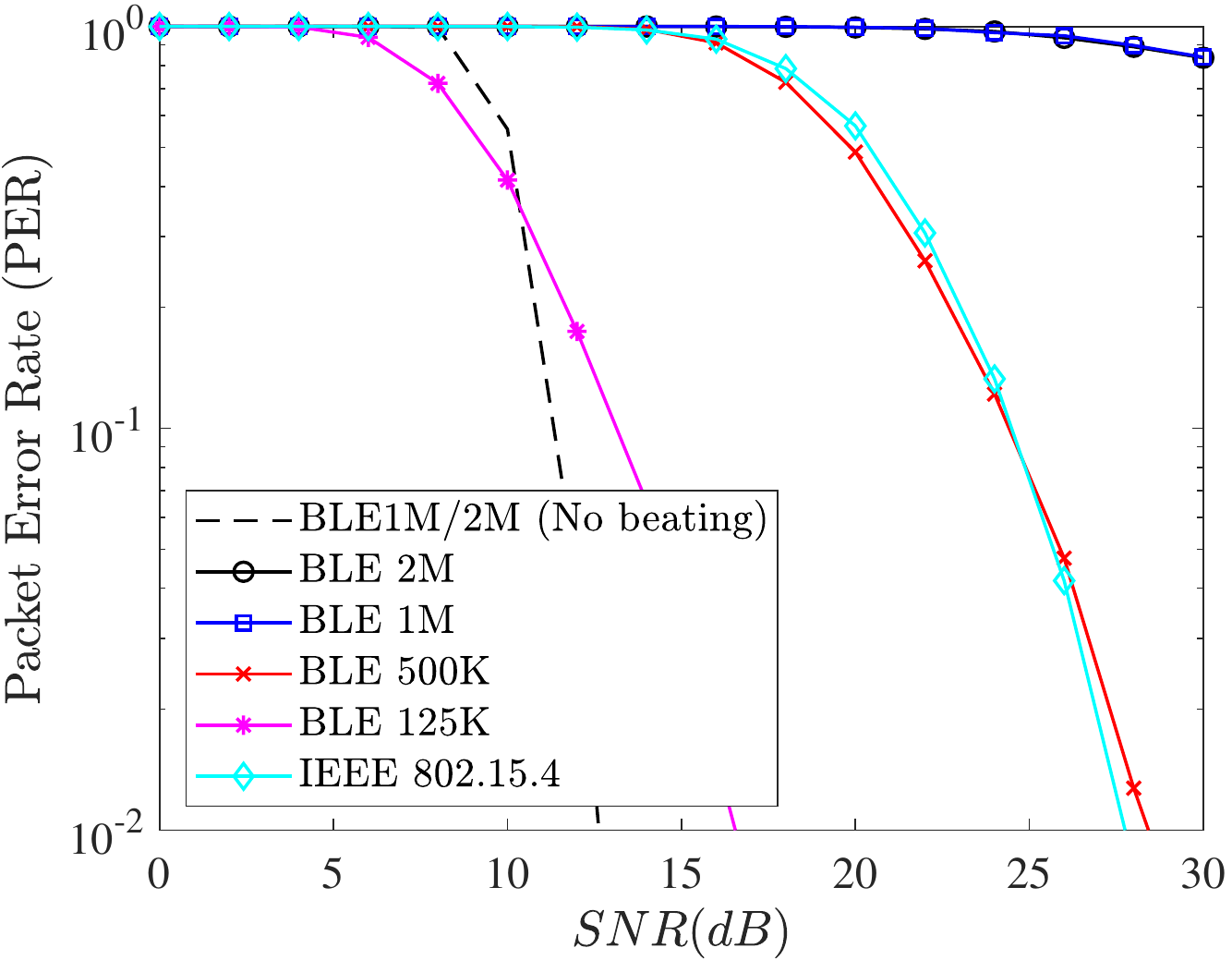}
		\vspace{-1.85mm}
		\caption{Narrow \& strong beating (CT\_2, RFO=10kHz, $\Delta$P=0dB)}
		\label{fig:rednode_sim3}		
	\end{subfigure}%
    \hfill
	\begin{subfigure}[t]{0.485\columnwidth}
		\centering
		\includegraphics[width=0.875\columnwidth]{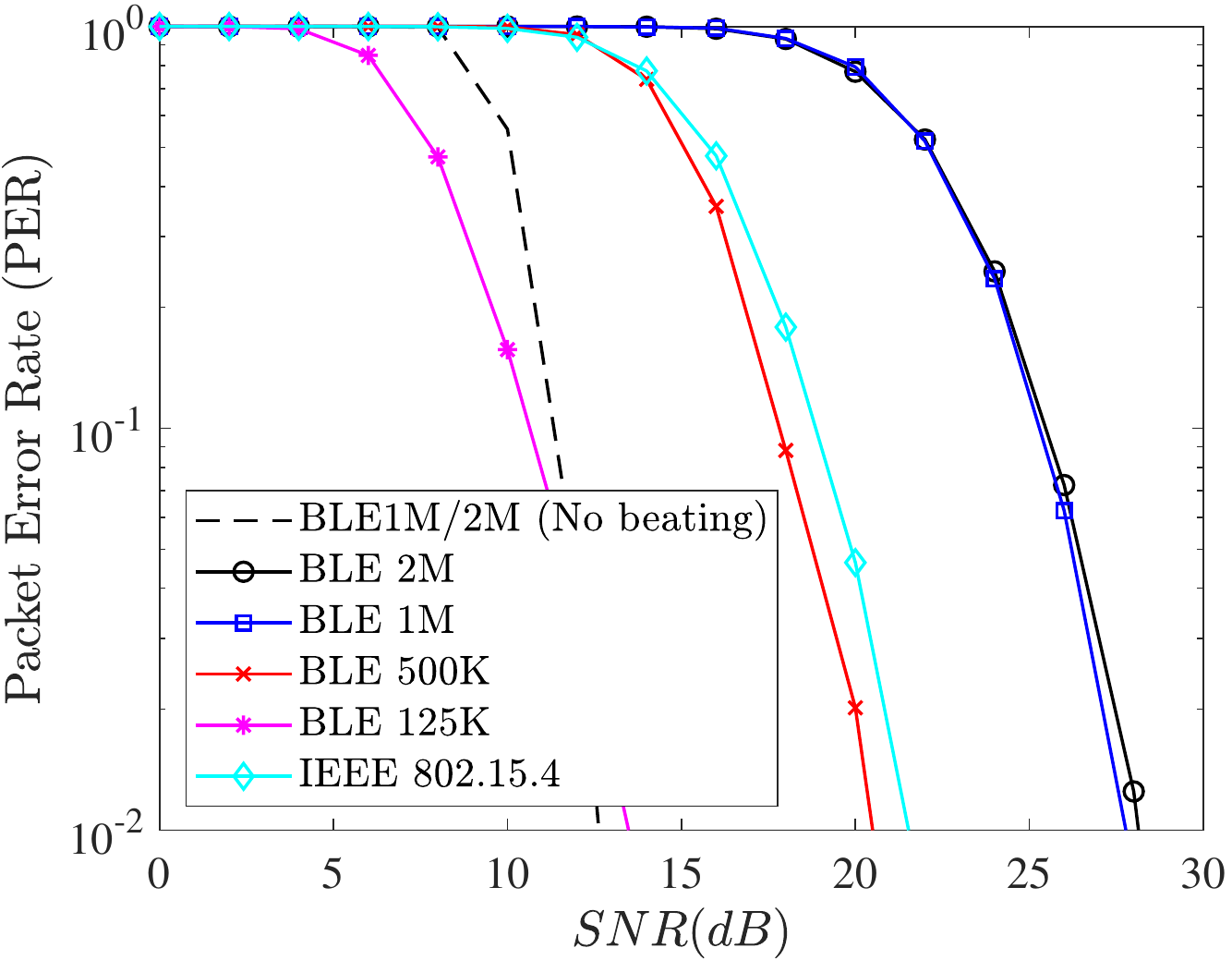}
		\vspace{-1.85mm}
		\caption{Narrow \& weak beating (CT\_2, RFO=10kHz, $\Delta$P=1dB)}
		\label{fig:rednode_sim4}		
	\end{subfigure}%
	\vskip\baselineskip
	\vspace{-2.00mm}
	\begin{subfigure}[t]{0.485\columnwidth}
		\centering
		\includegraphics[width=0.875\columnwidth]{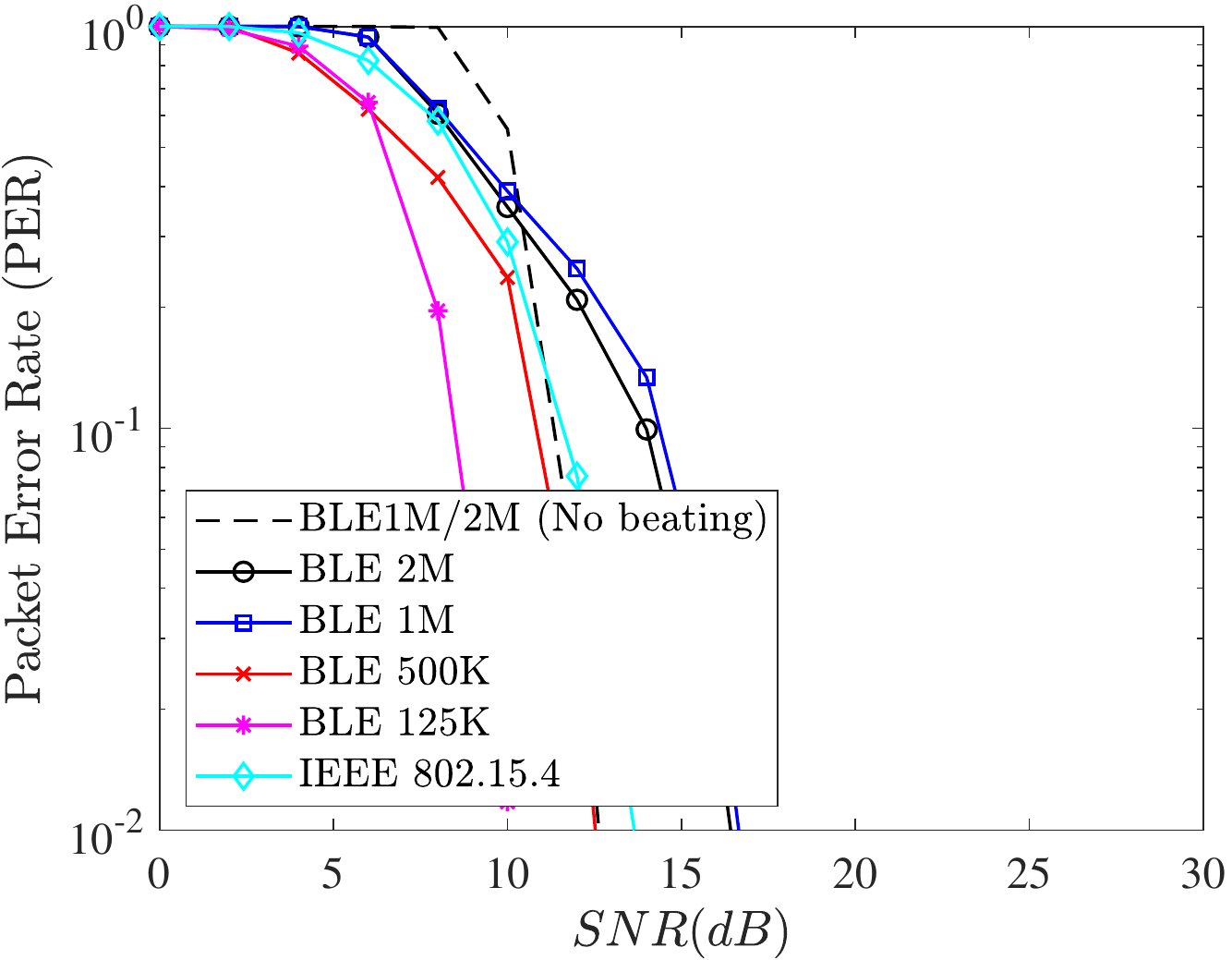}
		\vspace{-1.85mm}
		\caption{Wide\,beat.\,\&\,capture\,effect\,(CT\_2,\,RFO=500Hz,\,$\Delta$P=6dB)}
		\label{fig:rednode_sim5}		
	\end{subfigure}%
	\hfill
	\begin{subfigure}[t]{0.485\columnwidth}
		\centering
		\includegraphics[width=0.875\columnwidth]{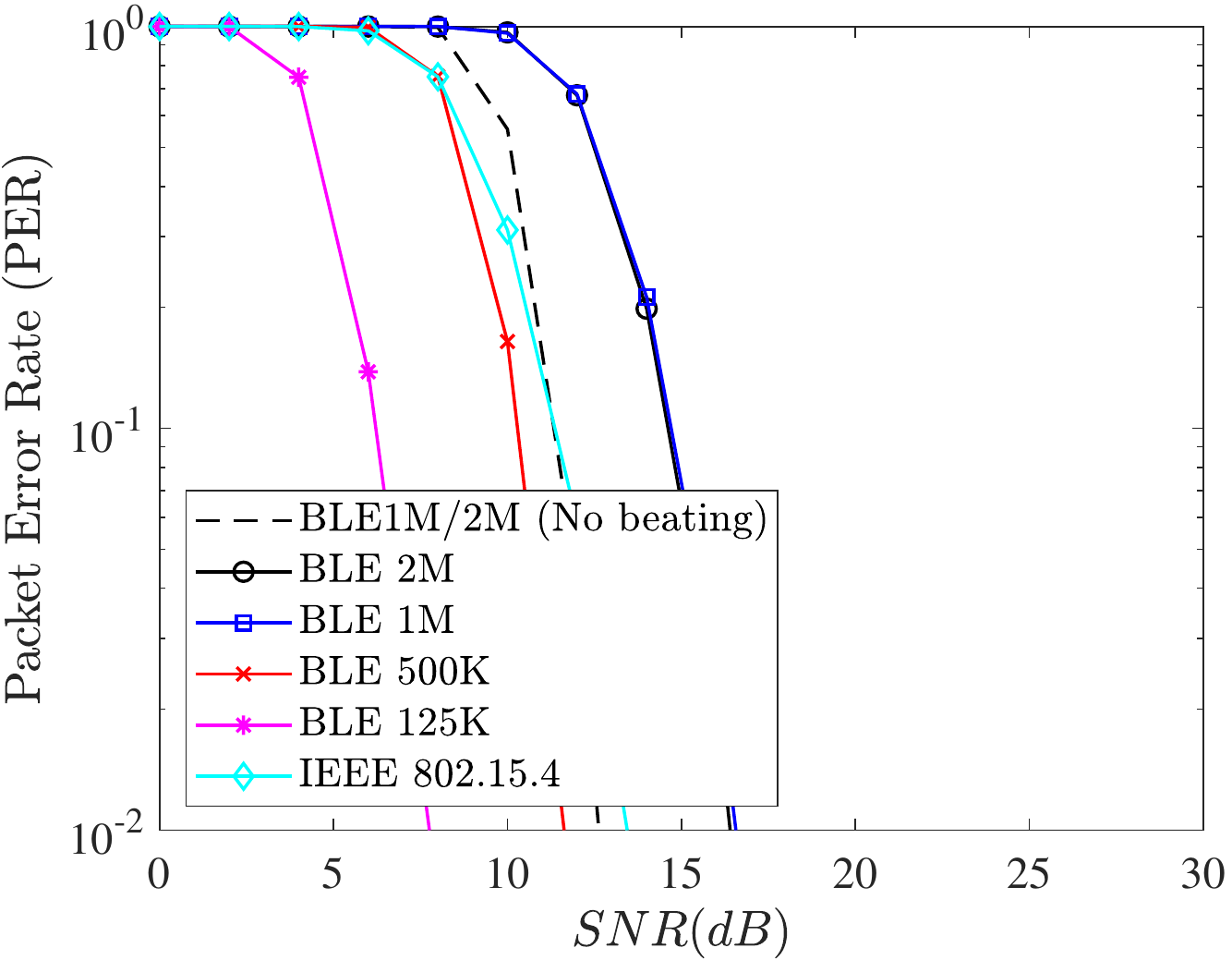}
		\vspace{-1.85mm}
		\caption{Narrow\,beat.\,\&\,capture\,effect\,(CT\_2,\,RFO=10kHz,\,$\Delta$P=6dB)}
		\label{fig:rednode_sim6}		
	\end{subfigure}%
	\vspace{-1.50mm}
	\caption{\textbf{Simulation results showing the impact of beating on the PER}. Simulating two concurrent transmitters with 30 byte packets. PER of a single transmitter (no beating) is also displayed for comparison. Differences in beating periods (wide or narrow) and power deltas have a significant impact on how beating affects performance. Different node pairs (with different RFOs) may therefore experience very different PERs even when operating with the same PHY and under similar conditions.}
	\label{fig:rednode_sim}
	\vspace{-4.75mm}
\end{figure*}

\begin{figure*}[!t]
	\vspace{-1.75mm}
	\begin{subfigure}[t]{0.485\columnwidth}
		\centering
		\includegraphics[width=0.875\columnwidth]{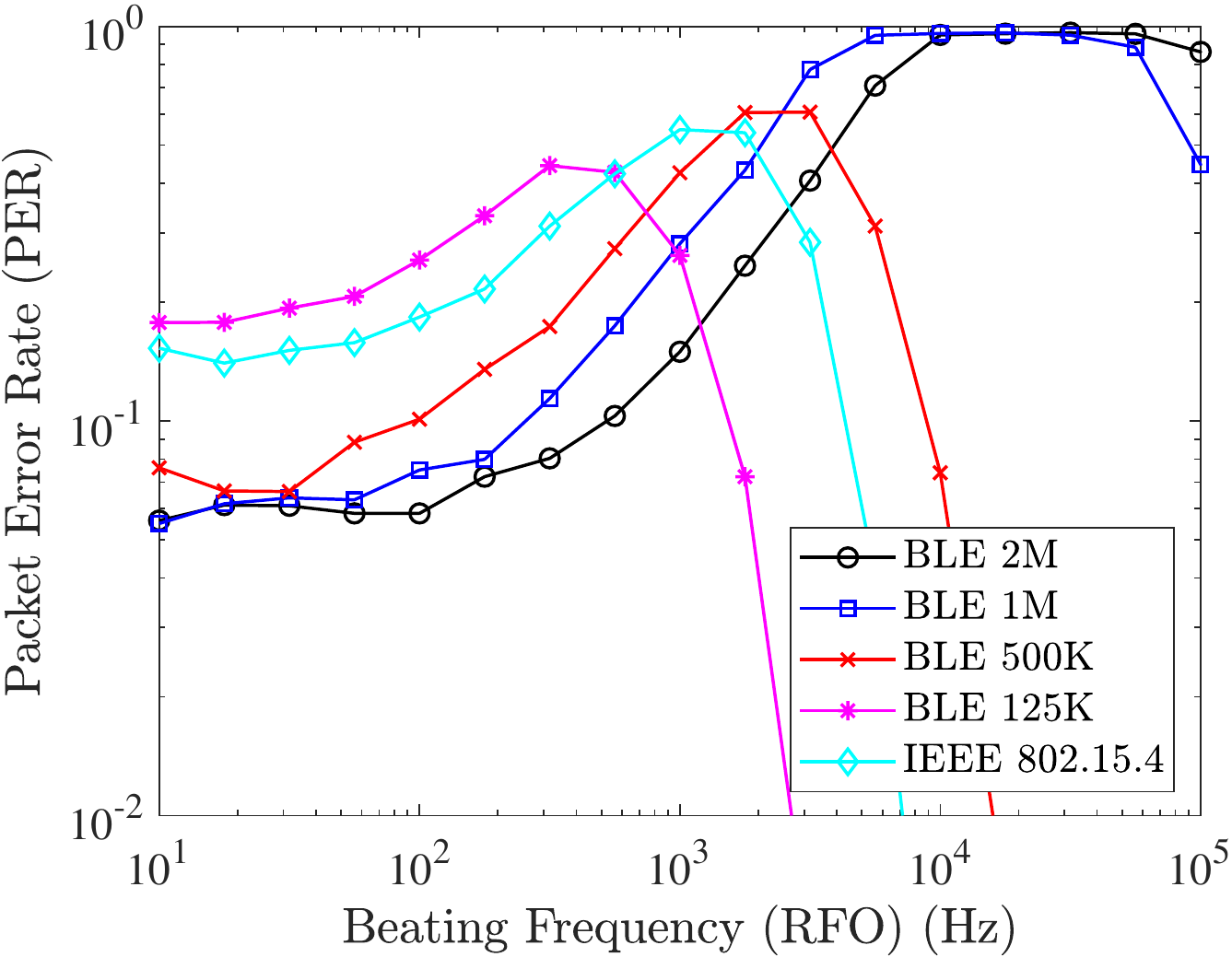}
		\vspace{-1.85mm}
		\caption{Strong beating (CT\_2, $\Delta$P=0dB, SNR=25dB)}
		\label{fig:rednode_sim1_freq}
	\end{subfigure}%
	\hfill
	\begin{subfigure}[t]{0.485\columnwidth}
		\centering
		\includegraphics[width=0.875\columnwidth]{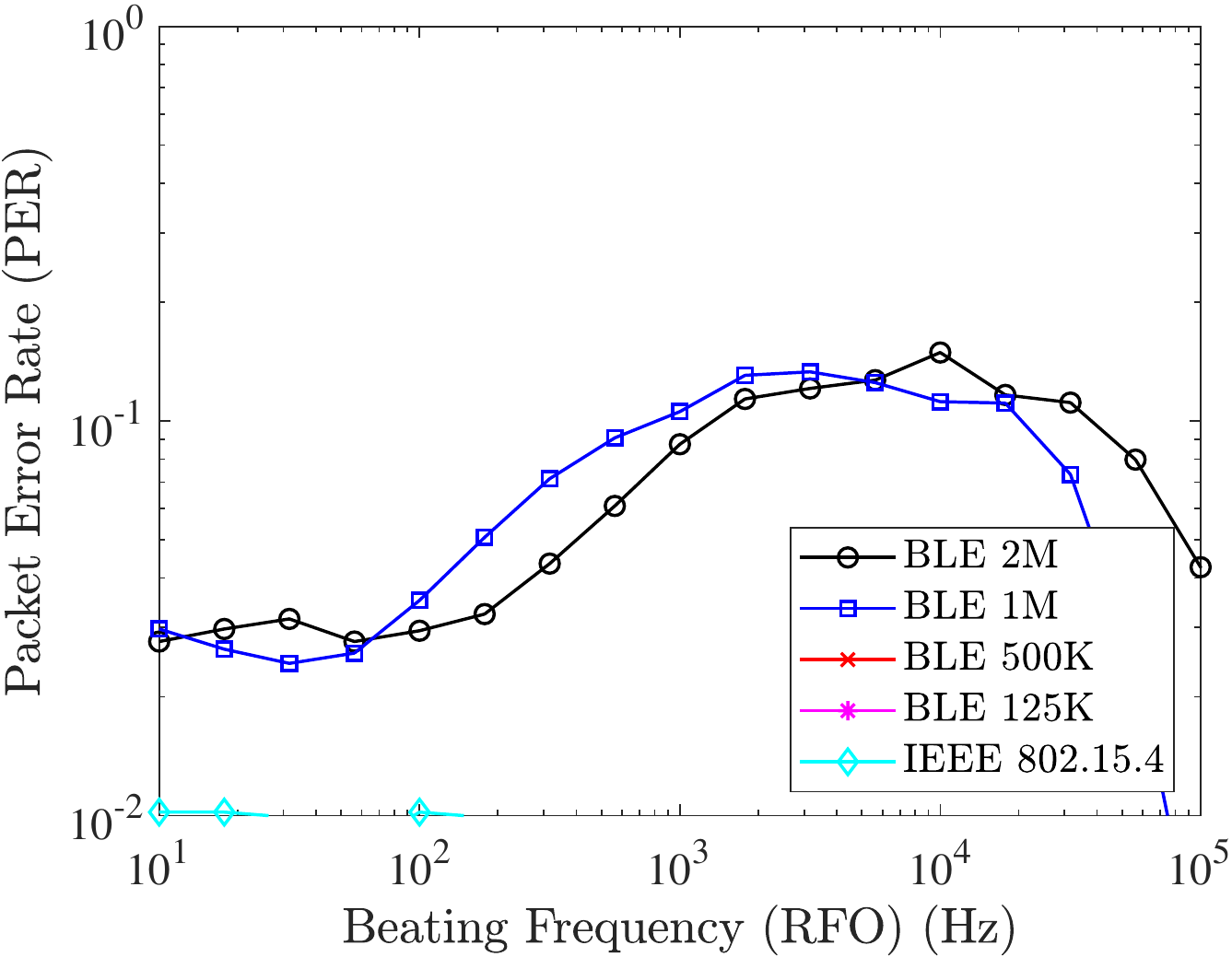}
		\vspace{-1.85mm}
		\caption{Weak beating (CT\_2, $\Delta$P=1dB, SNR=25dB)}
		\label{fig:rednode_sim2_freq}		
	\end{subfigure}%
	\vspace{-1.50mm}
	\caption{\textbf{Simulation results showing the impact of beating frequency (RFO) on the PER}. Simulating two concurrent transmitters with 30 byte packets. Uncoded PHYs (\blefive1M and 2M) perform better under low beating frequencies, while coded modes work optimally under high beating frequencies. Coded modes (\ieee, \blefive125K and 500K) experience an error peak in the middle of the frequency range, after which the coding becomes effective and the error decreases rapidly. Already with 1dB power delta (weak beating) the PER decreases greatly and it is constantly below 0.01 for the coded modes}
	\label{fig:rednode_sim_freq}
	\vspace{-1.75mm}
\end{figure*}

\noindent Fig.~\ref{fig:rednode_sim} and Fig.~\ref{fig:rednode_sim_freq} present the results of our simulation. \response{In general, beating increases the error rate in typical low-noise conditions; only in very harsh environments may reception chances improve due to the path diversity and energy gains during the peaks}. Based on these results, we derive the following observations: 

\begin{enumerate}[leftmargin=+5.75mm]
\item \textbf{Impact of beating.} We first compare the results obtained with concurrent transmissions from 2 nodes (CT\_2) with those obtained with a single transmitter (no beating). With low-noise (SNR\,$>$\,15\,dB), beating negatively affects packet reception and increases the PER. Only when operating in high-noise conditions (SNR\,$<$\,10\,dB), CT\_2 experience in some cases a PER lower than that of a single transmitter, due to the positive net effect of constructive interference intervals. CT are hence an optimal mechanism in harsh environments with high noise. 
Otherwise, the effect of destructive interference dominates and the PER increases.

\item \textbf{Wide ($\boldsymbol{T_{Beating}$\,$>$\,$T_{Packet}}$) and strong ($\boldsymbol{\Delta P$\,$\approx$\,0\,$dB}$) beating, Fig.~\ref{fig:rednode_sim1}}. In this case, $T_{Packet}$, which denotes the over-the-air time of a packet\footnote{$T_{Packet}$ depends on the PHY and it is computed as $T_{Packet} = B \cdot (1 / DR)$, with $DR$ being the effective data rate of the particular PHY and B being the length of the packet in bits.}, is the key factor dictating the PER. Indeed, the probability that the transmission spans a destructive interference interval is lower as the time the packet spends on the air decreases. Hence, when subjected to a same fixed $T_{Beating}$, uncoded PHYs (\blefive1M and 2M) perform better than coded ones (\ieee, \blefive125K and 500K), since the former benefit from faster transmissions. With wide energy valleys burst errors are frequent. The convolutional coding used by \blefive125K and 500K, and the DSSS used by \ieee are effective against discretely distributed one-bit errors, but not to correct contiguous errors. As a result, \blefive125K is the worst performing PHY, since it results in the longest packet duration.

\item \textbf{Narrow\,($\boldsymbol{T_{Beating}$\,$<$\,$T_{Packet}}$)\,and\,strong\,($\boldsymbol{\Delta P$\,$\approx$\,0\,$dB}$) beating, Fig.~\ref{fig:rednode_sim3}}. In this configuration, the packet transmission always spans one or more destructive valleys. \newtext{\blefive1M and 2M greatly struggle and feature a very high PER, since errors are frequent during the unavoidable narrow valleys, and since uncoded PHYs do not feature any built-in correction mechanisms. Contrarily, coding modes are very effective in narrow-beating conditions, since narrower valleys introduce shorter (potentially one-bit) error bursts that can be effectively corrected. \blefive125K is the best performing PHY, since it provides the most robust error-correction mechanism, while \blefive500K and \ieee perform similarly.}

\item \boldpar{Weak beating ($\boldsymbol{\Delta P}$\,$>$\,0\,dB), Fig.~\ref{fig:rednode_sim2} and~\ref{fig:rednode_sim4}} In real deployments, CT are normally received with dissimilar power levels. Under weak beating, the signal does not completely fade during the valleys, which greatly decreases the impact of beating on the PER, as can be seen in the simulations already with only 1\,dB difference.

\item \newtext{\textbf{Capture effect ($\boldsymbol{\Delta P}$\,$>$\,6\,dB), Fig.~\ref{fig:rednode_sim5} and~\ref{fig:rednode_sim6}}. For greater power dissimilarities, the well-known capture effect \cite{gezer2010capture}, i.e., the ability of the receiver to recover the message of the dominant transmission, typical of frequency modulations, kicks in. The PER tends to resemble the no-beating scenario as the power delta increases.}

\item \newtext{\textbf{Impact of the beating frequency, Fig.~\ref{fig:rednode_sim1_freq} and~\ref{fig:rednode_sim2_freq}}. Coded modes perform optimally for high beating frequencies (i.e., larger than 10\,kHz), since narrow valleys tend to create single-bit errors that can be properly corrected. In the low range, energy valleys span several symbols, errors cannot be corrected, and the uncoded modes (with shorter symbol durations) perform better. For weak beating, coded modes perform generally better for every beating frequency at the considered SNR of 25\,dB.}

\end{enumerate}

In real-world networks, all six scenarios depicted in Fig.~\ref{fig:rednode_sim} may simultaneously appear in different sections of a multi-hop network, depending on the practically unpredictable CFOs and power-level relationships between the concurrent transmitters. Even for a given link, conditions may change over time, since temperature alters the CFO and surrounding interference or multipath propagation cause dynamic fluctuations in the power deltas. It is hence desirable that an optimal PHY tailored to handle CT should have robust behavior in all scenarios, since controlling operating conditions within a wireless network is challenging. \newtext{As presented later with the experimental results, links can randomly have very different PERs depending on the strength and frequency of the beating they experiment; and the optimal PHY selection depends on the particular conditions of that link, like node layout, external noise level, and pairwise RFOs.}

\boldpar{Observations for beating mitigation} 
Based on these simulation results, we infer that beating can be mitigated by boosting energy diversity with proper node placement and techniques to dynamically control the transmission power. Nonetheless, this would affect the main advantages of using CT-based protocols: scalability and simplicity. \newtext{Uncoded PHYs are optimal in wide-beating conditions, since shorter packet durations combined with packet repetitions are likely to randomly trigger a successful transmission spanning an energy peak. Conversely, in narrow-beating conditions, coded PHYs behave optimally, since they are able to correct sparse errors. \blefive 1M and 2M are optimal for wide beating, but they behave poorly under narrow beating, while the opposite holds true for \blefive125K. \blefive500K and \ieee constitute an interesting trade-off between data rate and PER in both narrow and wide beating. A system optimally designed to operate under beating should feature burst error-correcting mechanisms, such as interleaved coding, which is missing in the analyzed low-power protocols.
}

\newtext{To focus the discussion on the beating itself, different noise patterns, such as intermittent jamming or additional source of errors, such as synchronization mismatches between the CT, have not been considered in our simulation, but they have an impact in the testbed results presented later, particularly if the retransmissions are not synchronized within half the symbol period~\cite{escobar19imprel}, e.g., 0.5$\,\mu$s for \blefive1M.}

\fakepar{}
\newtext{
Next, we confirm our simulation results on real hardware using \emph{over-the-air} experiments. Based on the simulations, there is the potential to experimentally identify if a link is operating under narrow or wide beating by comparing the PER performance in the different radio modes. Since the initial phase of the beating is random for each packet transmission, we expect the errors to be randomly scattered within a packet transmission. Surprisingly, experimental results show differently, and we experience a fluctuating error probability when we map the distribution of bit errors across a packet with a frequency matching the expected beating pattern, which we think is coming from the calibration that the receiver performs during the reception of the packet preamble. This is a simple and effective way to obtain and confirm the impact of the beating frequency without requiring additional hardware.}
%

\section{CT Performance over different PHY\lowercase{s}: \\Experimental Evaluation of Bit Error Distribution} \label{sec:beating}
Early CT literature has attributed gains seen at the receiver to constructive interference (CI)~\cite{ferrari2011efficient}. More recent works~\cite{liao2016revisiting, escobar19imprel}, in addition to this paper's analysis in Sect.~\ref{sec:simulation}, have proposed that, contrary to this assertion, instances of few concurrent transmitters will result in beating (observed as periodic peaks and valleys across a waveform) due to innate CFO inaccuracies between devices defined as RFO in Sect.~\ref{sec:simulation}. Rather than a CI gain, beating causes periods of \emph{both} constructive \emph{and} destructive interference across the packet, leading to errors during beating valleys, while beating peaks will benefit from a receiver gain. 
This has recently been demonstrated experimentally by observing the raw in-phase and quadrature~(IQ) samples when connecting a small number of CT devices to an SDR using coaxial cables~\cite{alnahas2020blueflood}. Although these efforts help to better explain some of the processes underpinning CT-based communication, it is hard to directly witness and evaluate how the occurrence of fundamental physical layer properties affect CT performance.
To date, there has been no \emph{over-the-air} testbed experiment able to demonstrate how PHY effects, such as RFO-induced beating and de-synchronization due to clock drift, directly affect the signal observed by a receiving node.

To address this gap, we present experiments that evaluate PHY effects on a 1-hop network of nodes communicating wirelessly by means of CT. 
Given that PHY properties cause the error probability to flux across the received packet (as shown in Sect.~\ref{sec:simulation}), we \emph{observe beating by mapping the distribution of bit errors across a packet} when considering a large transmission sample. 
Specifically, we study the CT performance across the multiple available PHYs supported by the nRF52840 devices in \dcubee. 
We observe both beating frequencies and de-synchronization effects through analysis of the received error distribution, and demonstrate that their impact on CT performance is highly dependent on the choice of the underlying PHY, on the RFO between transmitting devices, and on the number of concurrent transmitters.
\newtext{Furthermore we extend the \dcubee testbed with TempLab temperature control modules~\cite{boano2014templab}, and observe how changes in temperature at transmitters and receivers can affect the impact of beating.}
We perform all experiments using the \emph{Atomic-SDN} CT stack developed for the EWSN 2019 Dependability Competition~\cite{baddeley2019atomic, baddeley2020thesis, baddeley2019competition}.

\subsection{Experimental Setup}
\begin{figure}[t]
	\centering
	\includegraphics[width=0.5\columnwidth]{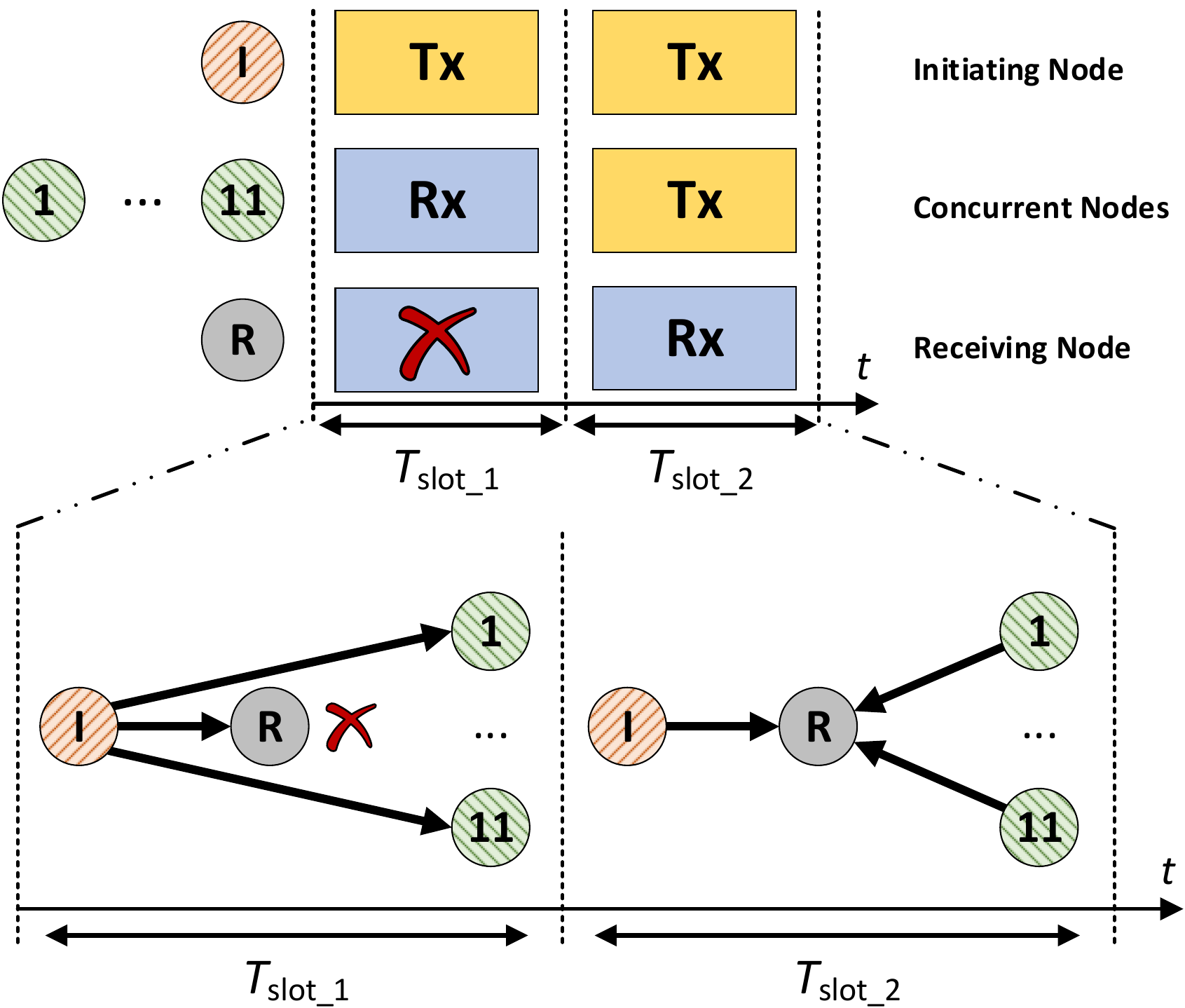}
	\vspace{-3.25mm}
	\caption{\textbf{Experimental setup.} An initiating node $I$ synchronizes transmitting nodes, allowing a receiving node $R$ to examine the 1-hop CT (CT\_2 -- CT\_12) transmissions.}
	\label{fig:beating_experimental_setup}
\end{figure}

We configure the \dcubee testbed in order to provide a single-hop scenario for up to $12$ concurrently-transmitting nodes and a fixed receiving node. 
Fig.~\ref{fig:beating_experimental_setup} demonstrates this setup and shows how the network is able to synchronize all transmitting nodes while limiting packet receptions at the receiver to only those that are a sum of signals from multiple concurrent transmitters. 
Node $R$ ignores the first transmission from the CT initiator $I$ in $T_{slot\_1}$, while allowing concurrent nodes to receive and synchronize to $I$. In $T_{slot\_2}$ all concurrent nodes synchronously transmit, including the initiator, and are observed at $R$. 

\newtext{
\boldpar{Over-the-air experiments}
}
We firstly perform over-the-air experiments to compare the effect of beating on different PHYs, as well as the number (density) of concurrent transmitters. Node $I$ is configured to periodically generate and transmit a pseudo-random payload every 250\,ms, which is logged before each transmission in $T_{slot\_1}$, alongside an 8-byte CT header. When using \ieee, this corresponds to a 119\,B payload (due to the 127\,B maximum transmission unit limitation), while a payload of 200\,B is used 
\newtext{when comparing beating across the various \blefive PHYs, which allows sufficient time to capture beating effects while reasonably limiting transmission time when using the \blefive125K PHY.} 
At the receiving node $R$, the byte arrays of all correctly and incorrectly received packets are logged in $T_{slot\_2}$. A direct comparison of transmitted and received byte arrays subsequently allows observation of the following metrics.

\vspace{+0.50mm}
\begin{enumerate}[leftmargin=+5.75mm]
	\setlength\itemsep{-0.10em}
	\item\emph{Bit error distribution} -- error distribution across a packet exposes periods of gain and interference over time. 
	\item\emph{Packet Reception Ratio (PRR)} -- indicating the ability of the physical layer to recover from CT interference. 
	\item\emph{Packet Error Ratio (PER)} -- indicating CT interference with a packet after the correct reception of the preamble.
	\item\emph{Packet Loss Ratio (PLR)} -- an energy minimum during a preamble results in the radio discarding the packet.
\end{enumerate}
\vspace{+1.00mm}

\response{Collection of the bit error distribution allows passive observation of beating-induced errors, and is the main focus of this section. Additionally, while PRR or PER are commonly used metrics for measuring protocol performance (typically, $PRR = 1 - PER$) it is useful to also consider PLR when comparing different physical layers. Specifically, periods of beating energy minima can potentially result in receivers unable to detect a packet preamble i.e., $PRR = 1 - (PER + PLR)$.

We run this setup across the \ieee and all the \blefive PHYs for concurrent transmissions from 2, 3, 4, 6, 8, 10, and 12 nodes at -8\,dBm\footnote{We select a transmission power of -8\,dBm, as it allows to capture sufficient bit errors within a reasonably short time window, while minimizing experimentation time.}, where we define the CT density as CT\_2, CT\_3,~...~CT\_12 respectively.}  Each experiment runs for $\approx$18K transmitted packets, representing over 100 total hours of experimental data. 

\begin{figure}[t]
	\centering
	\begin{subfigure}[t]{0.35\textwidth}
		\centering
		\includegraphics[width=1\columnwidth]{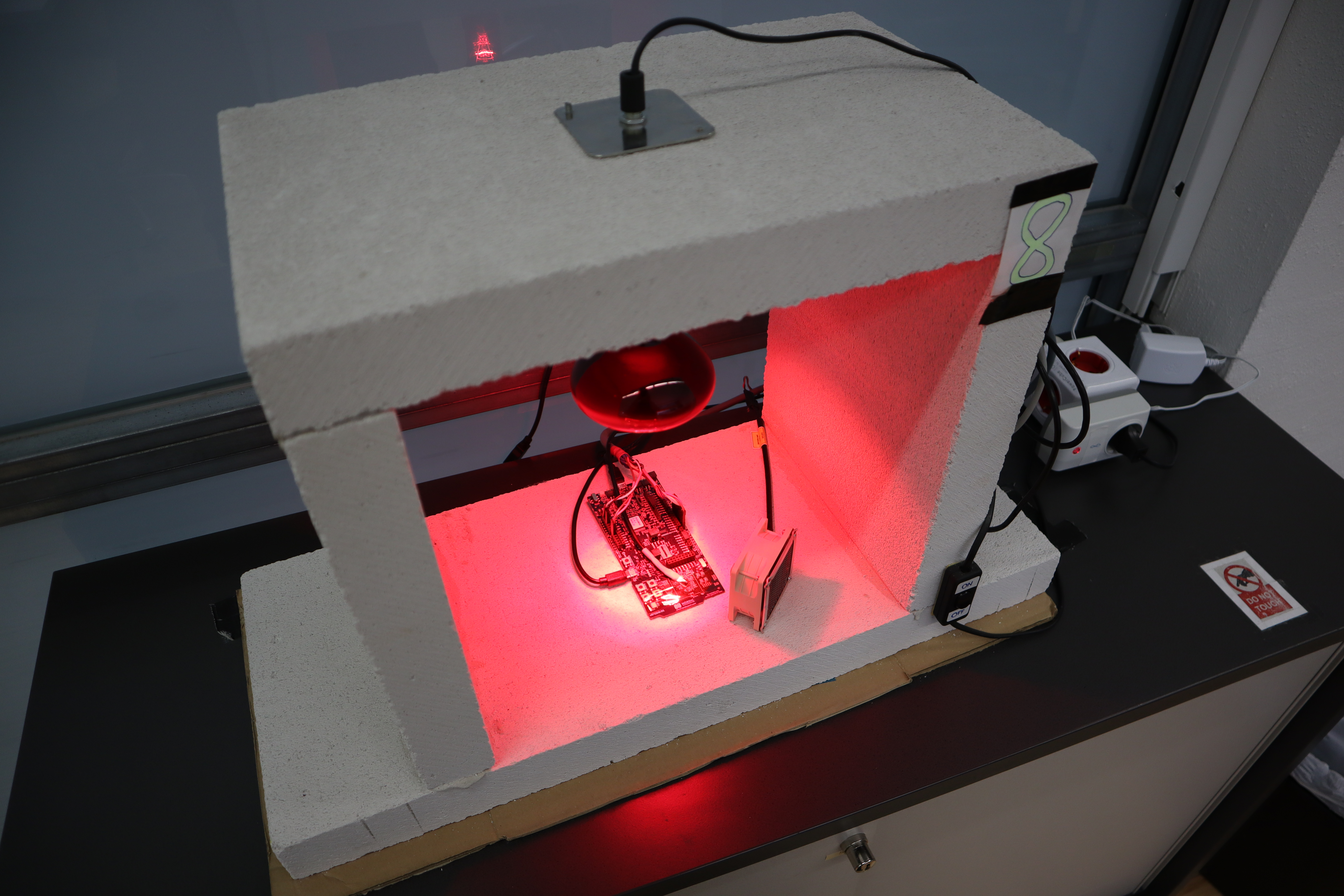}	
		\vspace{-5.75mm}
	\end{subfigure}%
	\hspace{+10.00mm}
	\begin{subfigure}[t]{0.42\textwidth}
		\centering
		\includegraphics[width=1\columnwidth]{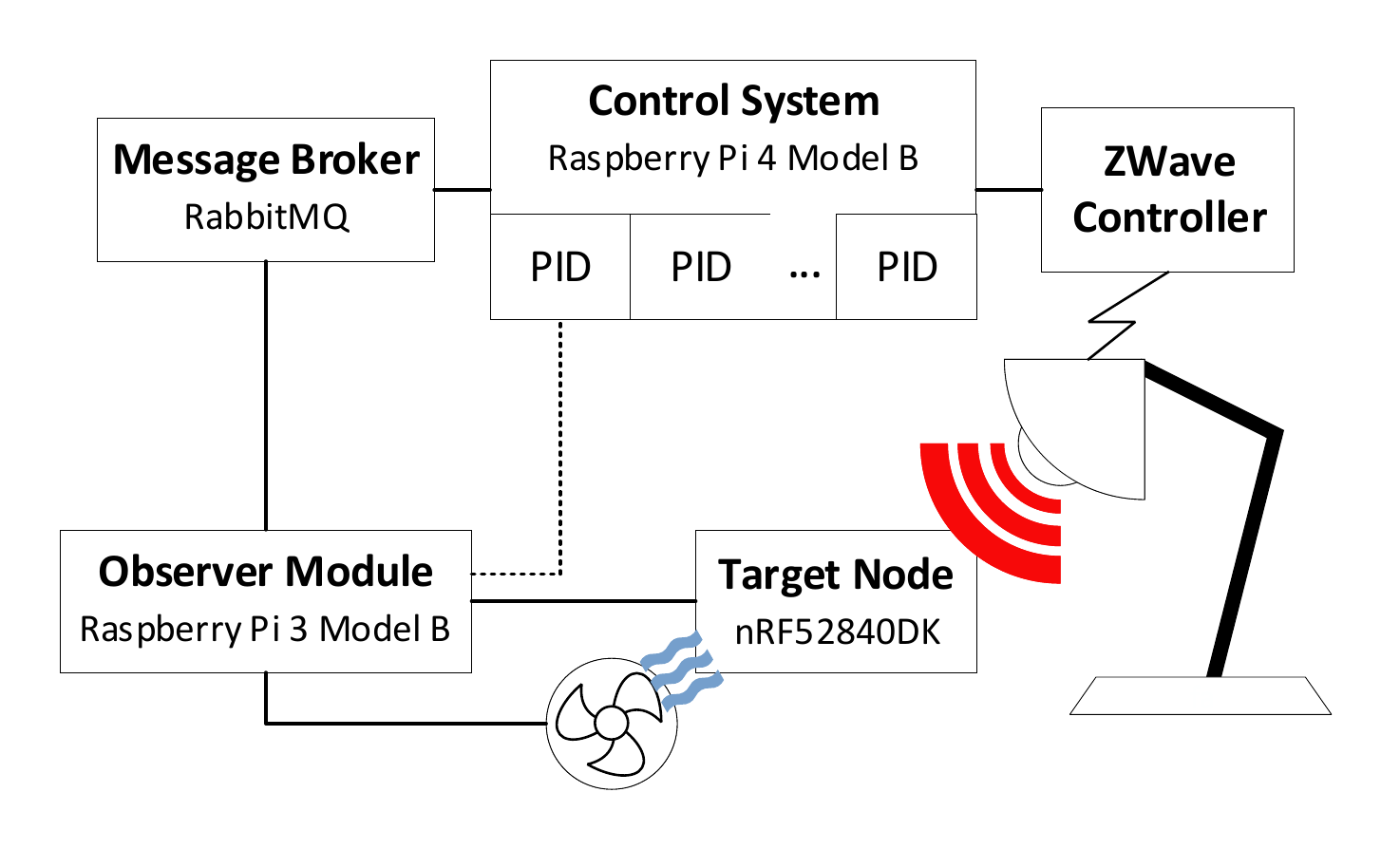}	
		\vspace{-5.75mm}
	\end{subfigure}
	\vspace{-2.75mm}
	\caption{\textbf{Enhanced observer module in \dcubee.} An \texttt{nRF52840DK} node equipped with TempLab's temperature sensor, heating lamp, and cooling fan. Each observer module is controlled by an individual PID instance on a central control system.}
	\label{fig:templab_node}
\end{figure}

\newtext{
\boldpar{Temperature-controlled experiments}
Furthermore, we perform experiments on temperature-controlled nodes, connected via coaxial cabling, to examine the impact of temperature on the beating effect (and, consequently, on CT performance). We utilize the TempLab design outlined in~\cite{boano2014templab}, which are equipped with an extended variant of \dcubee's observer modules -- with additional input for a \texttt{DS18B20+} temperature sensor as well a controllable fan. 
A Raspberry\,Pi\,4\,Model\,B implements a closed-loop control system with an individual PID controller per node directly actuating an incandescent heat lamps operated on a wireless dimmer circuit mounted directly above each node. 
The dimmer uses Z-Wave technology, which is based on a sub-GHz radio, and thus does not affect the communication of the nodes. Fig.~\ref{fig:templab_node} shows a single observer module enhanced with TempLab, where the porous concrete isolation retains little heat and serves as fire retardant.
Each observer module reports its temperature measurements from the \texttt{DS18B20+} temperature sensor to a \texttt{templab} topic on the testbed's message broker. The Raspberry\,Pi\,4\,Model\,B running the control system subscribes to these measurements from the message broker and feeds them into the PID controller responsible for the adjustment of the corresponding observer modules' temperature. 
Using the temperature profile as set value, the PID either turns on the heating lamp using the dimmer in order to increase the temperature, or sends back commands through the broker to the observer module so it actuates the attached fan in order to decrease the nodes' temperature.}
\newtext{
We subsequently use this setup to measure the impact of temperature on beating patterns between node pairs: we do so by varying the temperature at a \emph{single} transmitter between 30$^{\circ}$C and 75$^{\circ}$C at 1$^{\circ}$C intervals. We transmit over \blefive500K
as the beating sensitivity of this PHY provides clearer beating patterns within the histogram (as we show in Fig.~\ref{fig:tosh_layoutcomp}) and is thus more quantifiable by an FFT. Each 1$^{\circ}$C interval lasts 10 minutes, where we send 255 byte packets every 333ms\footnote{In contrast to experiments comparing beating over different PHYs, using \emph{only} \blefive500K ensures that use of the full MTU doesn't take excessive time, and allows collection of a greater number of samples.}, for a total of $\approx$1800 samples per interval. This experiment is run over all possible transmitting pair / destination combinations between four nodes $\{1, 2, 3, 4\}$, with each transmitting pair connected to the a single receiving node using coaxial cables to remove multipath effects, and we make these data publicly available\footnote{\url{https://zenodo.org/record/5882965}}.}


\subsection{Results}
\label{subsec:beating_results}

\begin{figure}[t]
	\centering
	\begin{subfigure}[t]{0.3\textwidth}
		\centering
		\includegraphics[width=1\columnwidth]{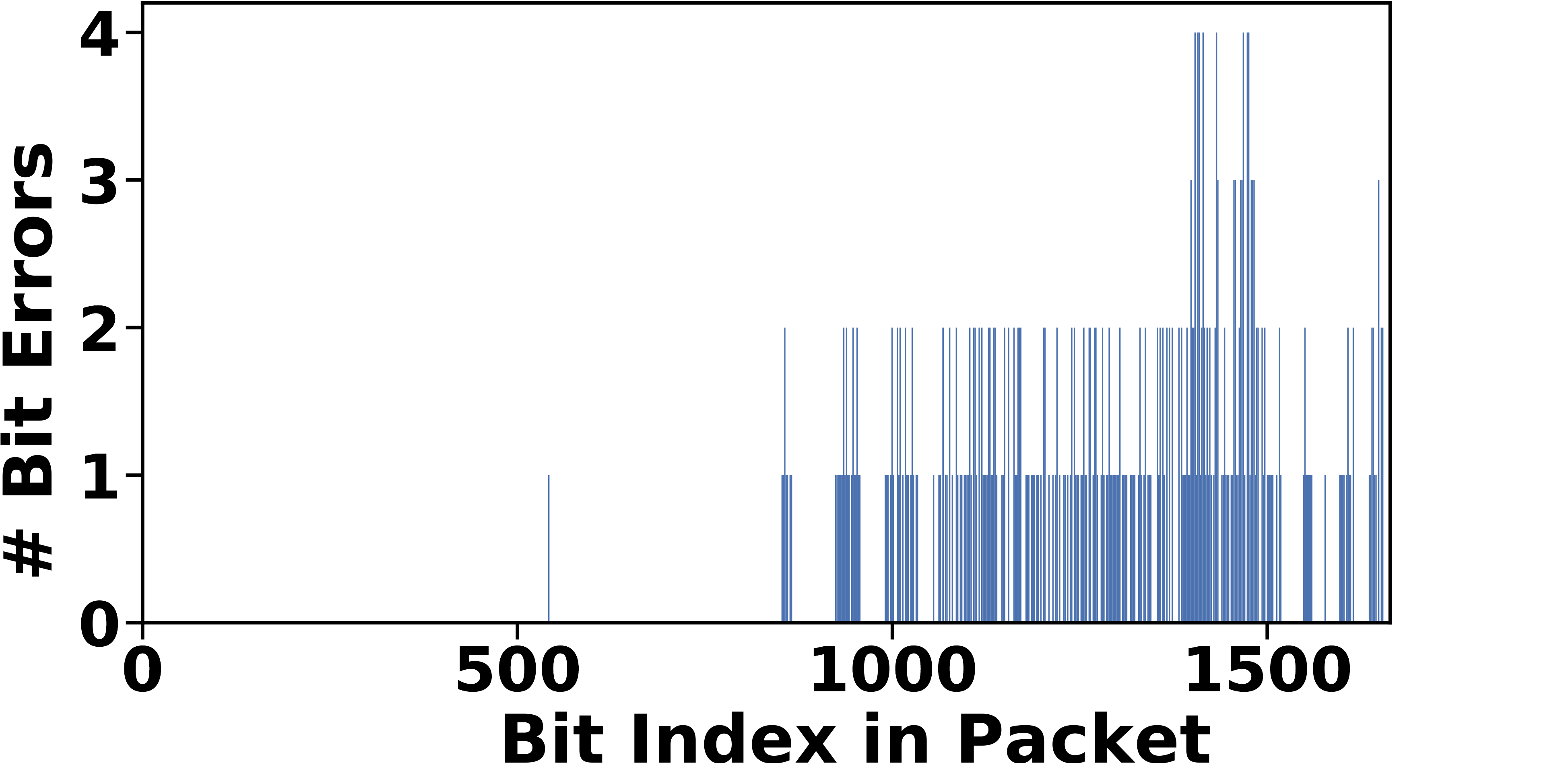}
		\vspace{-5.75mm}
		\caption{\blefive125K}
		\label{fig:tosh_phycomp_125k}
	\end{subfigure}%
	\begin{subfigure}[t]{0.32\textwidth}
		\centering
		\includegraphics[width=1\columnwidth]{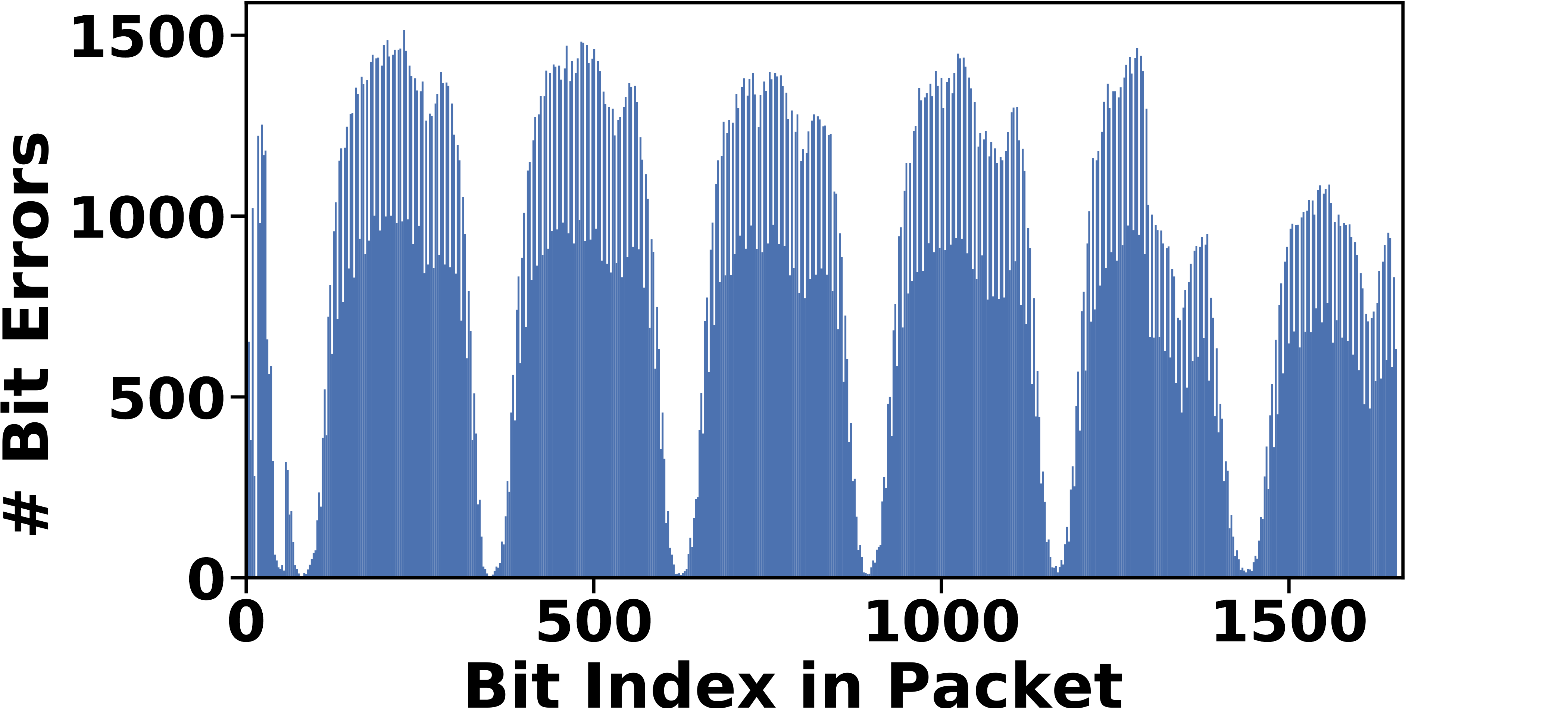}
		\vspace{-5.75mm}
		\caption{\blefive500K}
		\label{fig:tosh_phycomp_500k}
	\end{subfigure}
	\begin{subfigure}[t]{0.32\textwidth}
		\centering
		\includegraphics[width=1\columnwidth]{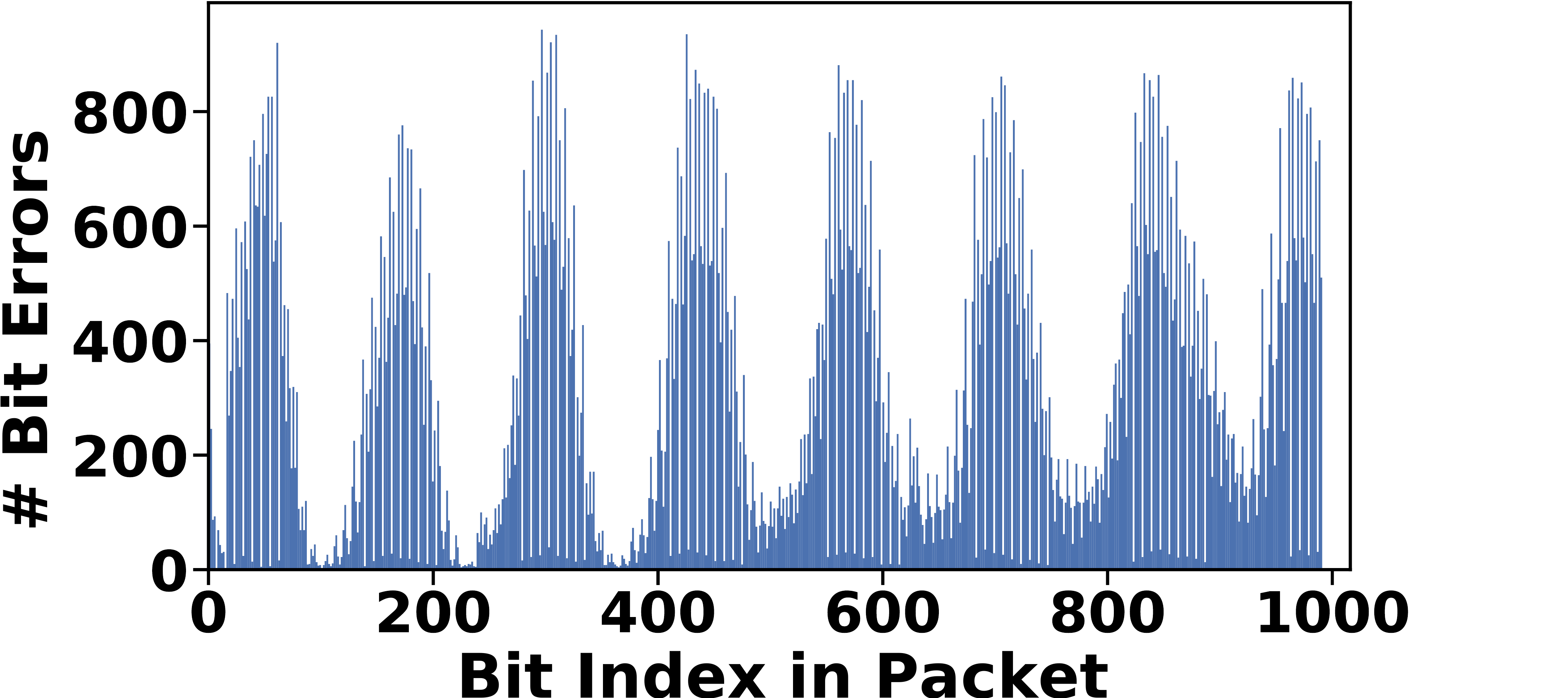}
		\vspace{-5.75mm}
		\caption{\ieee}
		\label{fig:tosh_phycomp_802154}
	\end{subfigure}
	\begin{subfigure}[t]{0.32\textwidth}
		\centering
		\includegraphics[width=1\columnwidth]{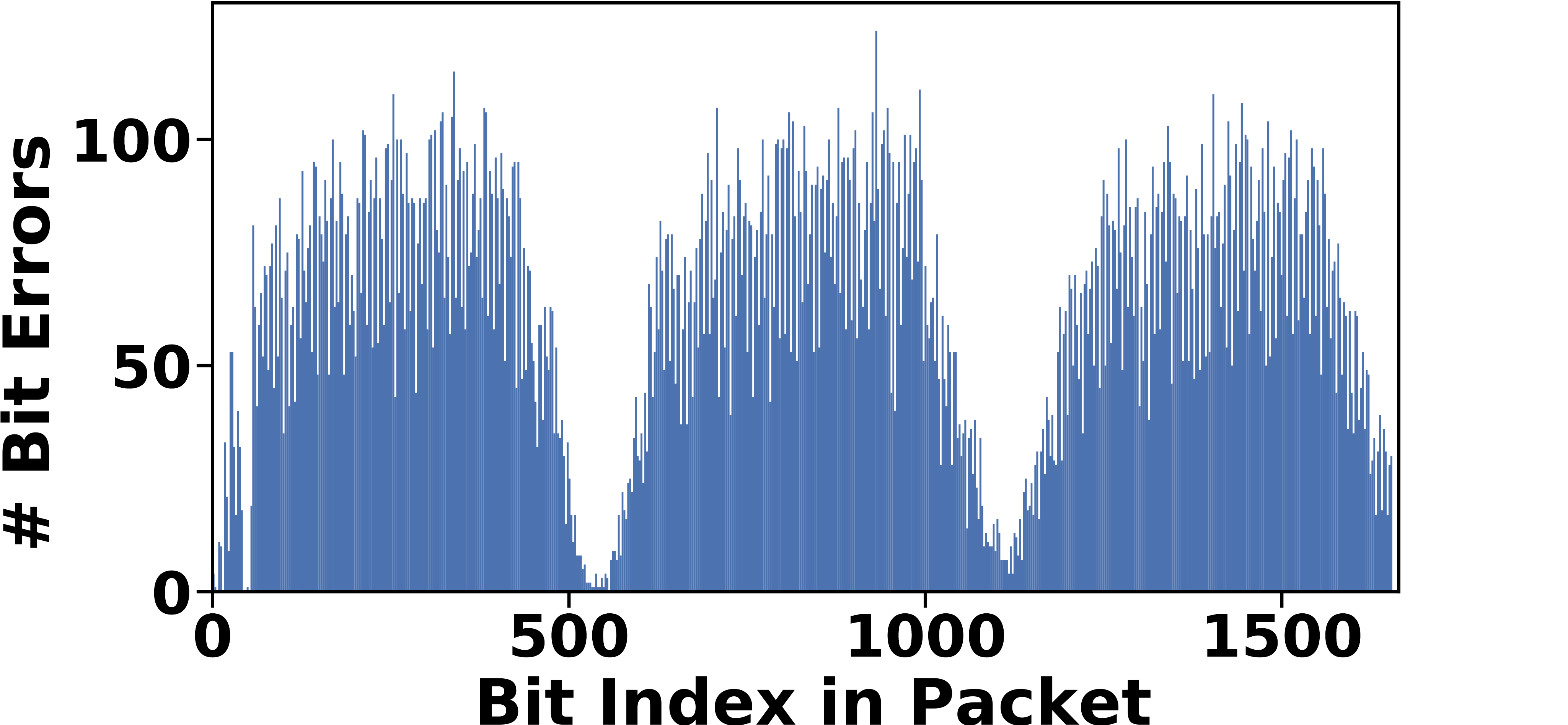}
		\vspace{-5.75mm}
		\caption{\blefive1M}
		\label{fig:tosh_phycomp_1m}
	\end{subfigure}
	\begin{subfigure}[t]{0.32\textwidth}
		\centering
		\includegraphics[width=1\columnwidth]{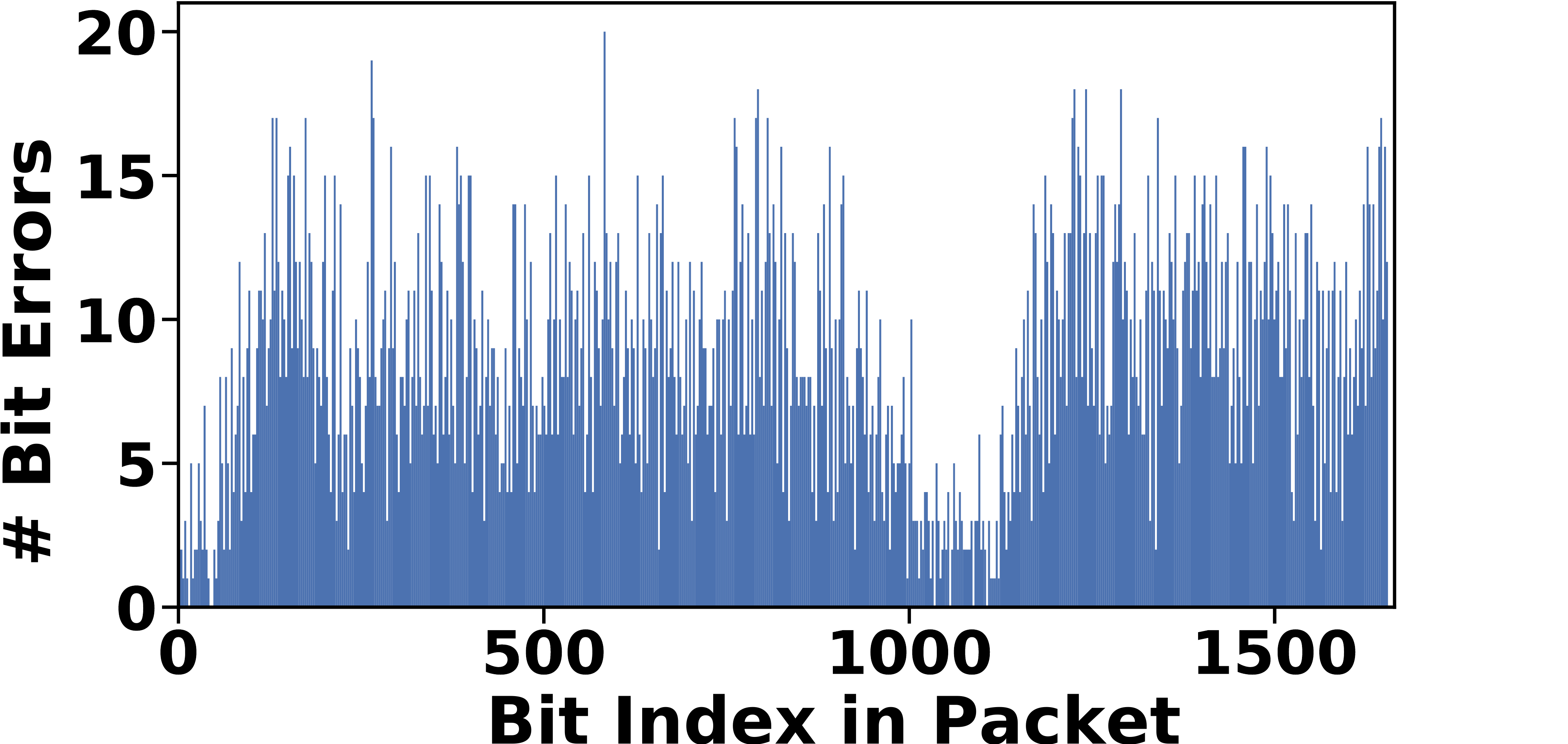}
		\vspace{-5.75mm}
		\caption{\blefive2M}
		\label{fig:tosh_phycomp_2m}
	\end{subfigure}
	\vspace{-3.75mm}
	\caption{\textbf{A 2\,kHz beating frequency can be seen from the \textit{same} CT pair across multiple \textit{different} PHY layers.} A histogram of bit errors across many packets shows the beating frequency. Low data-rate PHYs experience \emph{narrower} beating as they span (in time) a greater number of error peaks. The coding used in \blefive125K handles these errors.}
	\label{fig:tosh_phycomp}
\end{figure}

\begin{figure}[t]
	\centering
	\begin{subfigure}[t]{0.32\textwidth}
		\centering
		\includegraphics[width=1\columnwidth]{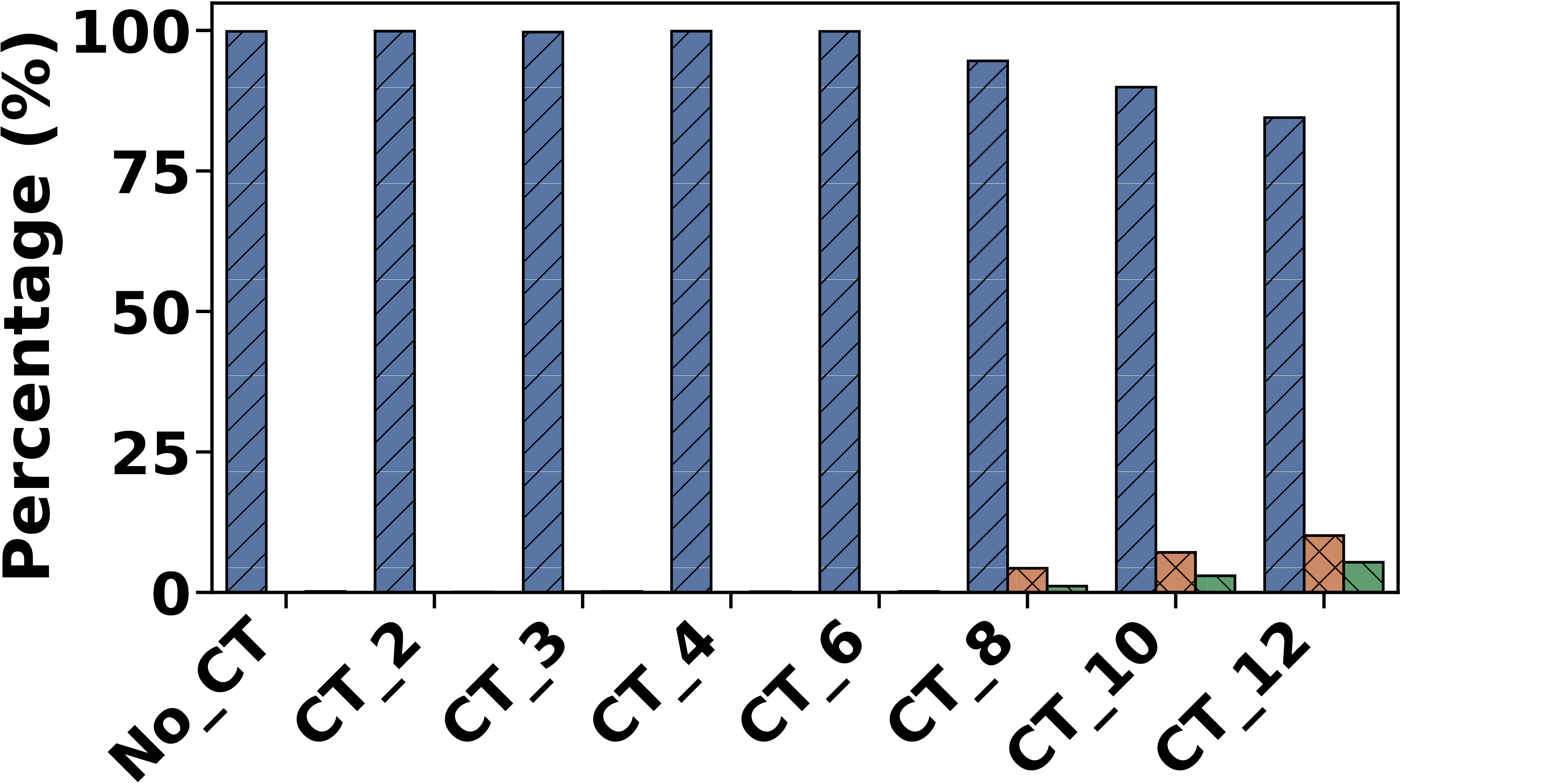}
		\vspace{-5.75mm}
		\caption{\blefive125K}
	\end{subfigure}%
	\begin{subfigure}[t]{0.32\textwidth}
		\centering
		\includegraphics[width=1\columnwidth]{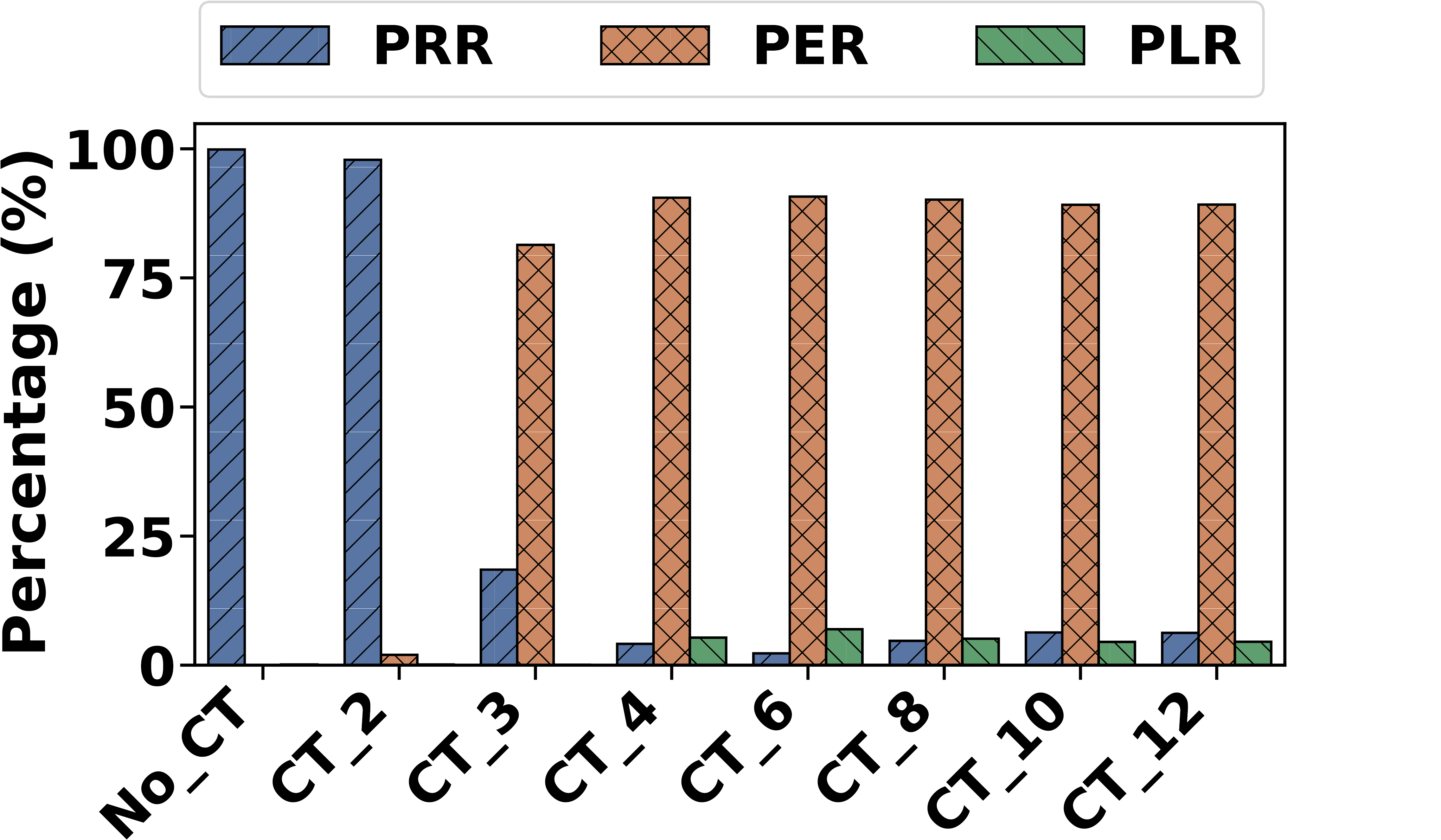}
		\vspace{-5.75mm}
		\caption{\blefive500K}
	\end{subfigure}
	\begin{subfigure}[t]{0.32\textwidth}
		\centering
		\includegraphics[width=1\columnwidth]{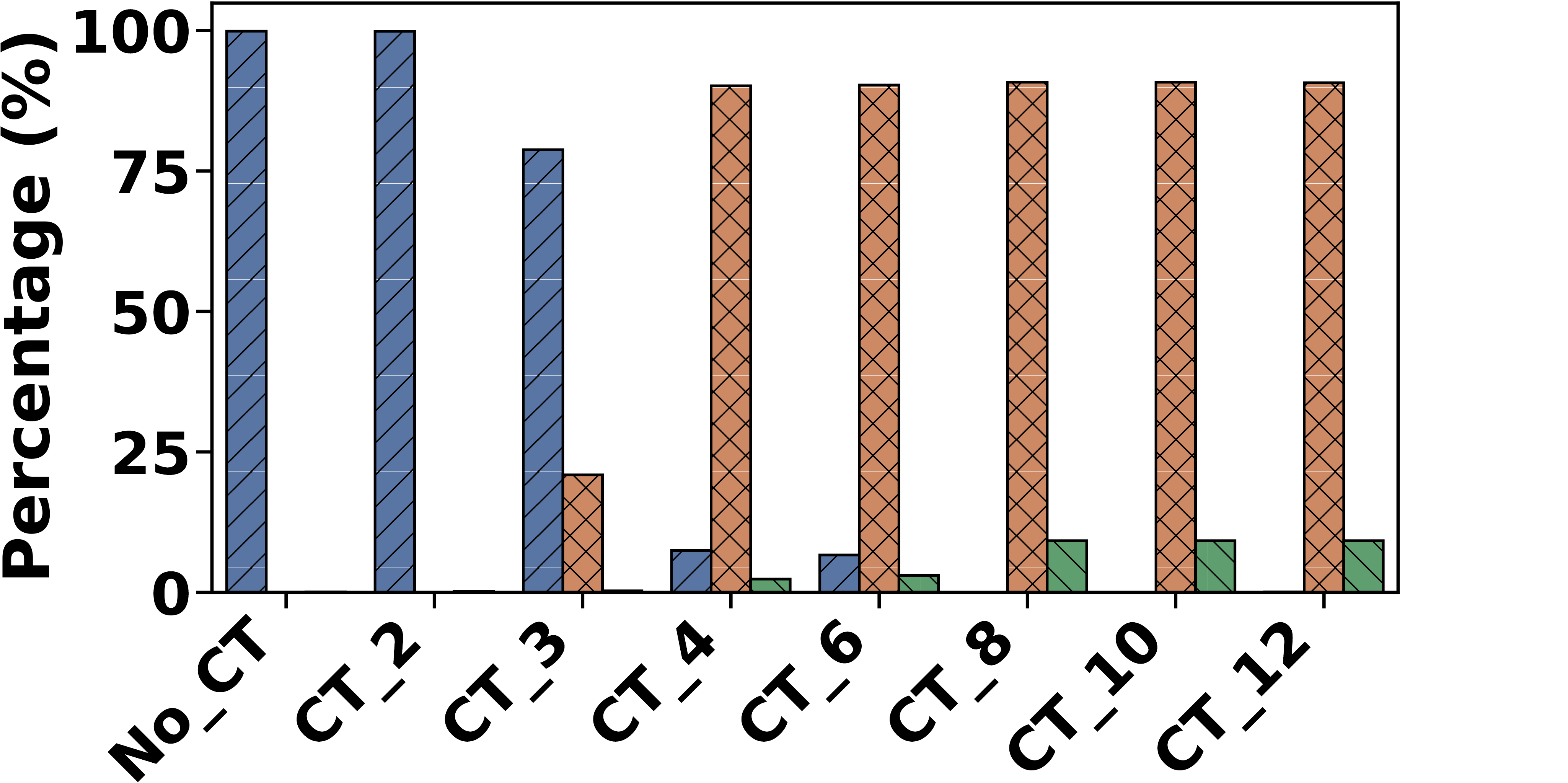}
		\vspace{-5.75mm}
		\caption{\ieee}
	\end{subfigure}	
	\begin{subfigure}[t]{0.32\textwidth}
		\centering
		\vspace{+2.00mm}
		\includegraphics[width=1\columnwidth]{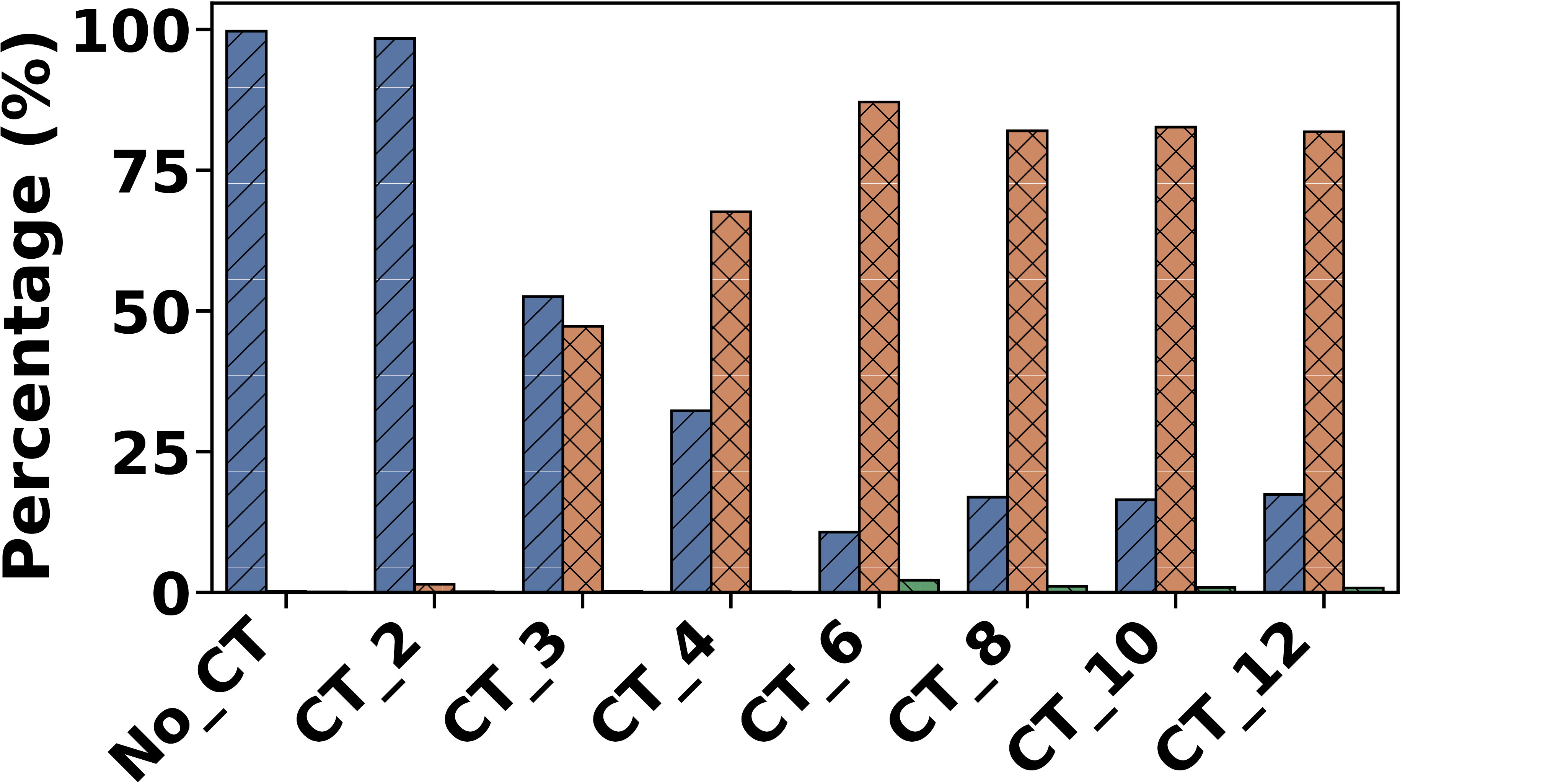}
		\vspace{-5.75mm}
		\caption{\blefive1M}
	\end{subfigure}
	\begin{subfigure}[t]{0.32\textwidth}
		\centering
		\vspace{+2.00mm}
		\includegraphics[width=1\columnwidth]{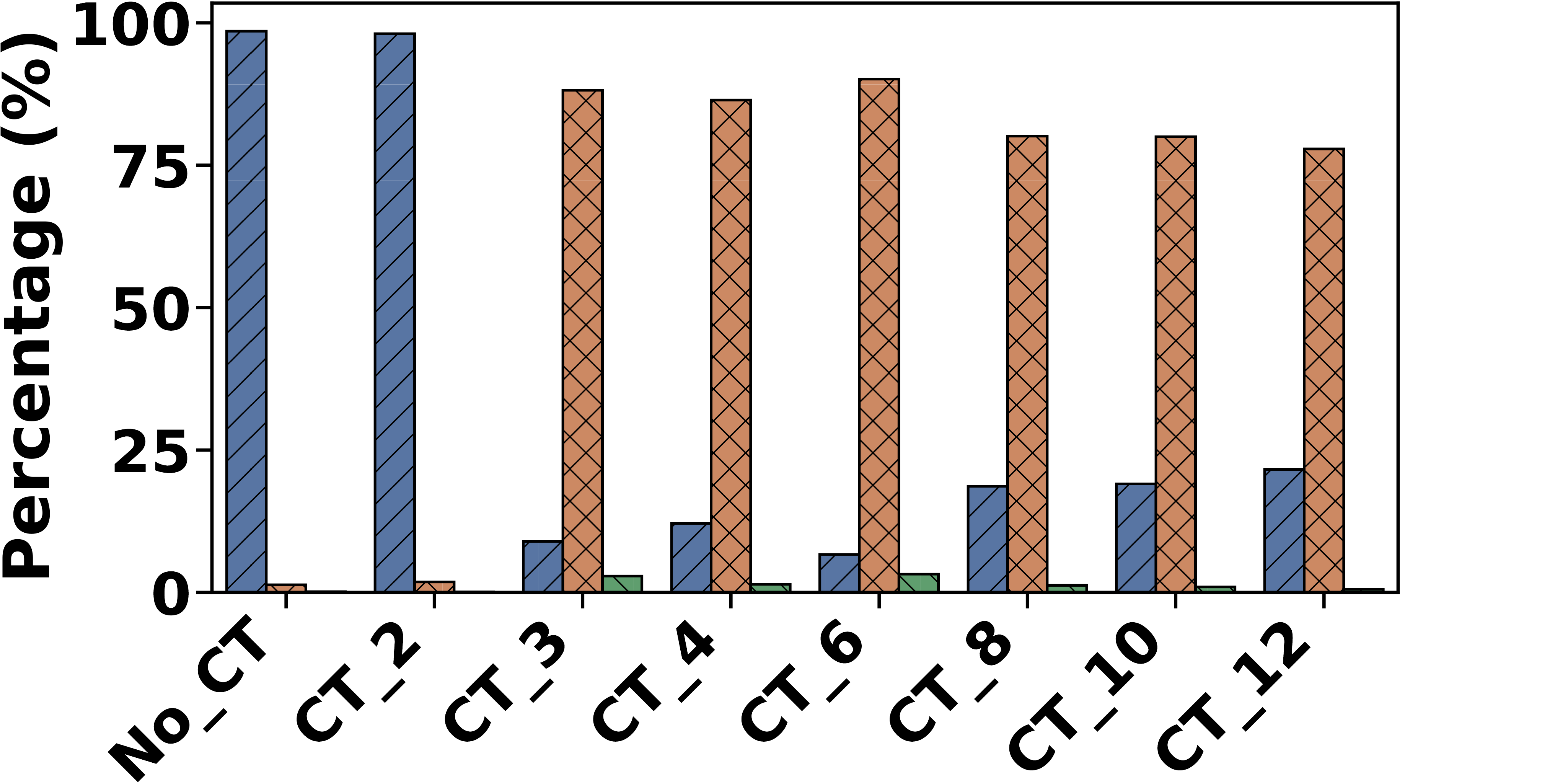}
		\vspace{-5.75mm}
		\caption{\blefive2M}
	\end{subfigure}
	\vspace{-3.75mm}
	\caption{\textbf{PRR, PER, and PLR as CT density increases from a single transmitter (No\_CT) to concurrent transmissions from twelve nodes (CT\_12).} Different PHYs handle the \textit{complex} beating frequencies (as described in Sect.~\ref{sec:background}) with varying degrees of success.}
	\label{fig:tosh_results_ct_comp}	
\end{figure}

\begin{figure}[t]
	\centering
	\includegraphics[width=0.4\columnwidth]{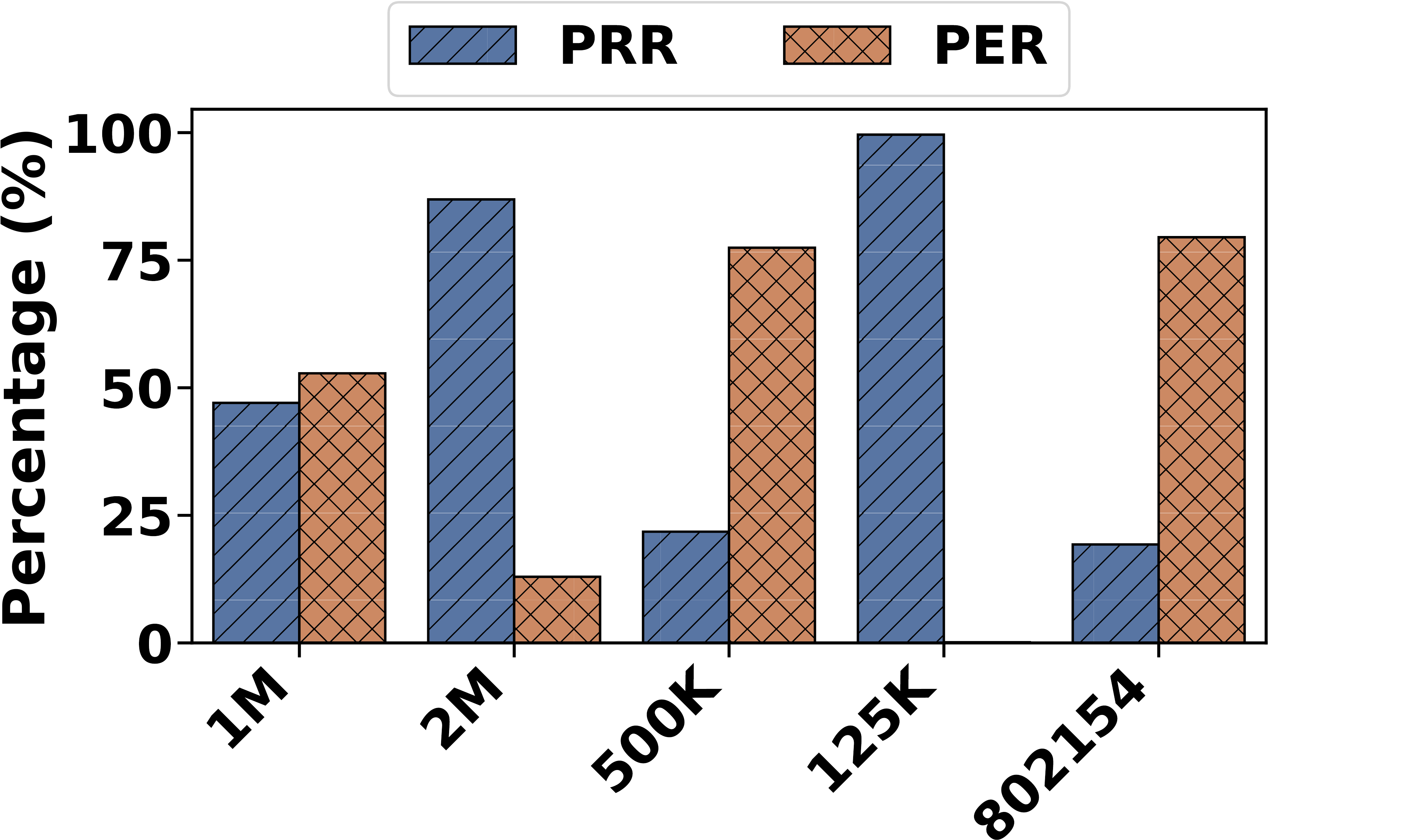}
	\vspace{-3.75mm}
	\caption{\textbf{PRR and PER for the subplots shown in Fig. \ref{fig:tosh_phycomp}.} The CT\_2 pair from this figure experiences \emph{significantly strong} beating over a 200 byte packet.}
	\label{fig:tosh_phycomp_ratio}
\end{figure}

\begin{figure}[t]
	\vspace{-1.75mm}
	\centering
	\begin{subfigure}[t]{0.32\columnwidth}
		\centering
		\includegraphics[width=1\columnwidth]{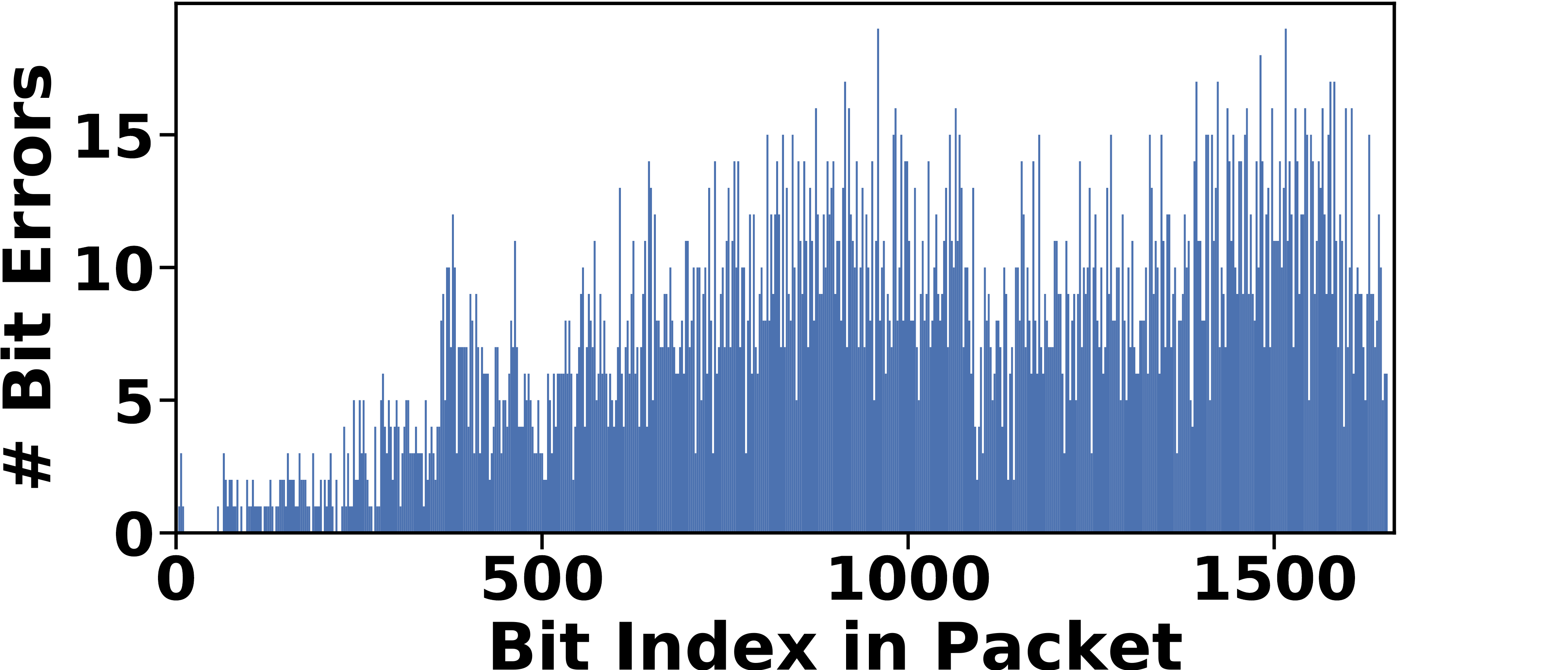}
		\vspace{-5.50mm}
		\caption{Pair 1}
		\label{fig:tosh_layoutcomp_1}
	\end{subfigure}%
	\begin{subfigure}[t]{0.32\columnwidth}
		\centering
		\includegraphics[width=1\columnwidth]{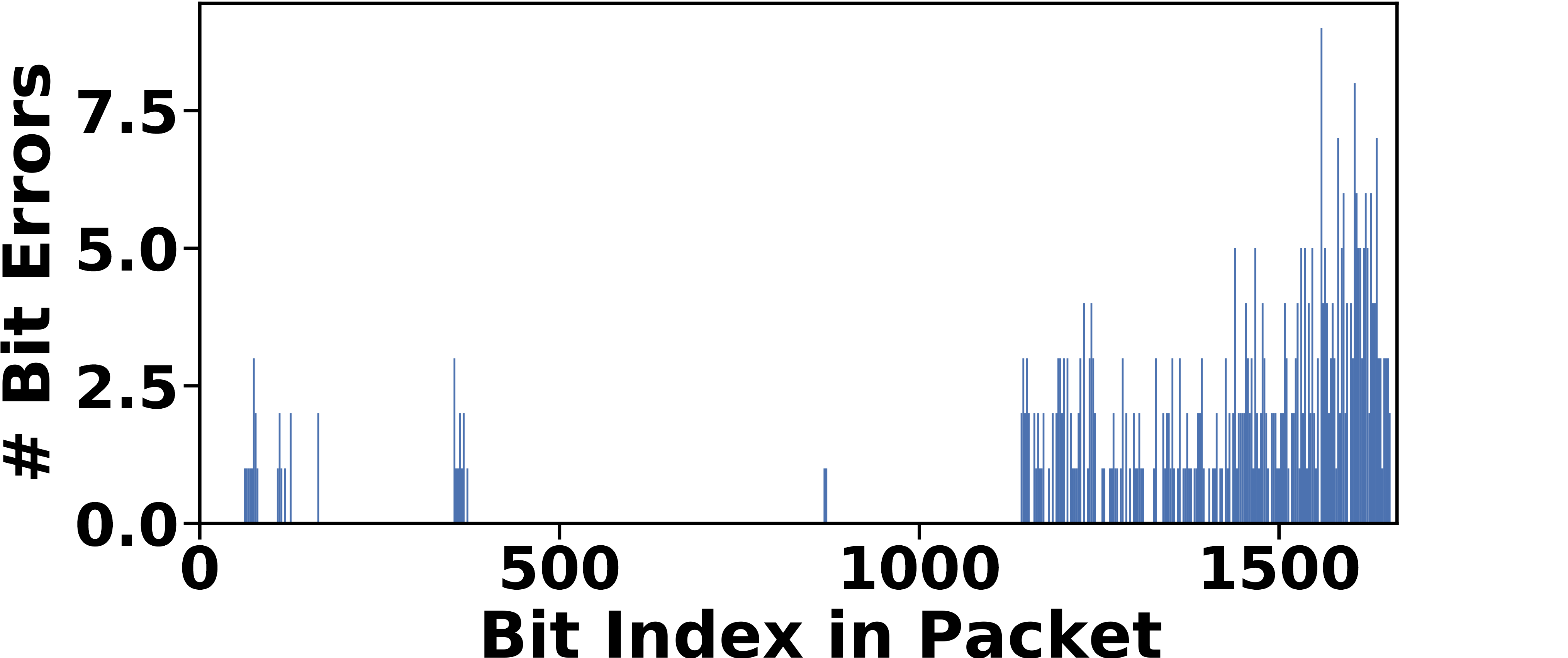}
		\vspace{-5.50mm}
		\caption{Pair 2}
		\label{fig:tosh_layoutcomp_2}
	\end{subfigure}
	\begin{subfigure}[t]{0.32\columnwidth}
		\centering
		\includegraphics[width=1\columnwidth]{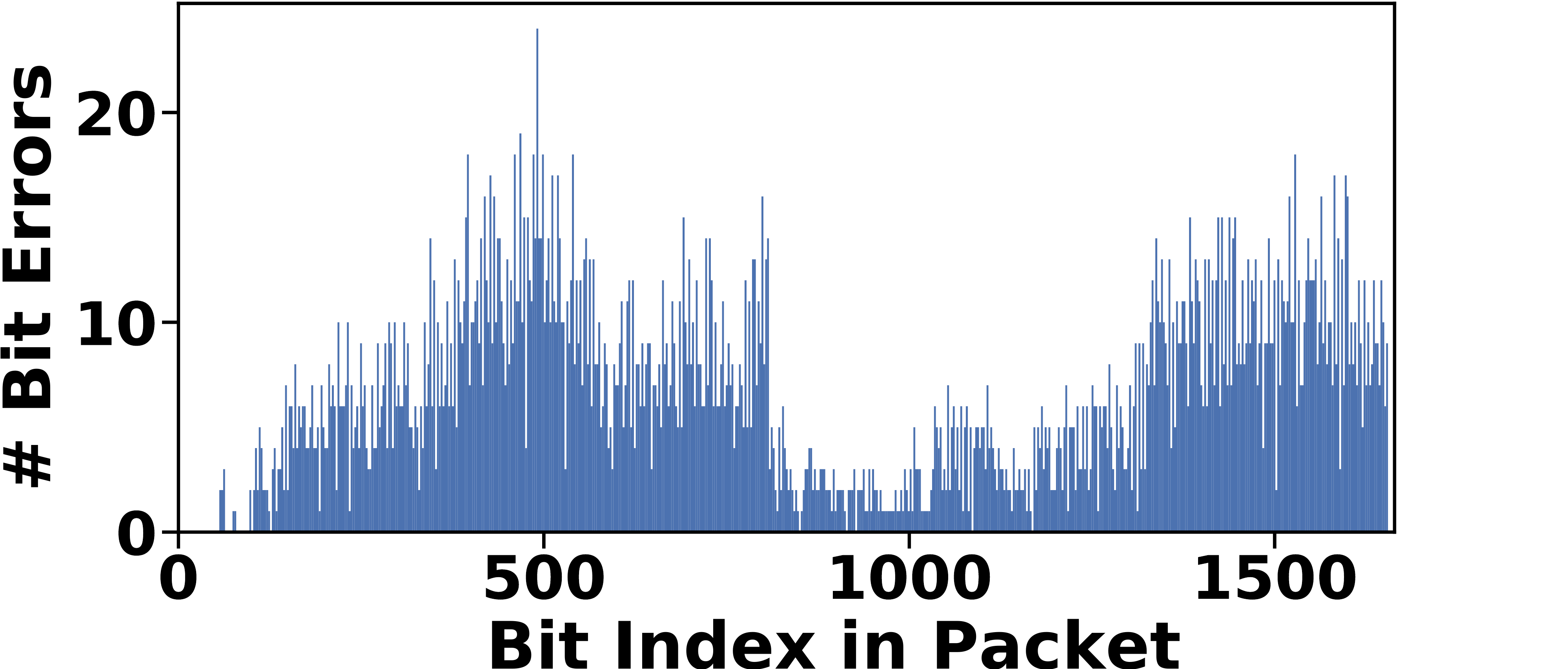}
		\vspace{-5.50mm}
		\caption{Pair 3}
		\label{fig:tosh_layoutcomp_3}
	\end{subfigure}
	\begin{subfigure}[t]{0.32\columnwidth}
		\centering
		\includegraphics[width=1\columnwidth]{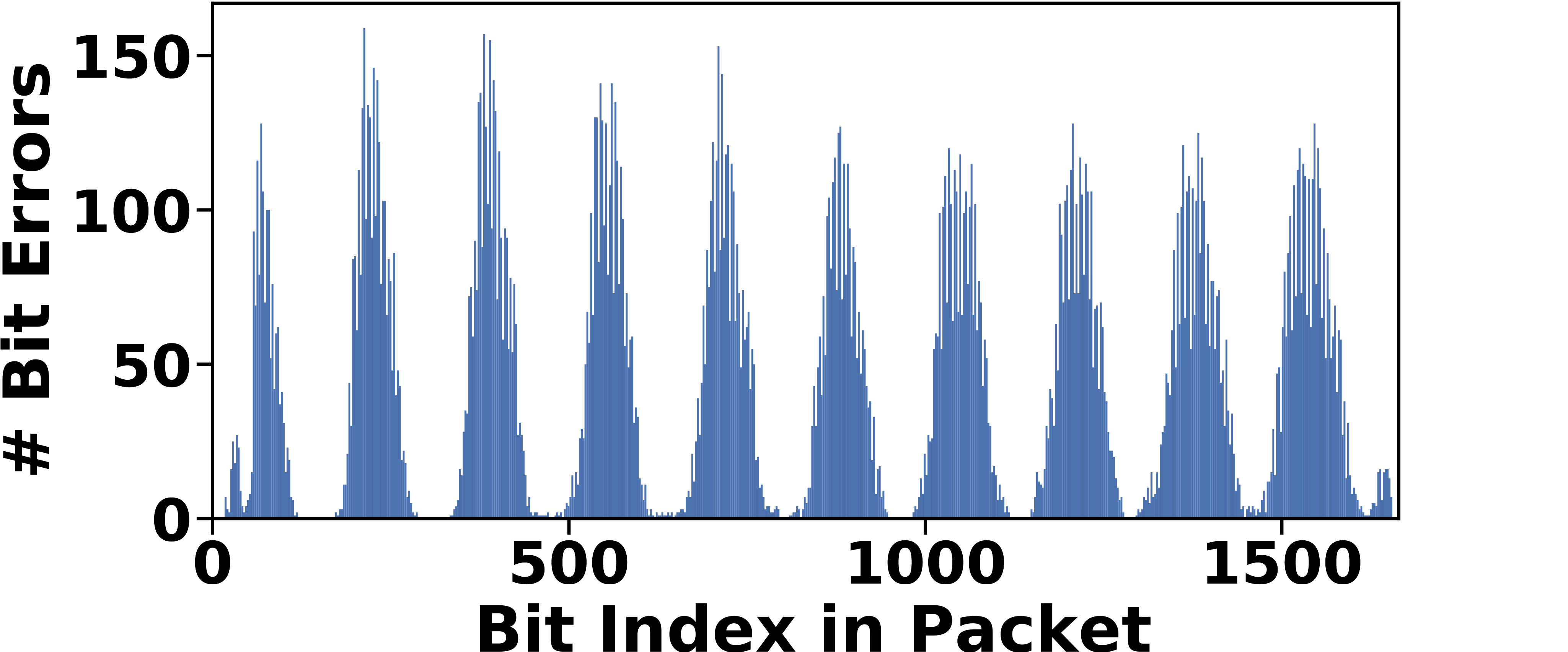}
		\vspace{-5.50mm}
		\caption{Pair 4}
	\end{subfigure}
	\begin{subfigure}[t]{0.32\columnwidth}
		\centering
		\includegraphics[width=1\columnwidth]{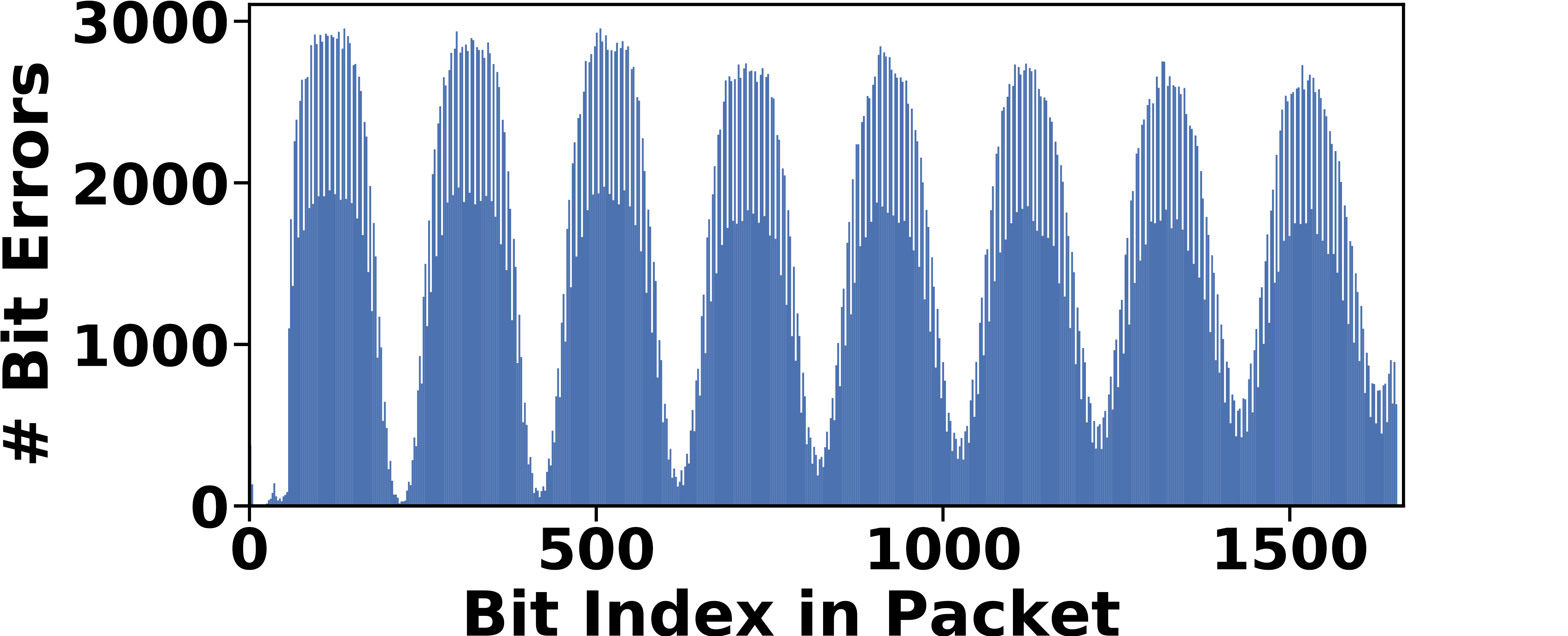}
		\vspace{-5.50mm}
		\caption{Pair 5}
		\label{fig:pair5}
	\end{subfigure}
	\begin{subfigure}[t]{0.32\columnwidth}
		\centering
		\includegraphics[width=1\columnwidth]{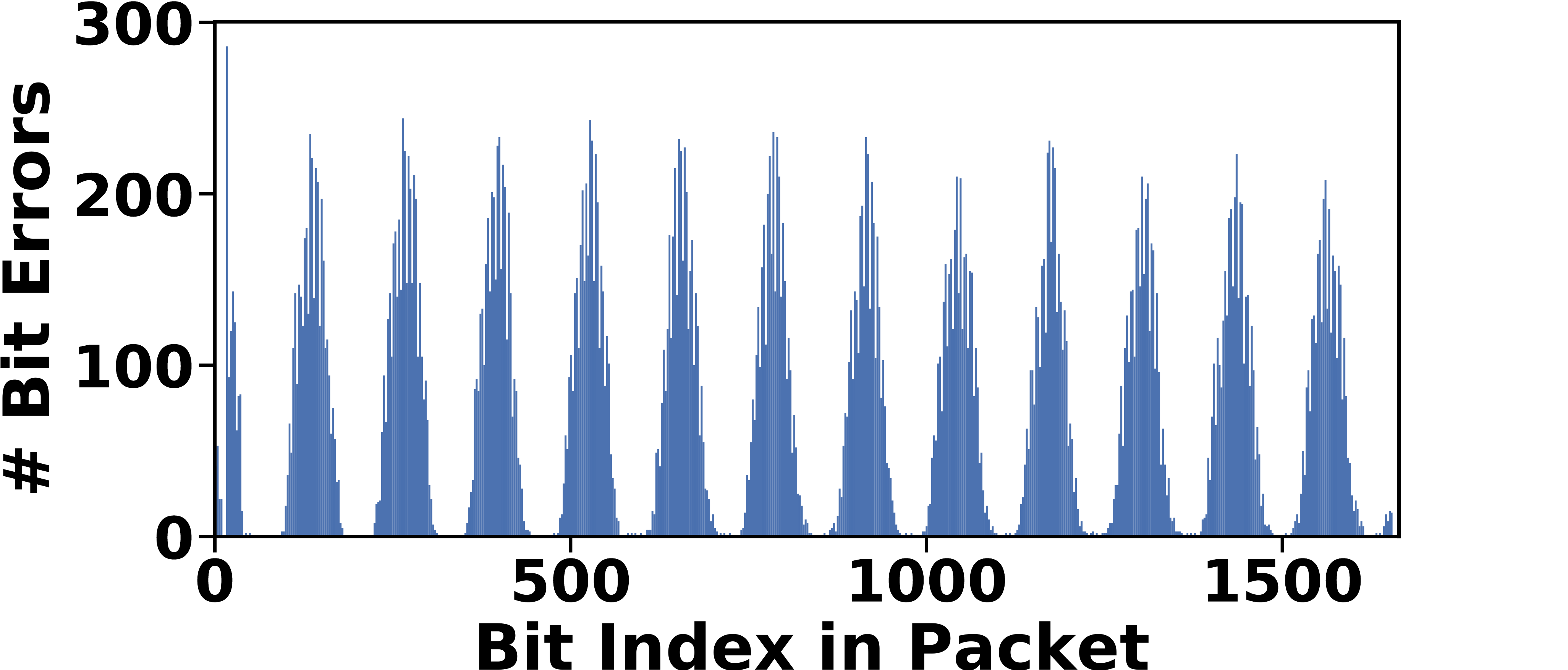}
		\vspace{-5.50mm}
		\caption{Pair 6}
	\end{subfigure}
	\begin{subfigure}[t]{0.32\columnwidth}
		\centering
		\includegraphics[width=1\columnwidth]{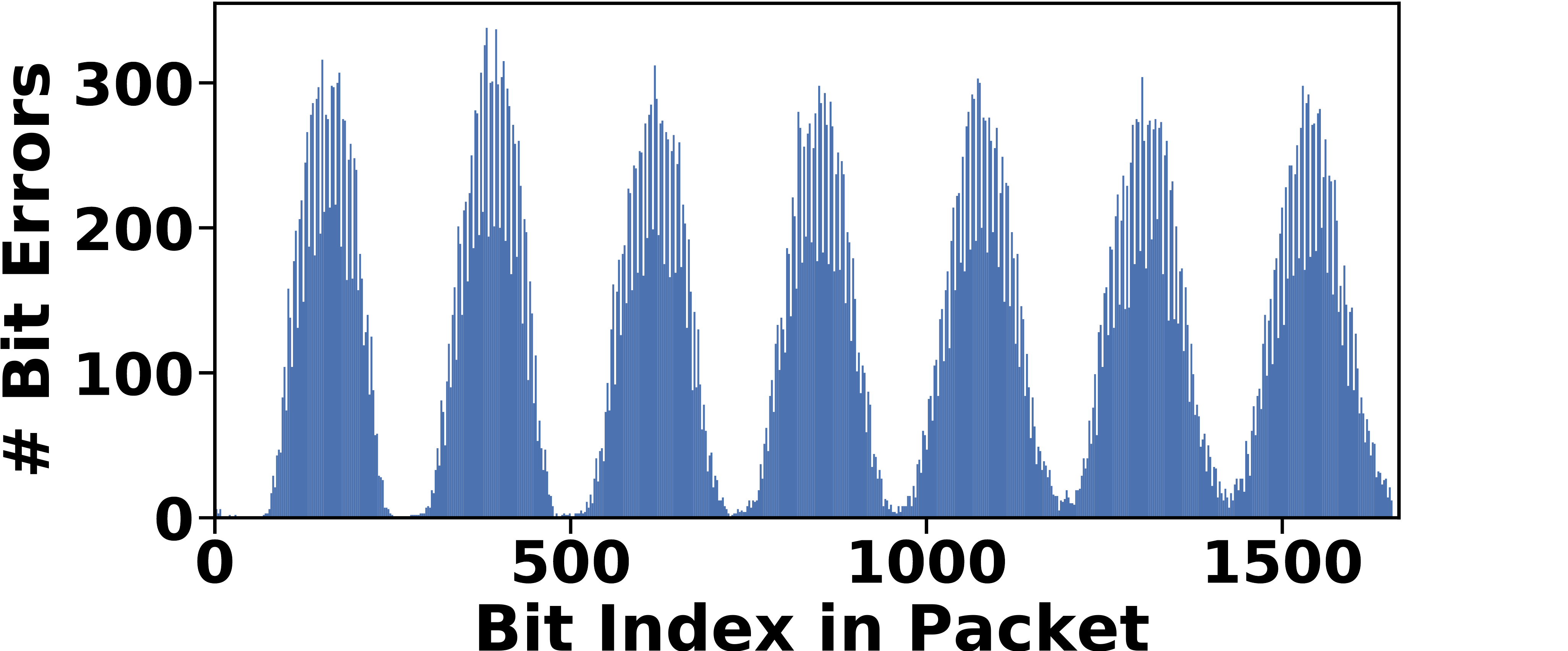}
		\vspace{-5.50mm}
		\caption{Pair 7}
	\end{subfigure}
	\begin{subfigure}[t]{0.32\columnwidth}
		\centering
		\includegraphics[width=1\columnwidth]{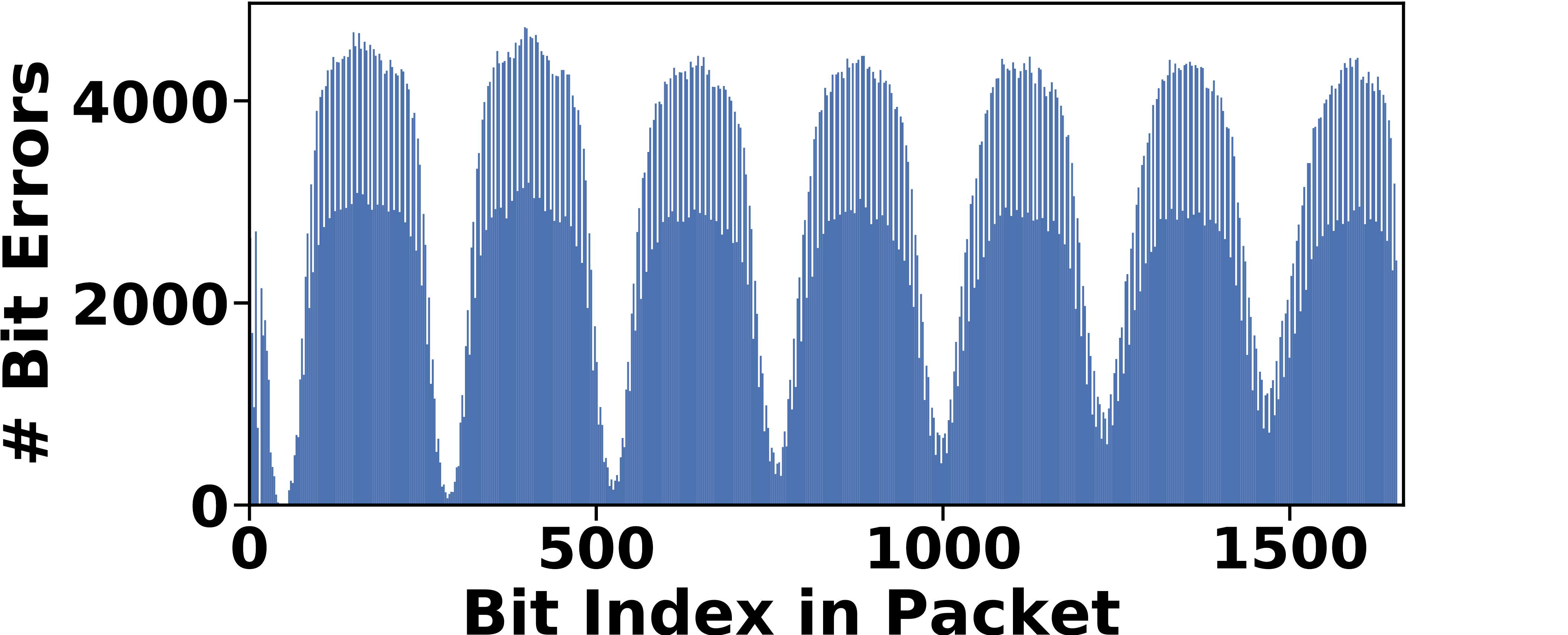}
		\vspace{-5.50mm}
		\caption{Pair 8}
		\label{fig:pair8}
	\end{subfigure}
	\begin{subfigure}[t]{0.32\columnwidth}
		\centering
		\includegraphics[width=1\columnwidth]{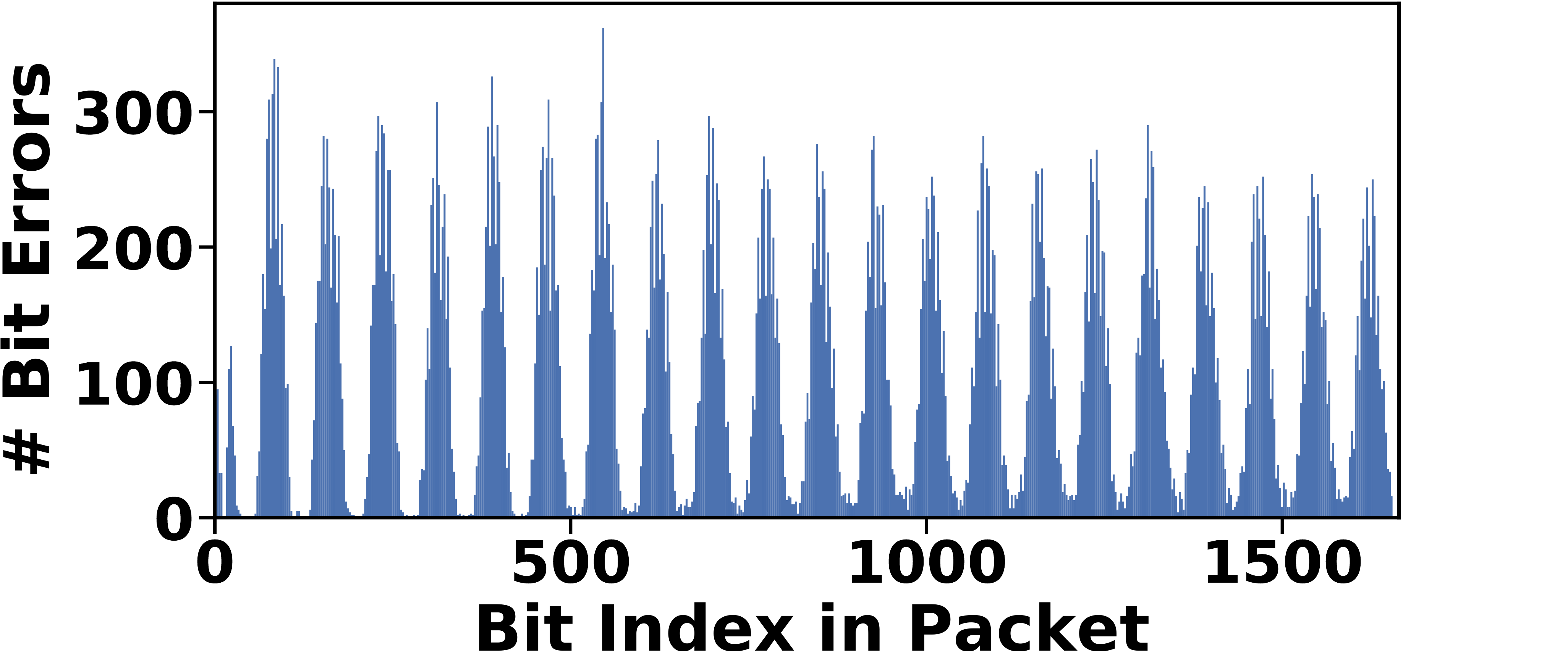}
		\vspace{-5.50mm}
		\caption{Pair 9}
	\end{subfigure}
	\caption{\textbf{Individual CT\_2 pairs experience \textit{different} beating frequencies depending on their RFOs.} This can be seen from examination of the bit error distribution across nine \textit{different} CT\_2 pairs when using the \blefive500K PHY.}
	\label{fig:tosh_layoutcomp}
\end{figure}

\begin{figure}[t]
	\vspace{-1.75mm}
	\centering
	\includegraphics[width=0.475\columnwidth]{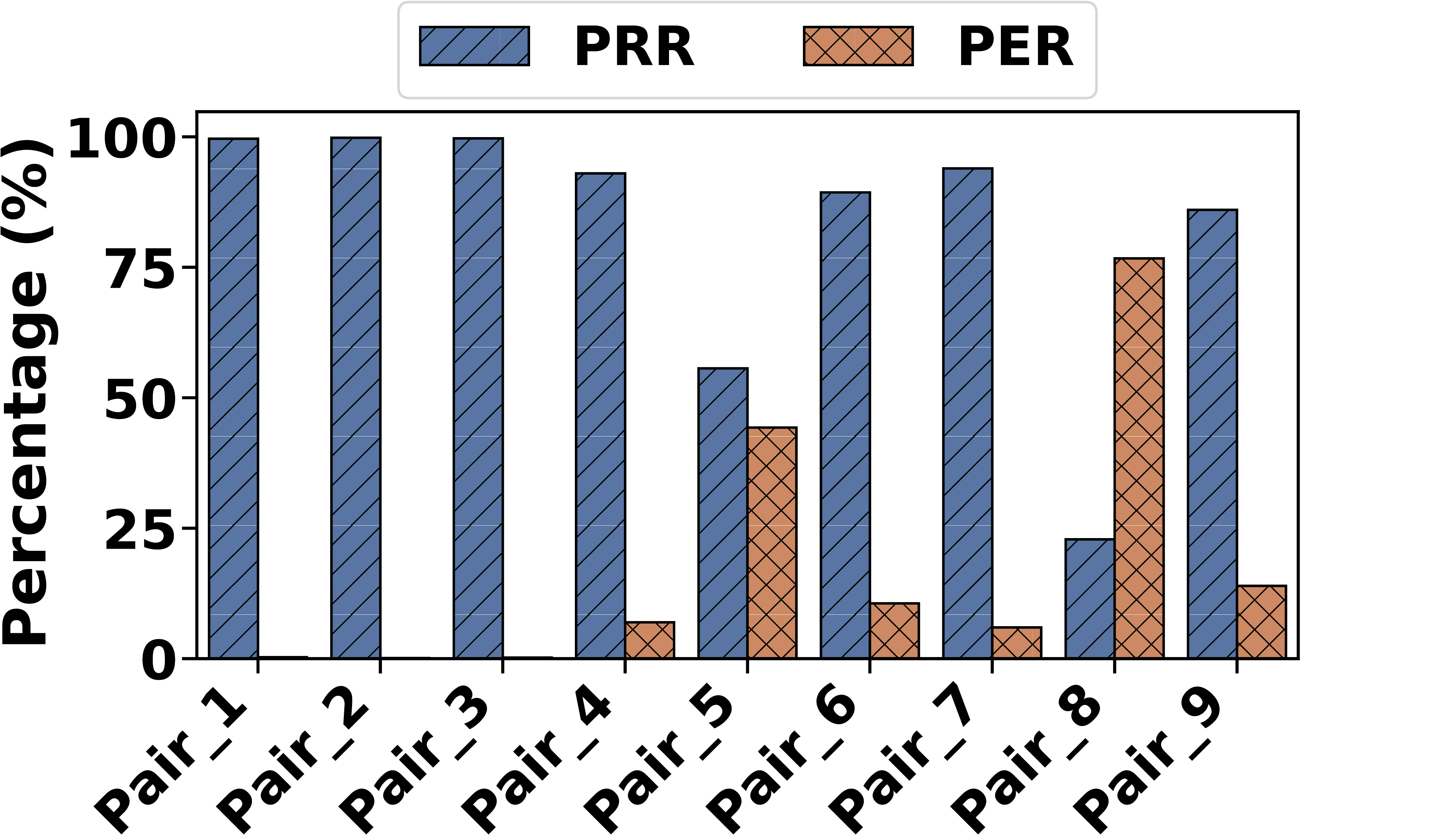}
	\vspace{-3.75mm}
	\caption{\textbf{\textit{Wide and strong} beating frequencies result in considerable losses.} CT reliability (PRR and PER) for the nine CT\_2 pairs shown in Fig.\,\ref{fig:tosh_layoutcomp} shows the coded \blefive500K PHY struggles when experiencing \textit{wide and strong} beating.}
	\label{fig:tosh_layoutcomp_prr}
\end{figure}

\boldpar{Beating effect on different PHYs} 
Fig.~\ref{fig:tosh_phycomp} shows the beating frequency for a single CT\_2 pair. The beating frequency remains consistent across all subfigures but, depending on the underlying PHY rate, 
the time window spanned by the packet is smaller or larger, resulting in \emph{narrower} or \emph{wider} beating 
over the same fixed packet length (i.e., high data-rates will suffer from fewer peaks than low data-rates). 
Taking into account the underlying bit duration and the bit distance between error peaks, it is possible to discern that the beating period for this \emph{specific} CT\_2 pair is $\approx$ 0.5\,ms (a frequency of 2\,kHz). 

Fig.~\ref{fig:tosh_phycomp_ratio} shows how the PRR and PER are closely linked to the way in which beating manifests across the various PHYs, and supports the findings previously shown through simulation in Fig.~\ref{fig:rednode_sim3} (\emph{narrow and strong} beating), where it is likely that this pair experiences \emph{very low noise} (SNR\,$>$\,25\,dB). As discussed in Sect.~\ref{sec:simulation}, the higher data rate PHYs (\blefive2M and 1M) experience fewer beating valleys. While these uncoded PHYs are unable to recover errors if they fall within a valley, the repetition commonly employed in CT protocols (\texttt{TX\_N\,=\,4} in these experiments) means it is likely that a retransmission will successfully fall between valleys and allow a successful reception of the preamble. This can be observed in the higher error rate for the 1M\,PHY, which experiences additional beating valleys as opposed to 2M. 
Furthermore, while it would be natural to assume that the redundancy employed in coded PHYs helps them to better recover from beating errors, our results show that the \blefive convolutional coding is unable to cope with significant beating (as seen from the high PER in the \blefive500K results shown in Fig.~\ref{fig:tosh_phycomp_ratio}). The same applies to the DSSS employed in \ieee, although it helps in recovering errors (despite the higher number of beating valleys caused by the lower data rate). On the other hand, the addition of the Manchester pattern mapper in the \blefive125K PHY provides sufficient gain to survive beating. 

\boldpar{Increasing CT density}
The effect of increasing CT density is explored in Fig.~\ref{fig:tosh_results_ct_comp}. Experiments were run across all PHYs for a single transmitter (\emph{no CT}), as well as increasing CT density from CT\_2 to CT\_12. Plots represent an average of multiple experiments run with randomly selected CT forwarders per experiment, while the same pseudo-random forwarding set remains consistent across each PHY. This averaging eliminates bias due to \emph{narrow and strong} beating experienced by CT\_2 pairs such as Figs.~\ref{fig:pair5}~and~\ref{fig:pair8}. Reliability drops at CT\_3 due to the high data rate of the \blefive2M PHY, which requires a large difference in received power between signals to experience the capture effect. This is consistent with recent literature~\cite{alnahas2020blueflood} and with our analysis in Sect.~\ref{sec:simulation}. While the \ieee PHY 
still performs well at CT\_3, its PRR also drops significantly at CT\_4. Interestingly, the \blefive1M PHY shows a gradual drop in performance at mid-level CT densities, while both \blefive uncoded PHYs (2M and 1M) experience a PRR `rally' at high CT density. We hypothesize that this is due to the increased diversity (i.e., additional paths and better chance of capturing dominant signals), or to the additional CT converging around an average RFO and spreading the effects of beating. 

\boldpar{Beating effect across different CT pairs}
Fig.~\ref{fig:tosh_layoutcomp} examines the bit error distribution for nine different CT pairs on the \blefive500K PHY -- \newtext{chosen due to its beating sensitivity, which allows clear beating patterns within the figures}. 
The absence of the Manchester pattern mapper (as used in the 125K PHY) results in a high degree of corruption across the packet, meaning that beating errors are more prominent. Beating is therefore clearly seen across almost all pairs, with the exception of Figs.~\ref{fig:tosh_layoutcomp_1}~and~\ref{fig:tosh_layoutcomp_2}, where the RFO was not significant enough to result in observable beating (i.e., these pairs experience wide beating greater than the packet's transmission period). Fig.~\ref{fig:tosh_layoutcomp_prr} provides additional information about the PRR and PER for each pair, further showing how the RFO between pairs affects the beating width and, consequently, the performance of CT.

\begin{figure}[t]
	\centering
	\begin{subfigure}[t]{0.24\textwidth}
		\centering
		\includegraphics[width=1\columnwidth]{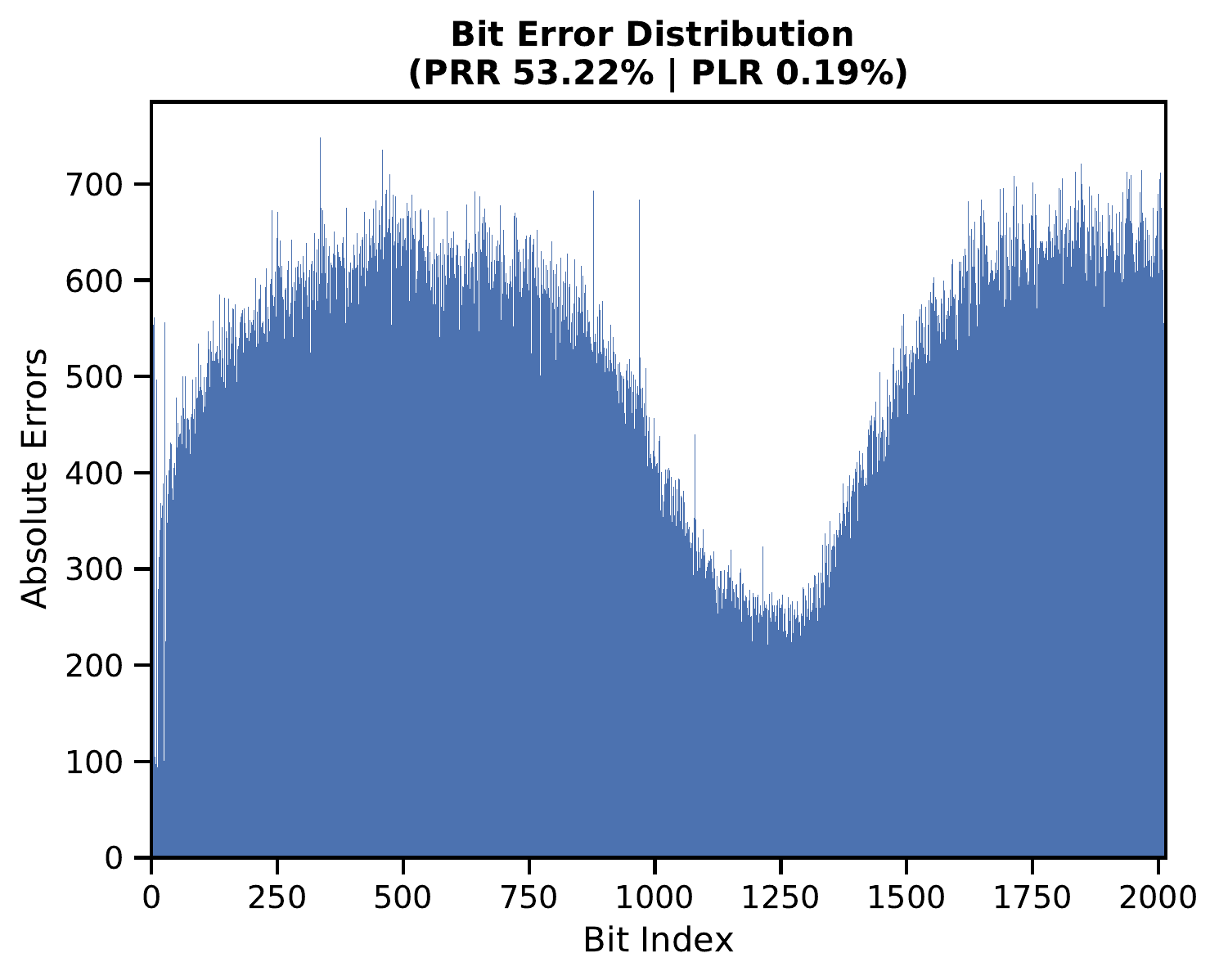}
		\vspace{-5.75mm}
		\caption{CT\_2 Pair 1}
		\label{fig:tii_beating_hist_496hz}
	\end{subfigure}%
	\begin{subfigure}[t]{0.24\textwidth}
		\centering
		\includegraphics[width=1\columnwidth]{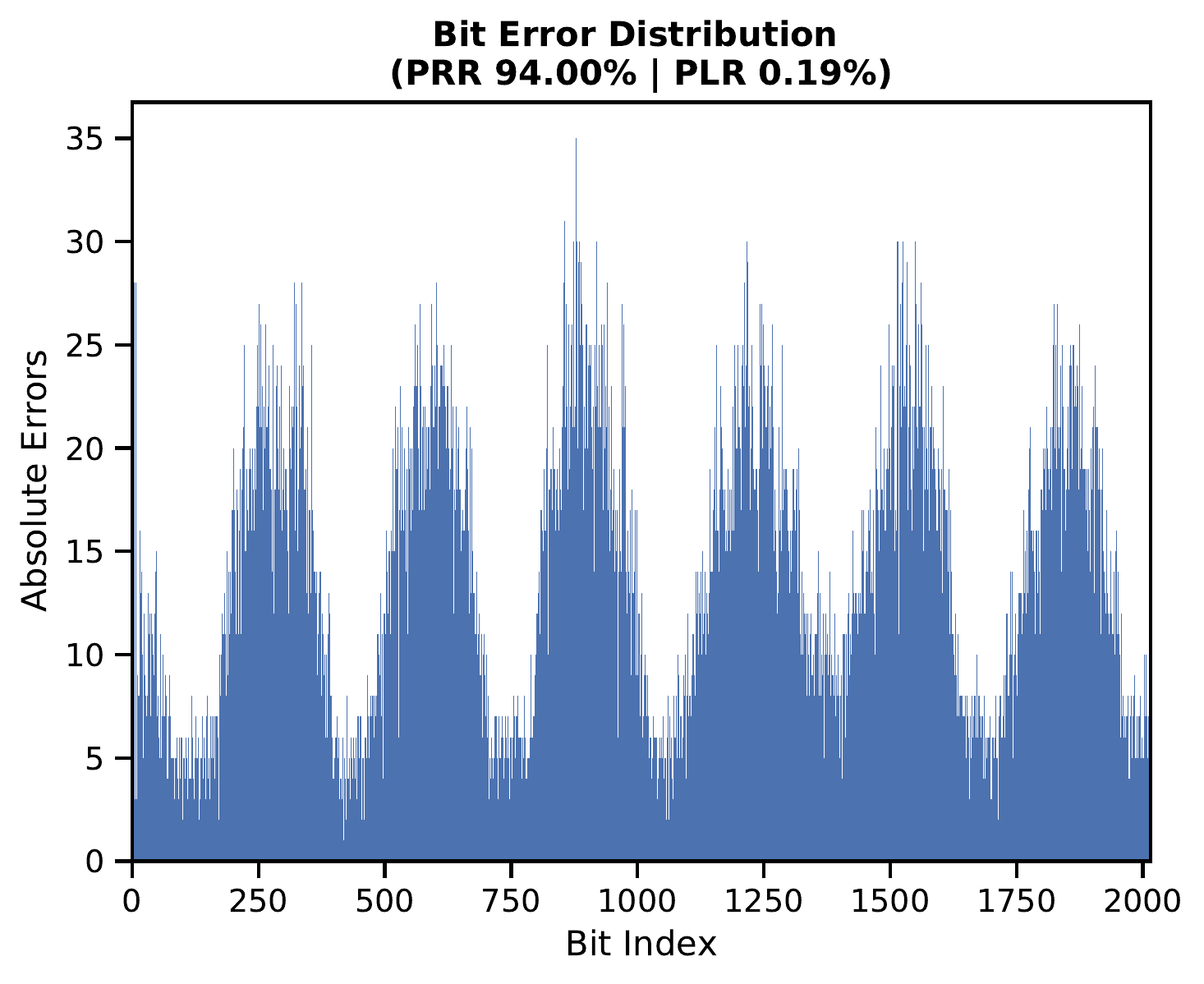}
		\vspace{-5.75mm}
		\caption{CT\_2 Pair 2}
		\label{fig:tii_beating_hist_1488hz}
	\end{subfigure}
	\begin{subfigure}[t]{0.24\textwidth}
		\centering
		\includegraphics[width=1\columnwidth]{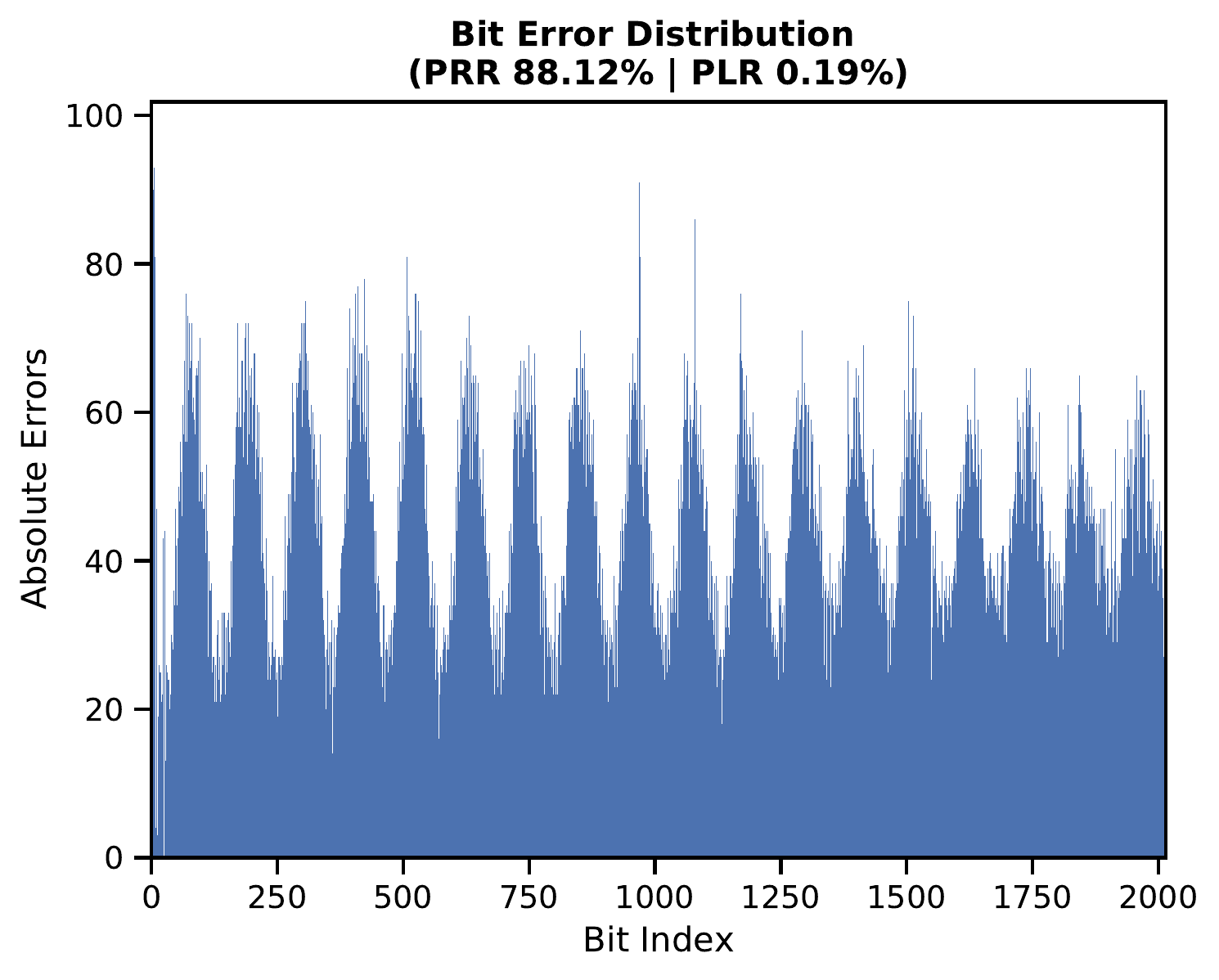}
		\vspace{-5.75mm}
		\caption{CT\_2 Pair 3}
		\label{fig:tii_beating_hist_4468hz}
	\end{subfigure}
	\begin{subfigure}[t]{0.24\textwidth}
		\centering
		\includegraphics[width=1\columnwidth]{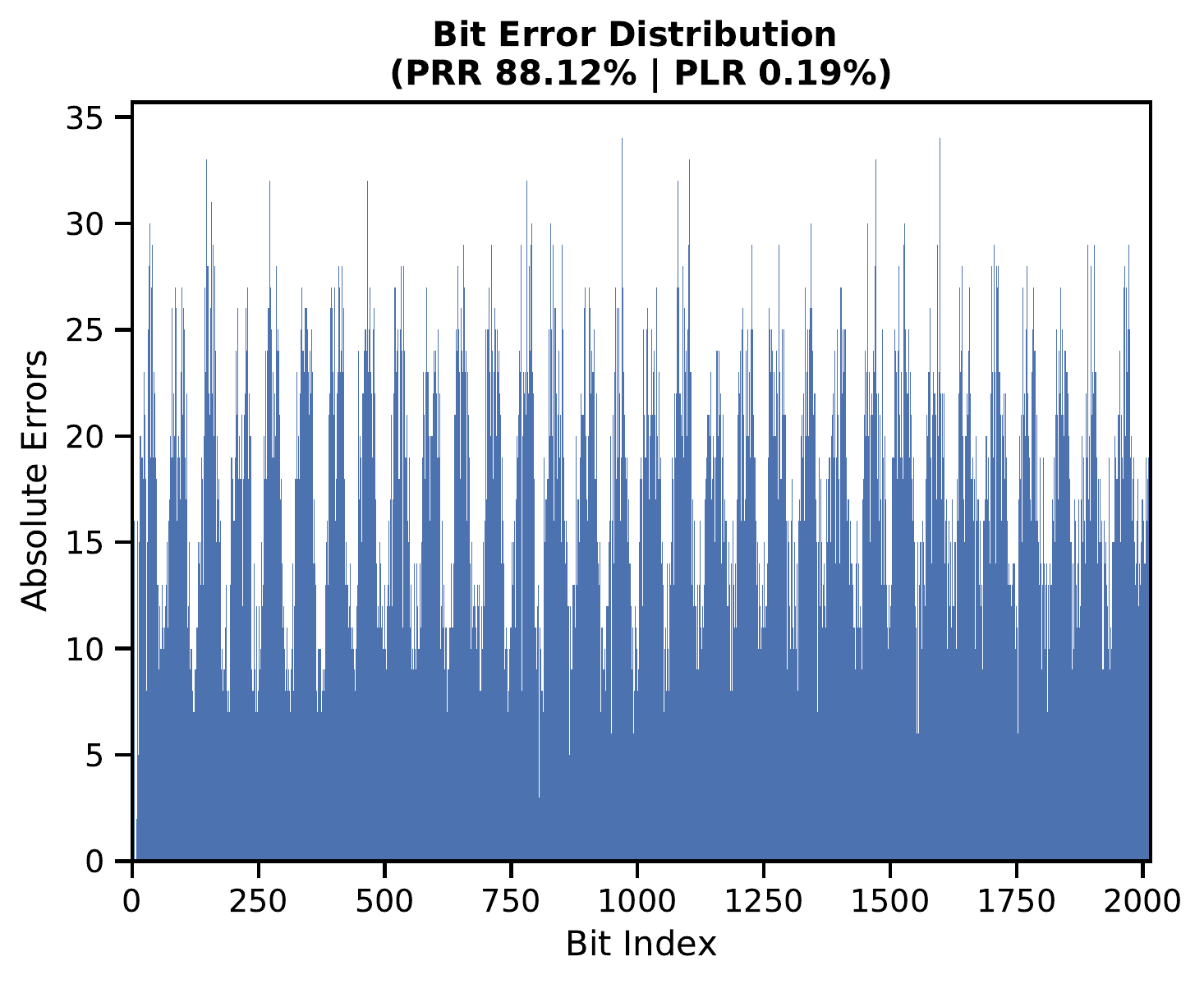}
		\vspace{-5.75mm}
		\caption{CT\_2 Pair 4}
		\label{fig:tii_beating_hist_7948hz}
	\end{subfigure}
	\begin{subfigure}[t]{0.24\textwidth}
		\centering
		\includegraphics[width=1\columnwidth]{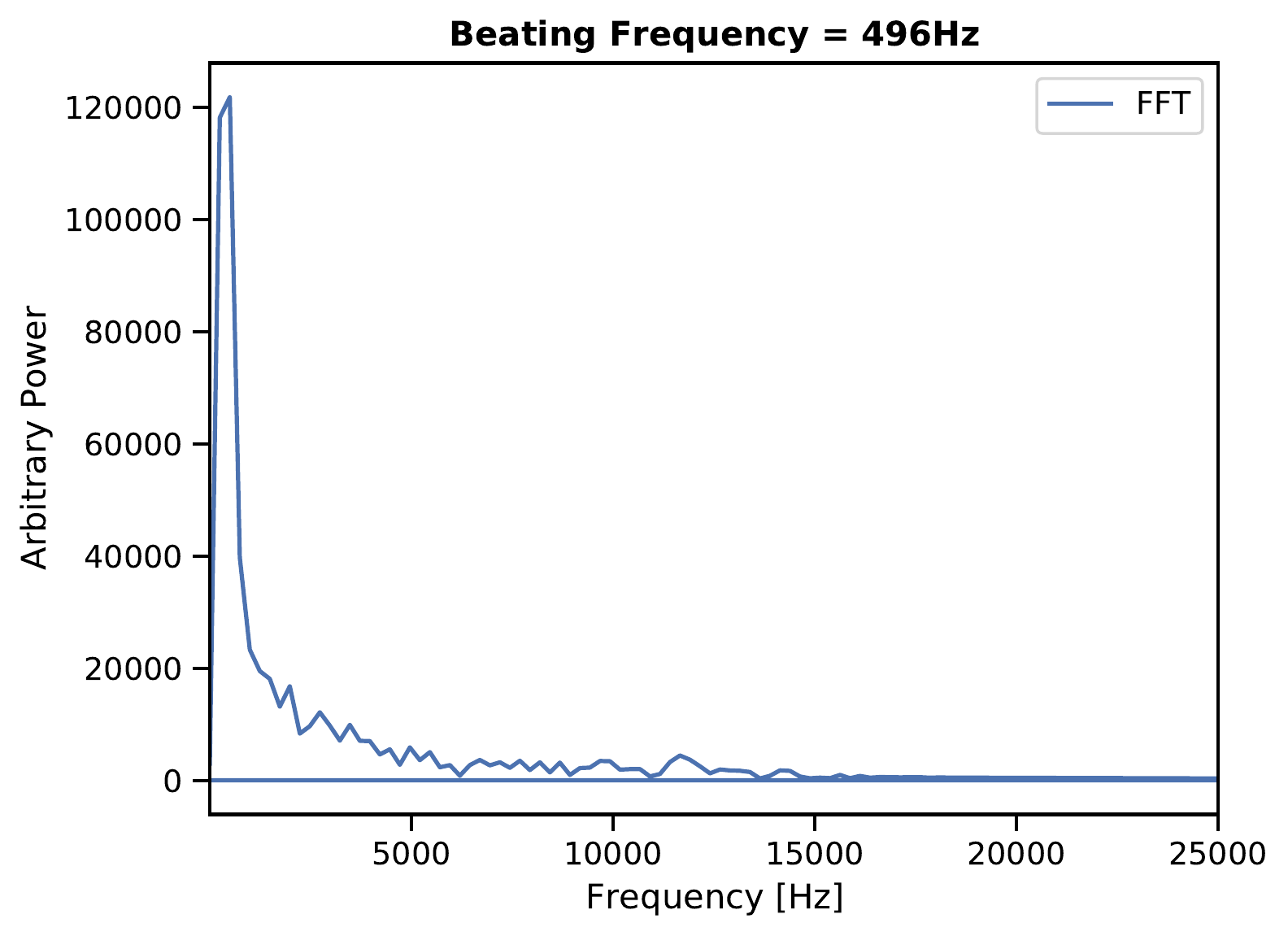}
		\vspace{-5.75mm}
		\caption{$RFO_{Pair1} = 496\,Hz$}
	\end{subfigure}%
	\begin{subfigure}[t]{0.24\textwidth}
		\centering
		\includegraphics[width=1\columnwidth]{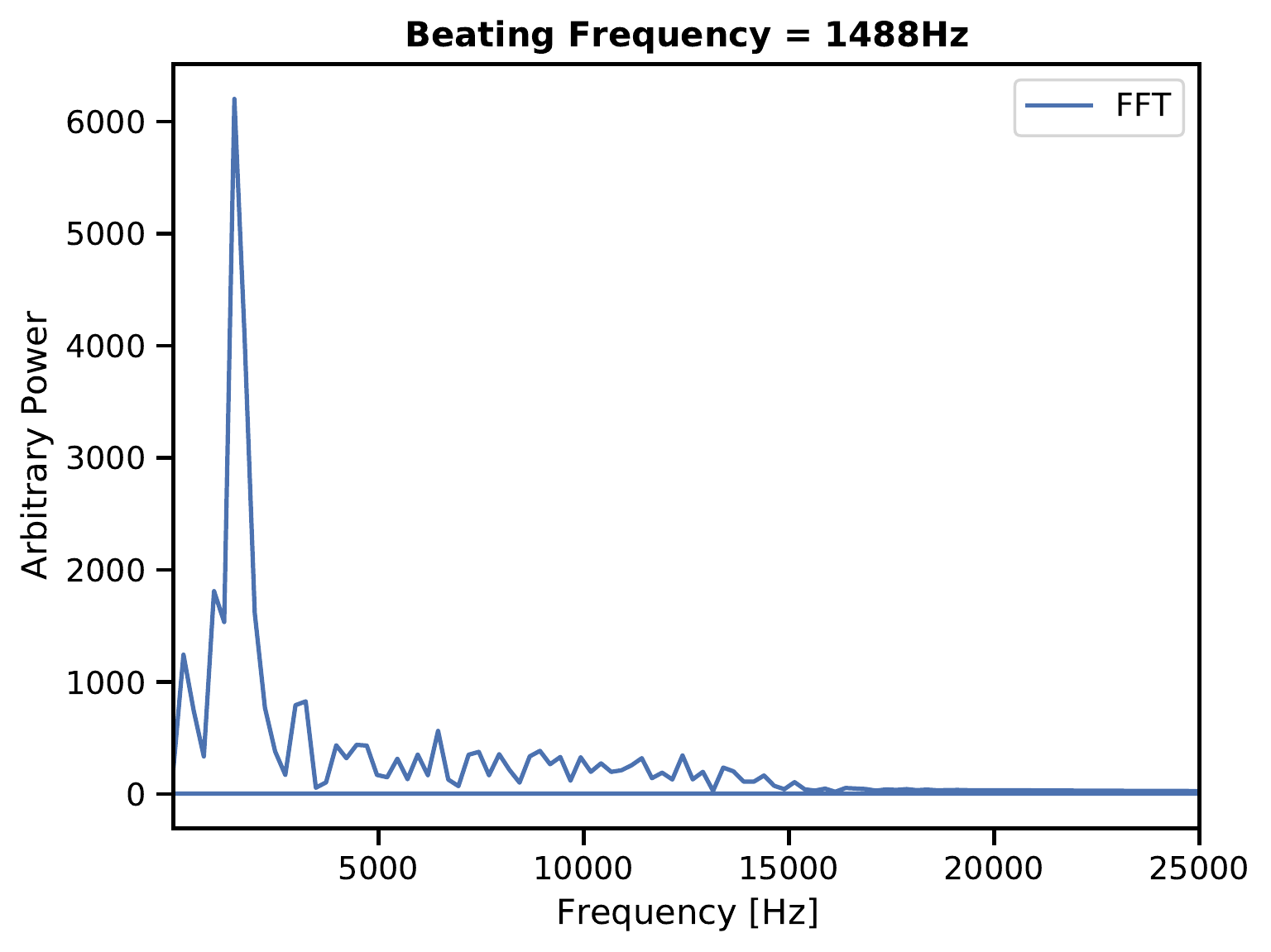}
		\vspace{-5.75mm}
		\caption{$RFO_{Pair2} = 1488\,Hz$}
	\end{subfigure}
	\begin{subfigure}[t]{0.24\textwidth}
		\centering
		\includegraphics[width=1\columnwidth]{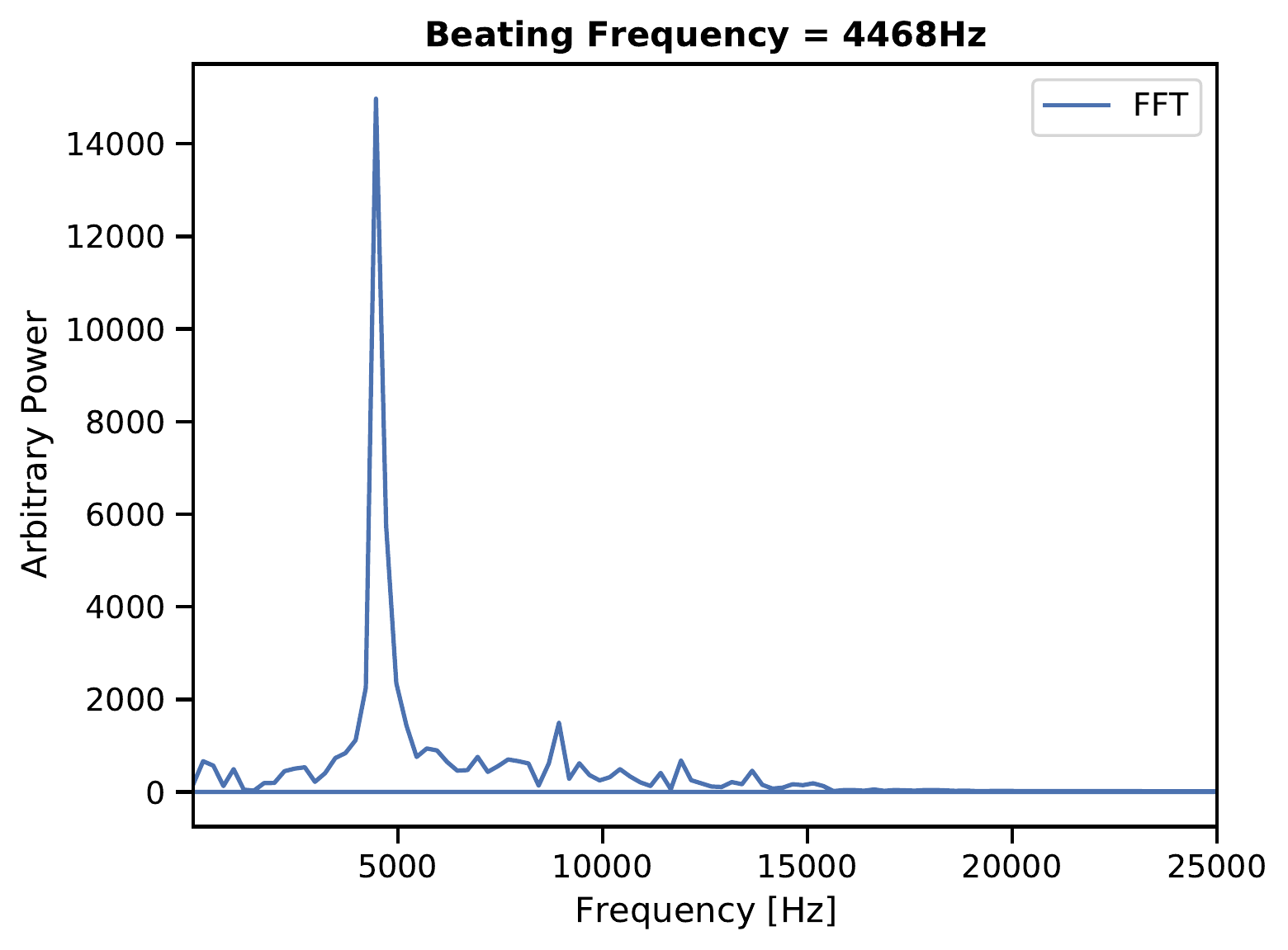}
		\vspace{-5.75mm}
		\caption{$RFO_{Pair3} = 4468\,Hz$}
	\end{subfigure}
	\begin{subfigure}[t]{0.24\textwidth}
		\centering
		\includegraphics[width=1\columnwidth]{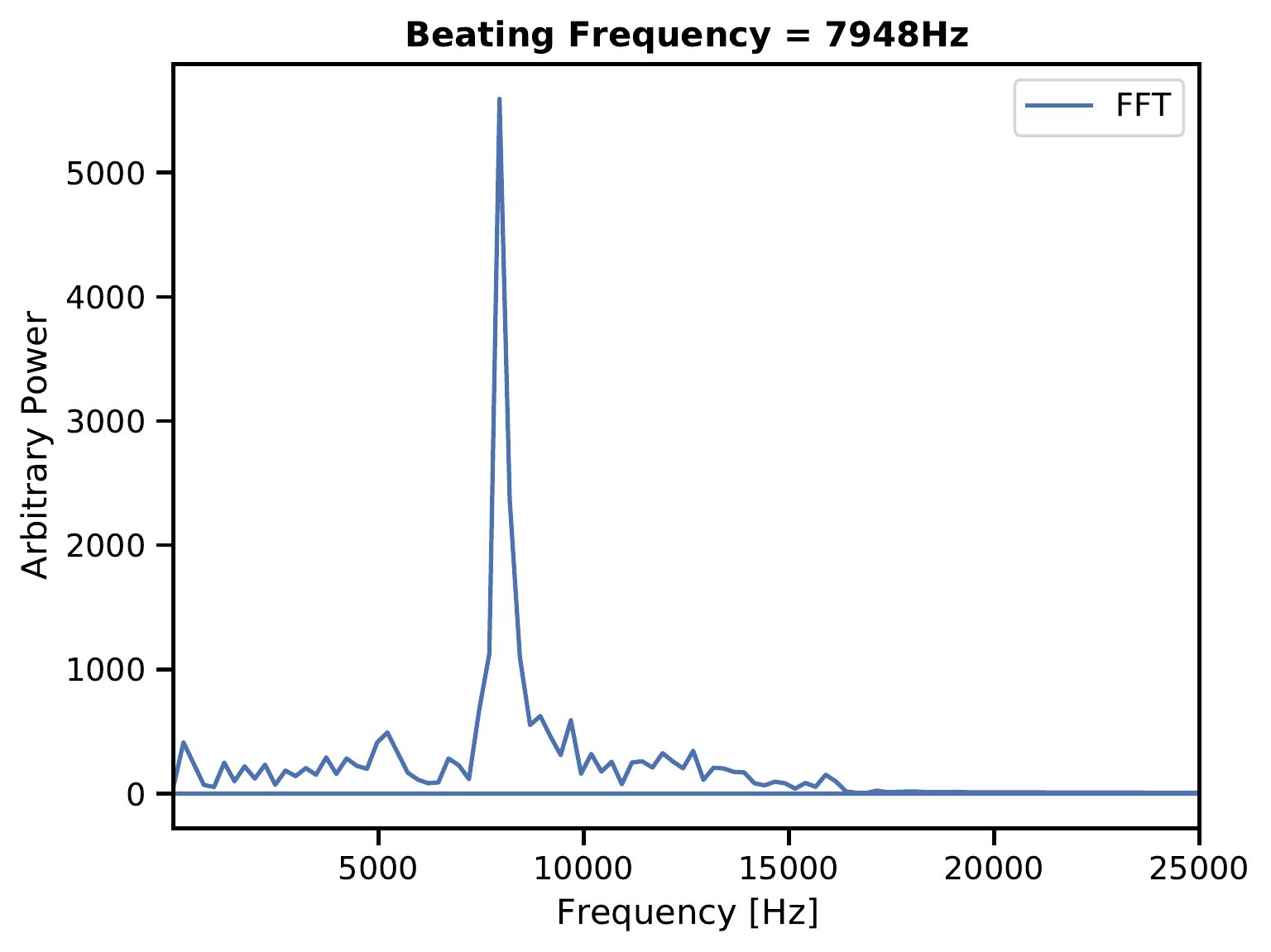}
		\vspace{-5.75mm}
		\caption{$RFO_{Pair4} = 7948\,Hz$}
	\end{subfigure}
	\vspace{-2.25mm}
	\caption{\textbf{The RFO between devices can be quantified with an FFT.} The beating frequencies of four \textit{different} CT\_2 pairs transmitting over the \blefive500K PHY are resolved by applying an FFT over the beating histogram.}
	\label{fig:tii_beating_hist}
		\vspace{-3.75mm}	
\end{figure}

\boldpar{Calculating RFO between devices} 
\newtext{Figs.~\ref{fig:tii_beating_hist}a--d present histograms of the bit errors generated by four independent CT\_2 device pairs transmitting over \blefive500K. A clear sinusoidal beating pattern emerges on each of the four pairs, with Fig.~\ref{fig:tii_beating_hist_496hz} exhibiting \emph{wide and strong} beating, while Fig.~\ref{fig:tii_beating_hist_7948hz} suffers from \emph{narrow and strong} beating (as categorized in Sect.\,\ref{sec:background}). Interestingly, by sampling the signal and applying an FFT across the bit error data, it is possible to calculate this beating frequency generated as the result of the RFO between the devices. 
Figs.~\ref{fig:tii_beating_hist}e--h show the FFT and the calculated RFO for each of the four CT\_2 pairs depicted in Fig.~\ref{fig:tii_beating_hist},
where we calculate a low frequency of 496Hz for $Pair1$, while $Pair4$ exhibits a relatively high beating frequency of 7946Hz.  
Note that, while this process has been performed statically for the purposes of this paper, the computing resources available in modern IoT chipsets (such as the nRF52840, or the dual-core nRF5340) introduce the possibility of performing such calculations on-the-fly during runtime.}



\begin{figure}[t]
	\vspace{-1.0mm}
	\centering
	\begin{subfigure}[t]{0.467\columnwidth}
		\centering
		\includegraphics[width=1\columnwidth]{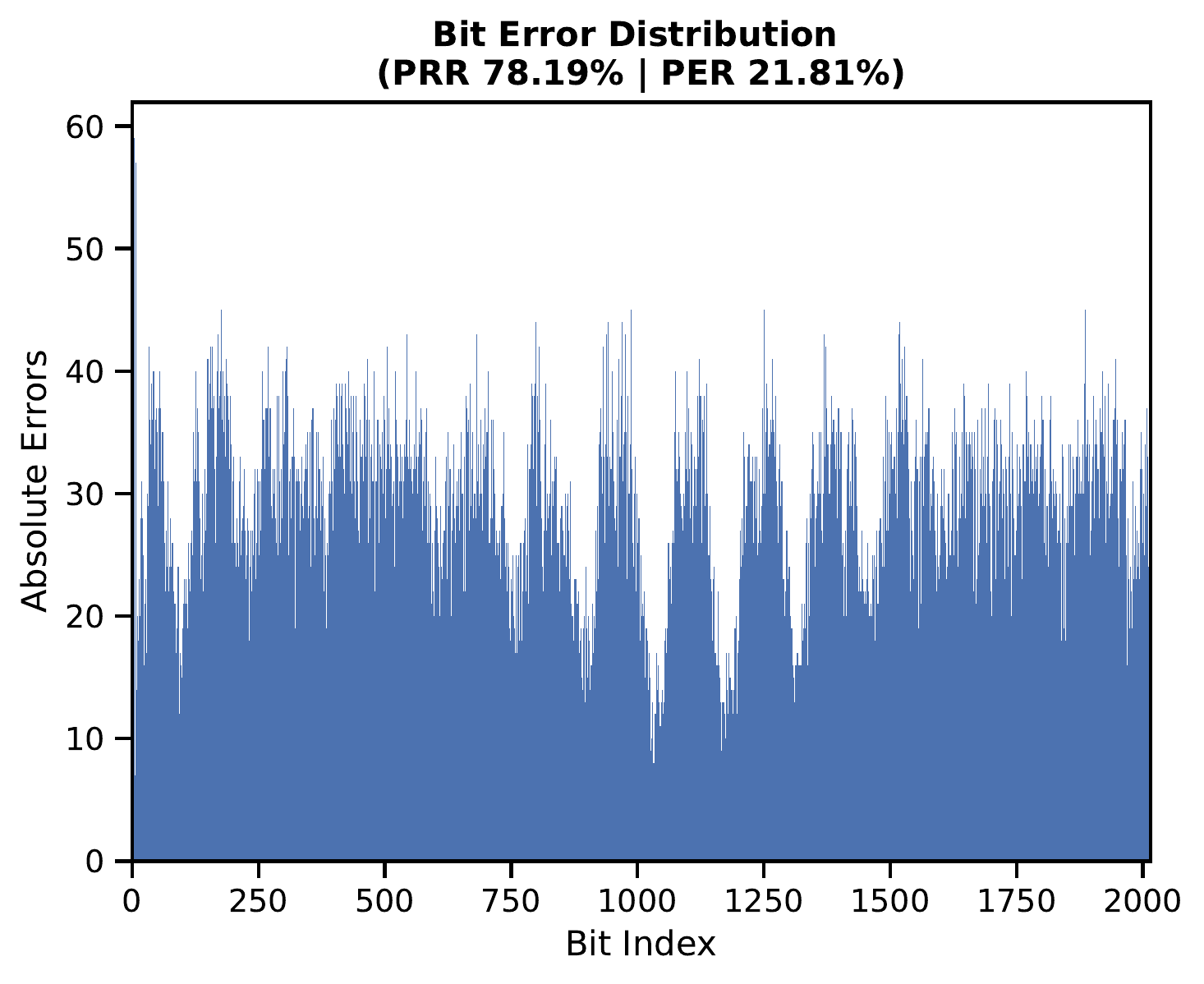}
		\vspace{-5.50mm}
		\caption{Histogram of complex beating pattern}
		\label{fig:tii_complex_hist}
	\end{subfigure}%
	\begin{subfigure}[t]{0.49\columnwidth}
		\centering
		\includegraphics[width=1\columnwidth]{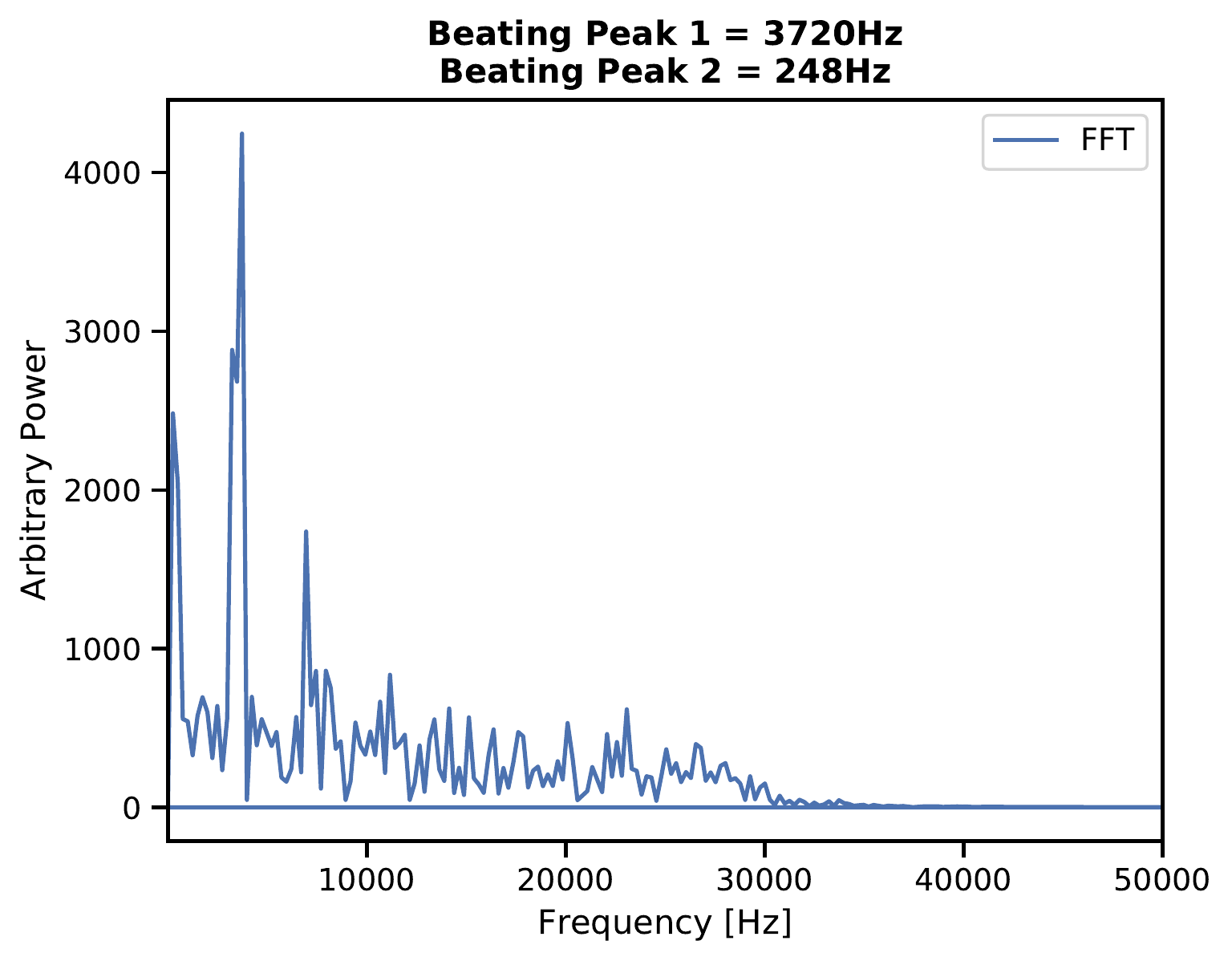}
		\vspace{-5.50mm}
		\caption{FFT of CT\_3 complex beating pattern}
		\label{fig:tii_complex_fft}
	\end{subfigure}
	\vspace{-2.50mm}
	\caption{\textbf{RFOs can be discerned even in \textit{complex} beating patterns.} Complex beating observed at a receiver as the result of oscillator offsets between three transmitter nodes (CT\_3). Multiple peaks in the FFT correspond to the frequencies observed in the histogram.}
	\vspace{-2.50mm}
	\label{fig:tii_complex}
\end{figure}

\boldpar{Complex beating resulting from multiple RFOs}
\newtext{Sect.~\ref{sec:background} introduced the concept of \emph{complex} beating when more than two devices are simultaneously transmitting. 
Fig.~\ref{fig:tii_complex_hist} shows the complex beating pattern generated by three transmitting devices (CT\_3), as observed through the histogram of bit errors at the receiver.
Fig.~\ref{fig:tii_complex_fft} shows the FFT of the treated signal, where the two highest peaks indicate the frequency of two waveforms at 248\,Hz and 3720\,Hz and correspond to the multiple RFOs between the devices. 
We observe that \emph{these two distinct sinusoidal beating patterns are clearly evident} as opposed to a single peak present in the CT\_2 FFTs in Fig.~\ref{fig:tii_beating_hist}. Analysis of these complex beating waveforms could therefore potentially indicate the number of concurrently transmitting nodes.}

\begin{figure}[t]
	\vspace{-1.75mm}
	\centering
	\begin{subfigure}[t]{0.42\columnwidth}
		\centering
		\includegraphics[width=1\columnwidth]{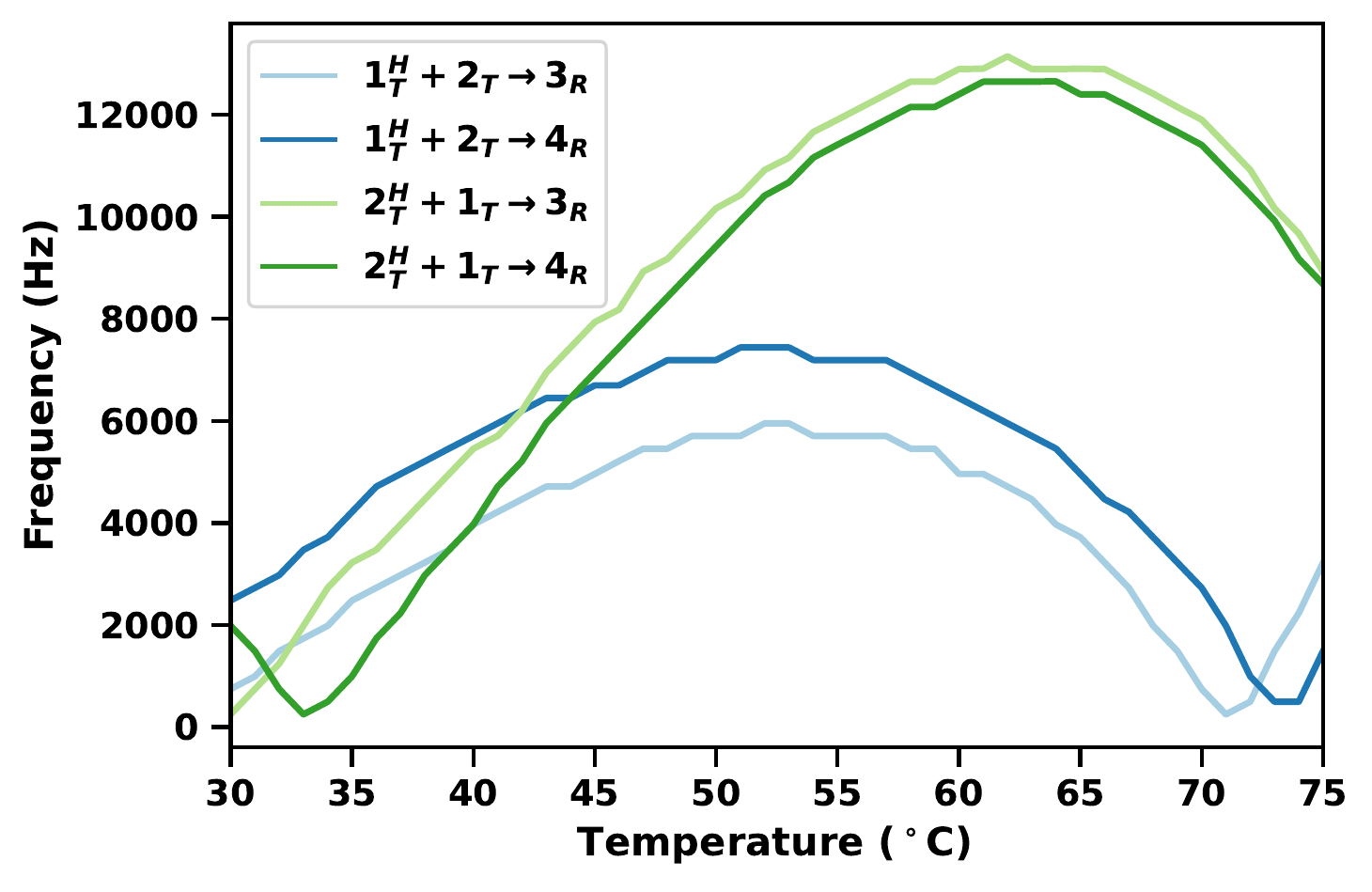}
		\vspace{-5.50mm}
		\caption{Transmission Pair Node 1 + Node 2}
		\label{fig:templab_n1n2_temp_freq}
	\end{subfigure}%
	\begin{subfigure}[t]{0.42\columnwidth}
		\centering
		\includegraphics[width=1\columnwidth]{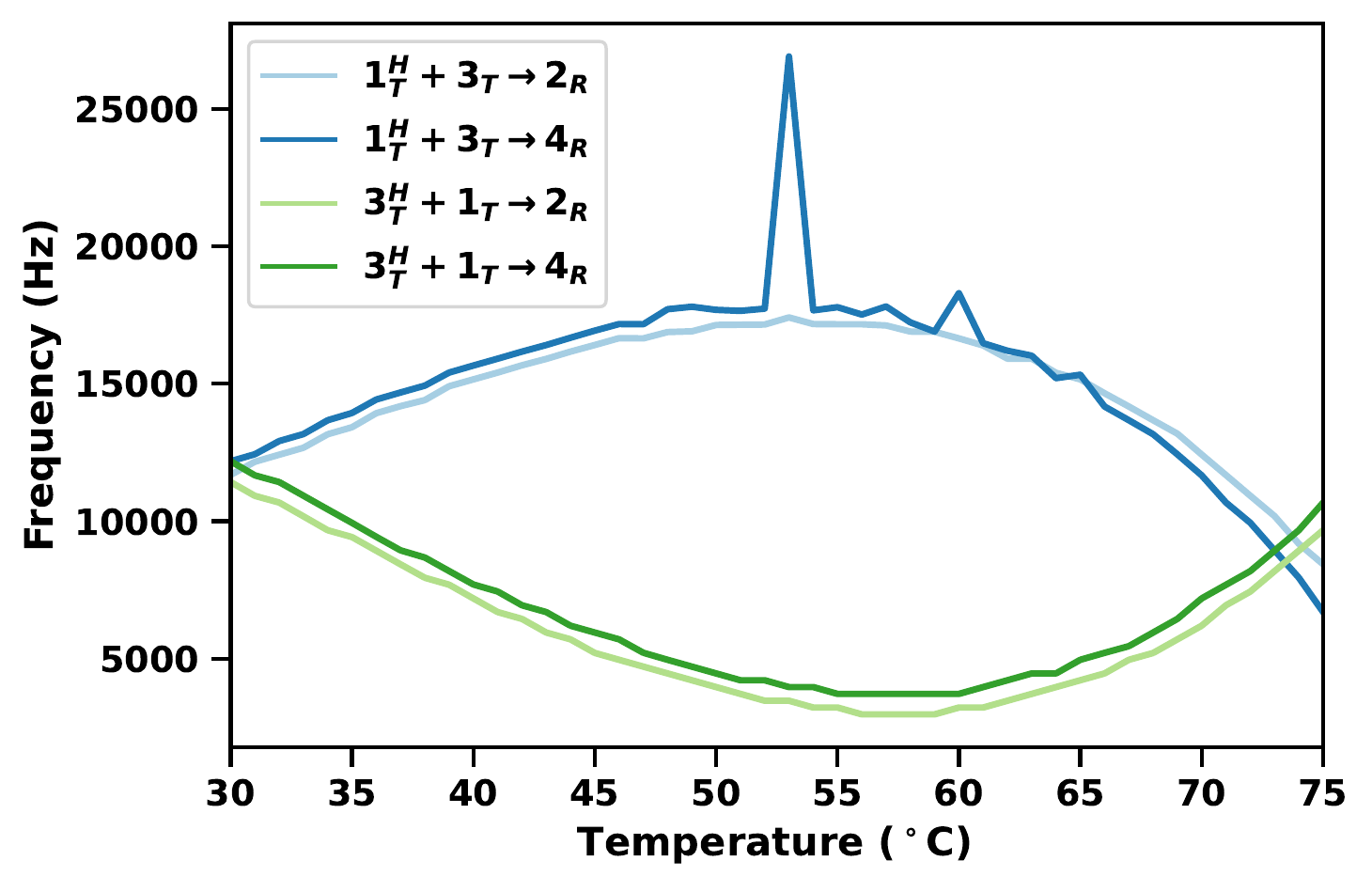}
		\vspace{-5.50mm}
		\caption{Transmission Pair Node 1 + Node 3}
		\label{fig:templab_n1n3_temp_freq}
	\end{subfigure}
	\begin{subfigure}[t]{0.42\columnwidth}
		\centering
		\includegraphics[width=1\columnwidth]{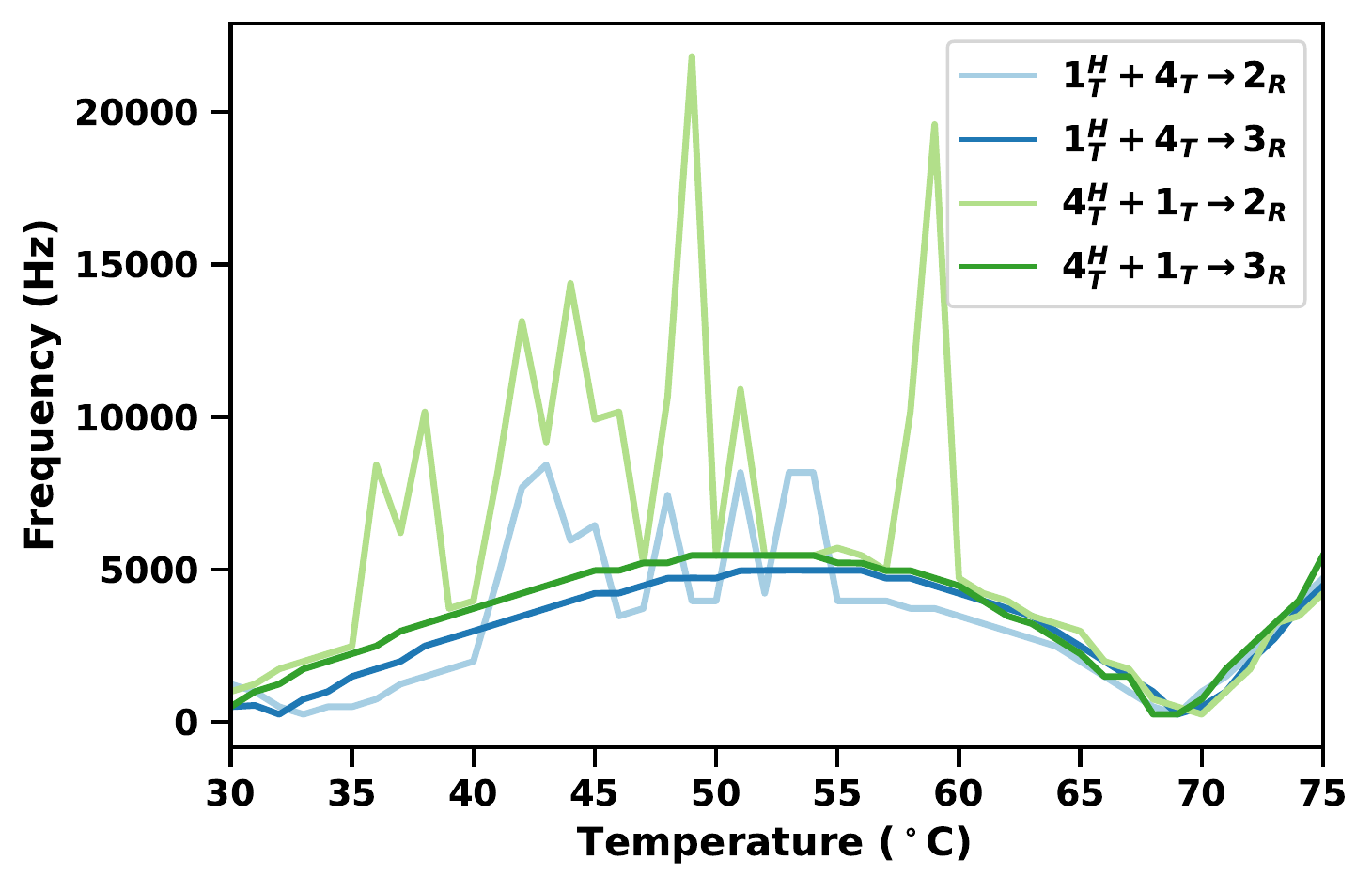}
		\vspace{-5.50mm}
		\caption{Transmission Pair Node 1 + Node 4}
		\label{fig:templab_n1n4_temp_freq}
	\end{subfigure}
	\begin{subfigure}[t]{0.42\columnwidth}
		\centering
		\includegraphics[width=1\columnwidth]{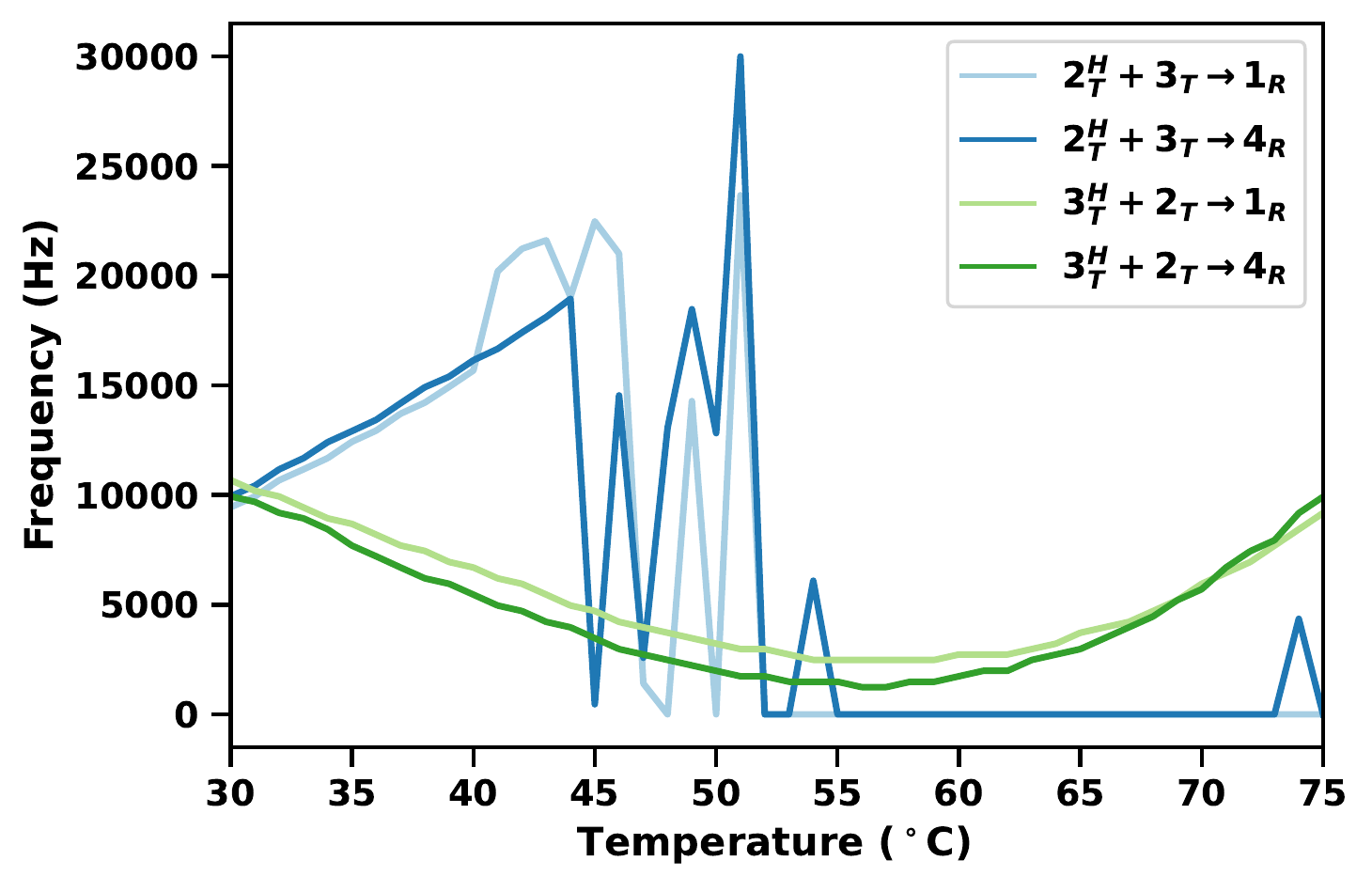}
		\vspace{-5.50mm}
		\caption{Transmission Pair Node 2 + Node 3}
		\label{fig:templab_n2n3_temp_freq}
	\end{subfigure}
	\begin{subfigure}[t]{0.42\columnwidth}
		\centering
		\includegraphics[width=1\columnwidth]{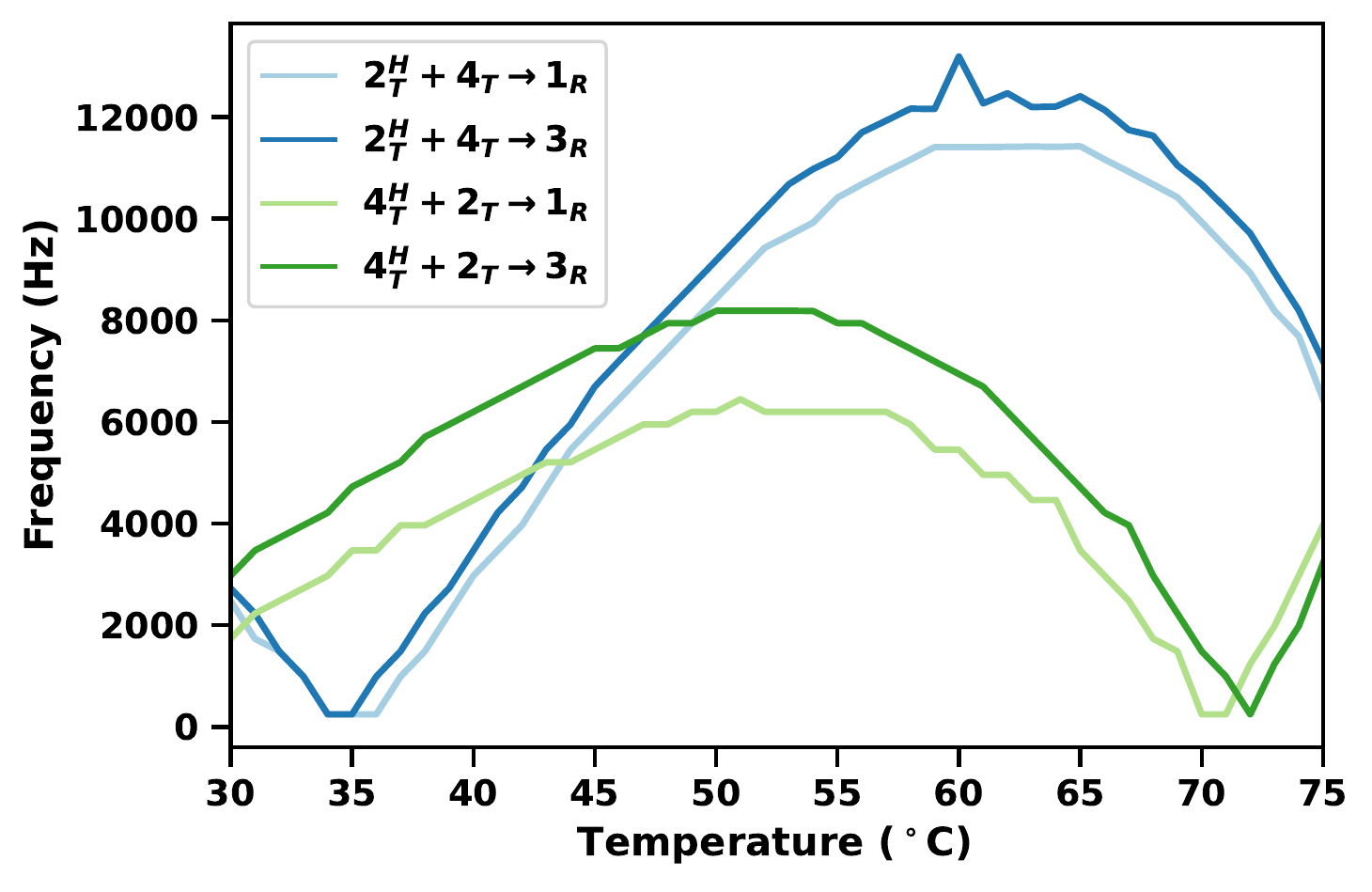}
		\vspace{-5.50mm}
		\caption{Transmission Pair Node 2 + Node 4}
		\label{fig:templab_n2n4_temp_freq}
	\end{subfigure}
	\begin{subfigure}[t]{0.42\columnwidth}
		\centering
		\includegraphics[width=1\columnwidth]{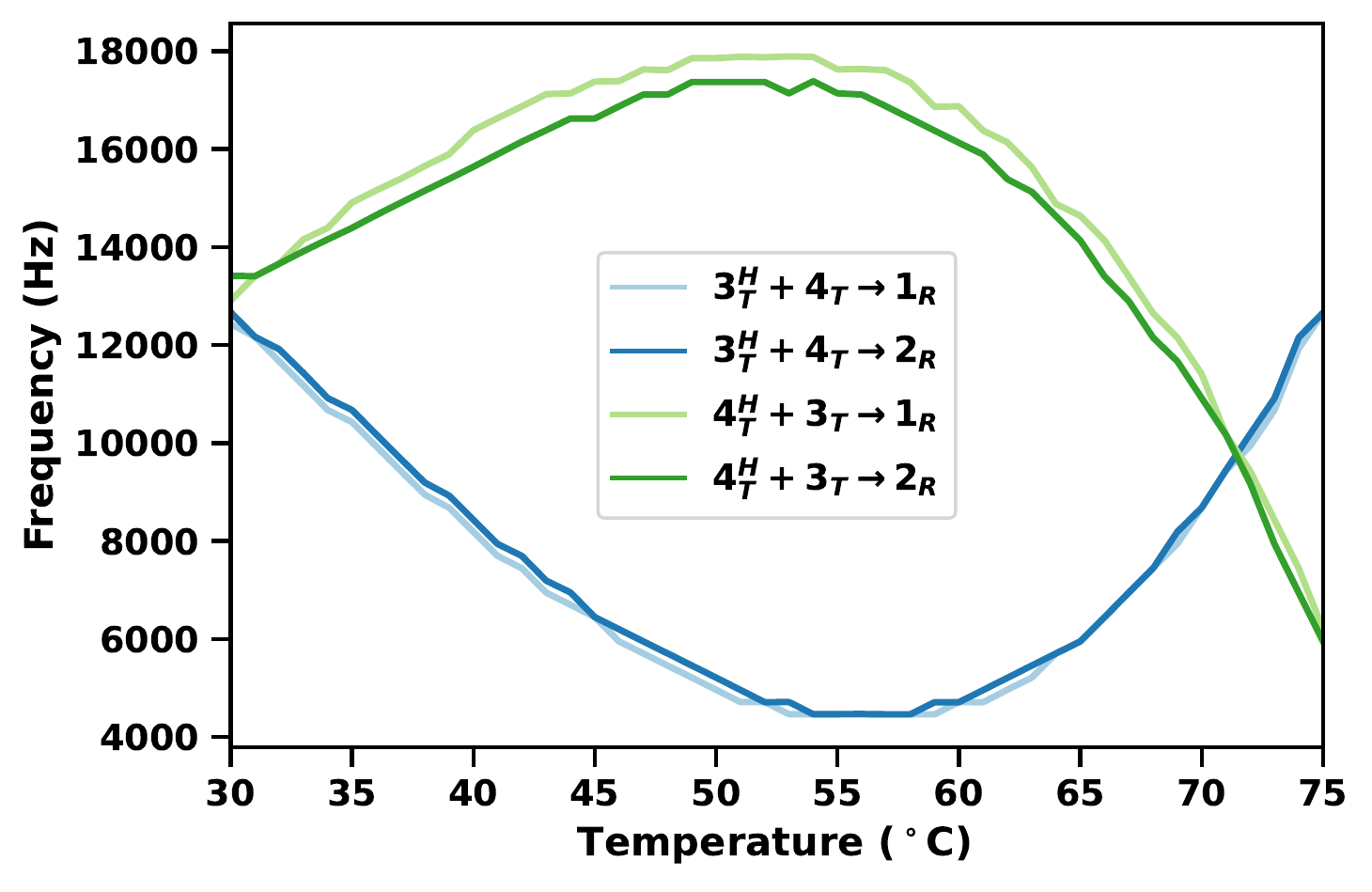}
		\vspace{-5.50mm}
		\caption{Transmission Pair Node 3 + Node 4}
		\label{fig:templab_n3n4_temp_freq}
	\end{subfigure}
	\caption{\textbf{Effect of temperature on beating frequency for node pairs in Fig.~\ref{fig:templab_all_freq_per}.} CFO changes induced by temperature are shown to affect beating frequency, regardless of the receiver.}
	\label{fig:templab_temp_freq}
	\vspace{-4.25mm}
\end{figure}

\begin{figure}[t!]
	\centering
	\begin{subfigure}[t]{0.42\columnwidth}
		\centering
		\includegraphics[width=1\columnwidth]{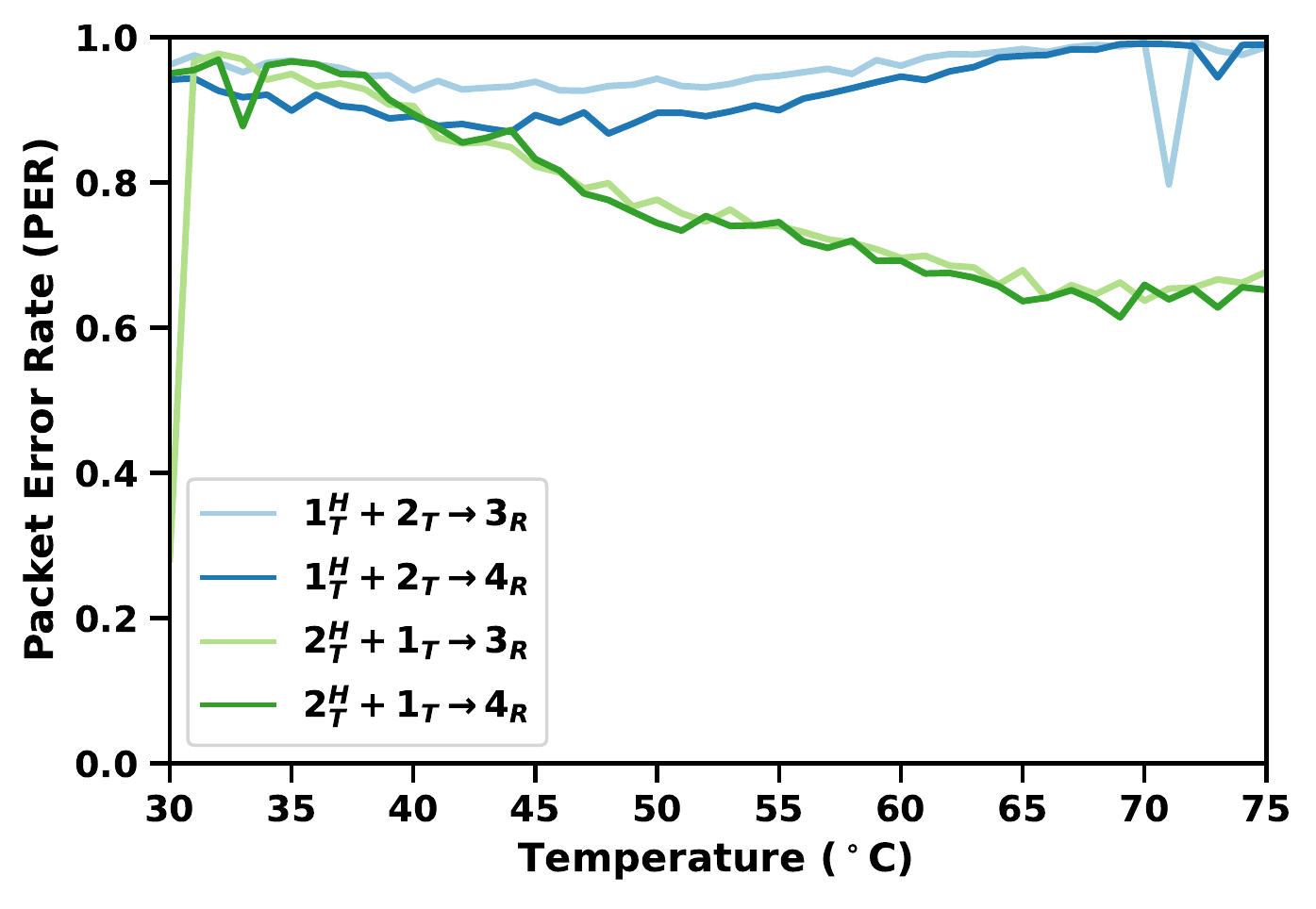}
		\vspace{-5.50mm}
		\caption{Transmission Pair Node 1 + Node 2}
		\label{fig:templab_n1n2_temp_per}
	\end{subfigure}%
	\begin{subfigure}[t]{0.42\columnwidth}
		\centering
		\includegraphics[width=1\columnwidth]{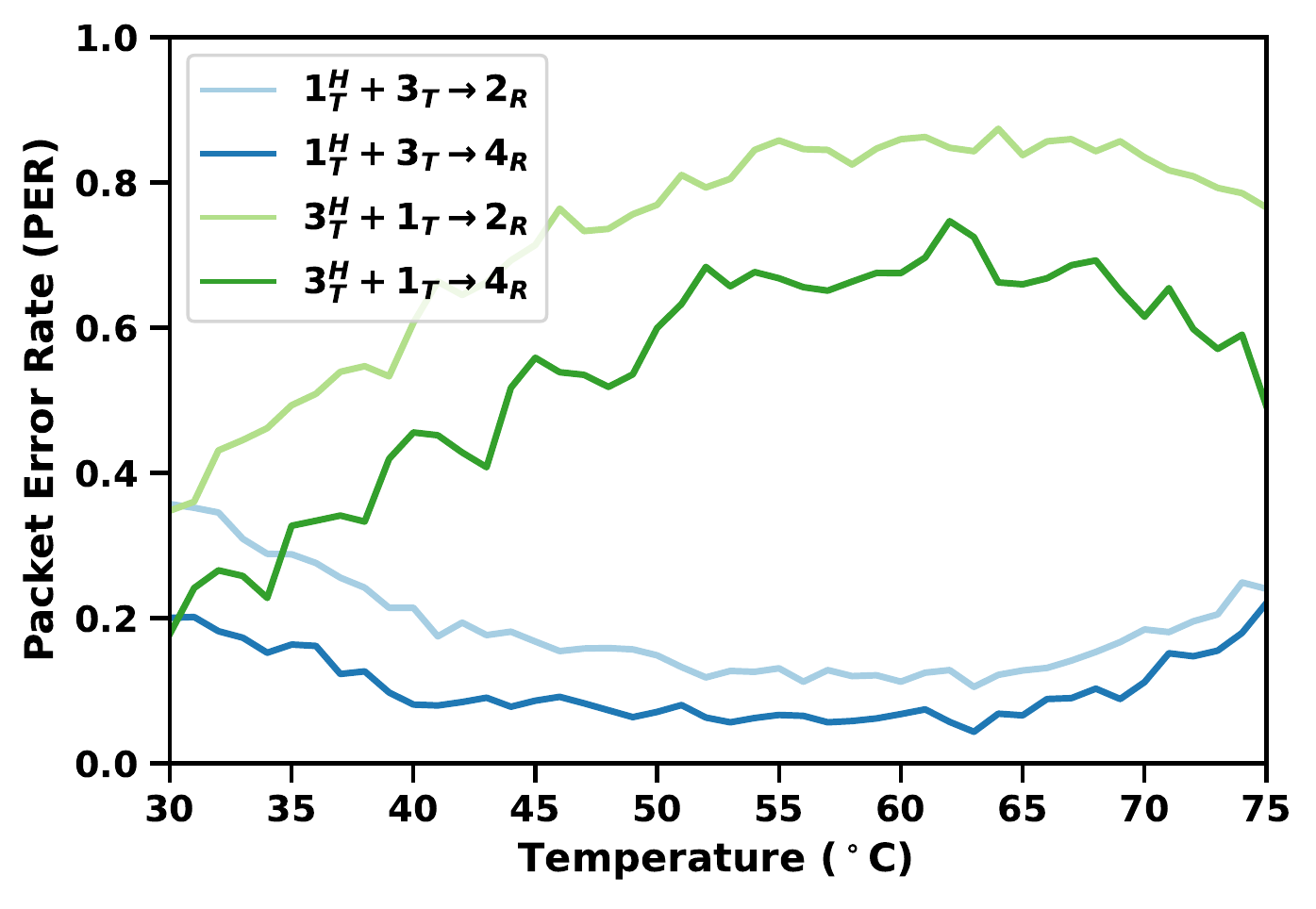}
		\vspace{-5.50mm}
		\caption{Transmission Pair Node 1 + Node 3}
		\label{fig:templab_n1n3_temp_per}
	\end{subfigure}
	\begin{subfigure}[t]{0.42\columnwidth}
		\centering
		\includegraphics[width=1\columnwidth]{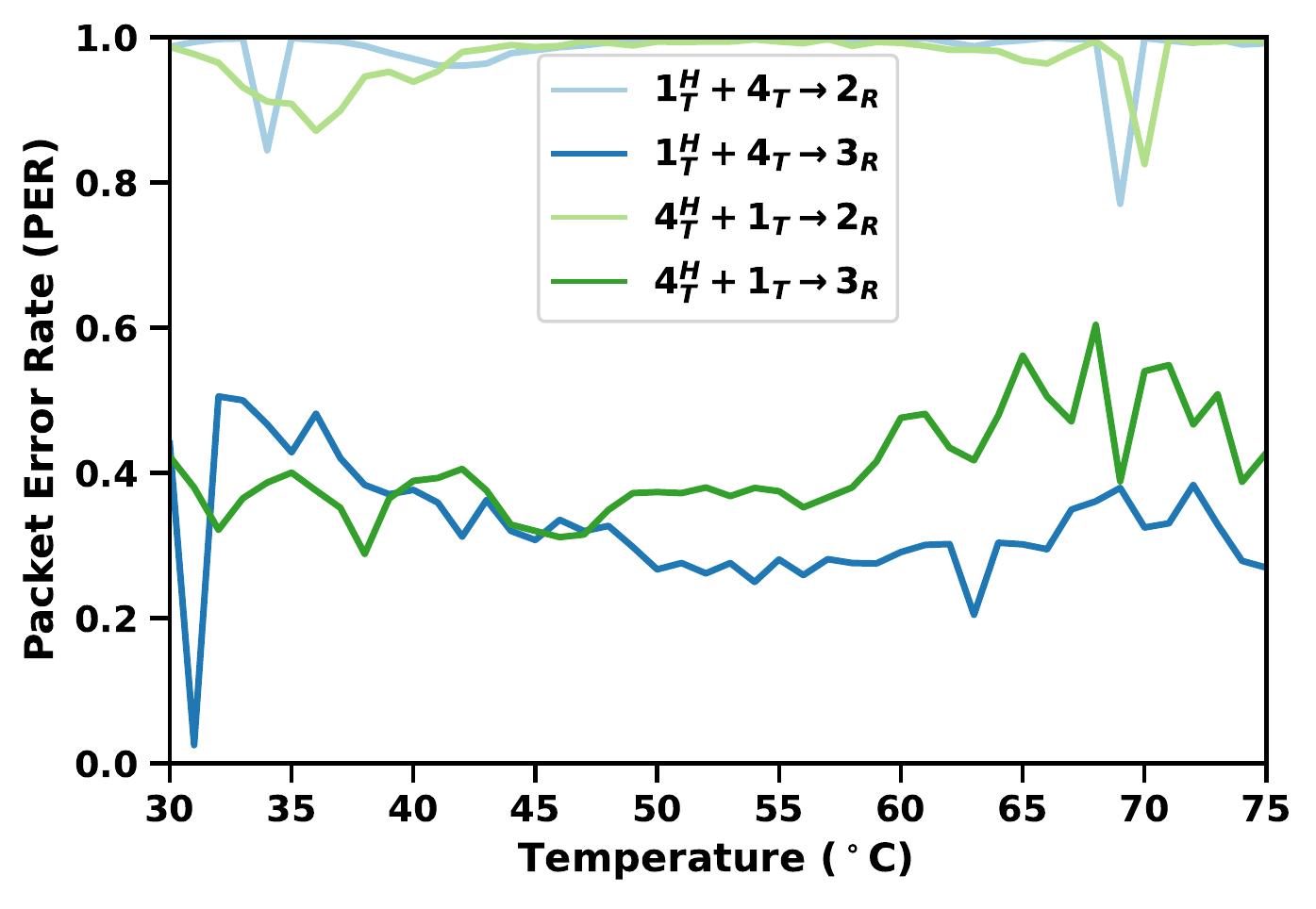}
		\vspace{-5.50mm}
		\caption{Transmission Pair Node 1 + Node 4}
		\label{fig:templab_n1n4_temp_per}
	\end{subfigure}
	\begin{subfigure}[t]{0.42\columnwidth}
		\centering
		\includegraphics[width=1\columnwidth]{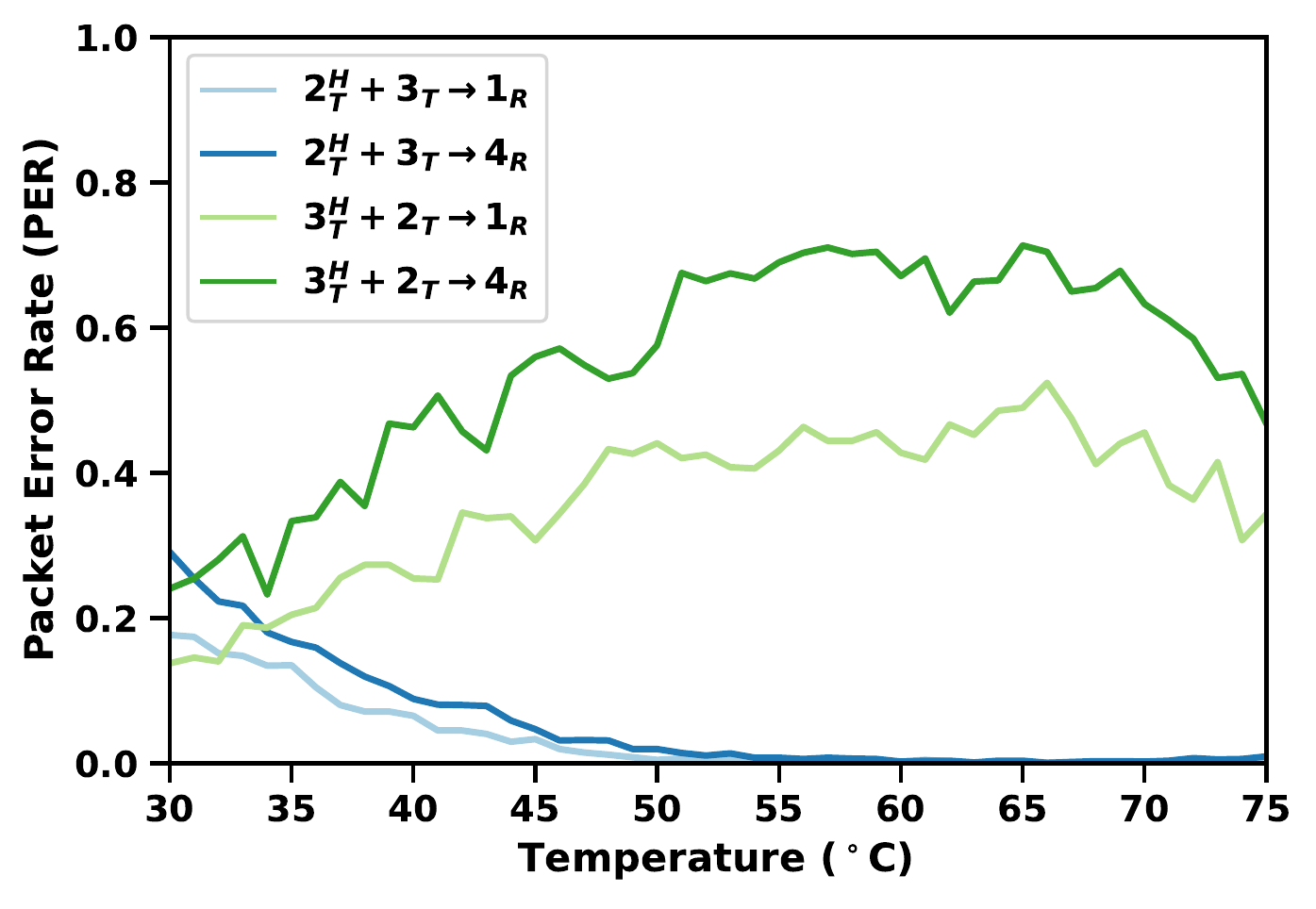}
		\vspace{-5.50mm}
		\caption{Transmission Pair Node 2 + Node 3}
		\label{fig:templab_n2n3_temp_per}
	\end{subfigure}
	\begin{subfigure}[t]{0.42\columnwidth}
		\centering
		\includegraphics[width=1\columnwidth]{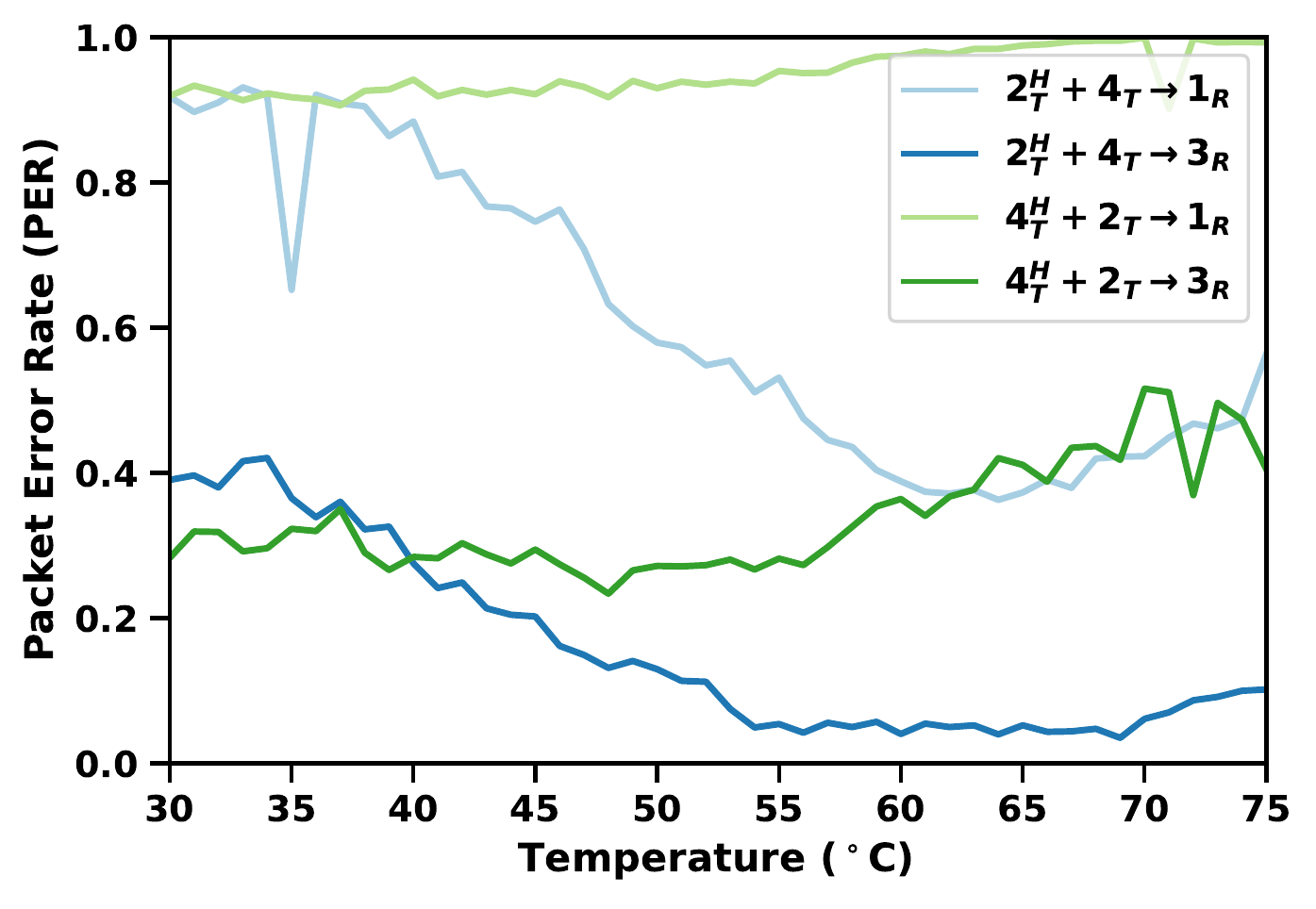}
		\vspace{-5.50mm}
		\caption{Transmission Pair Node 2 + Node 4}
		\label{fig:templab_n2n4_temp_per}
	\end{subfigure}
	\begin{subfigure}[t]{0.42\columnwidth}
		\centering
		\includegraphics[width=1\columnwidth]{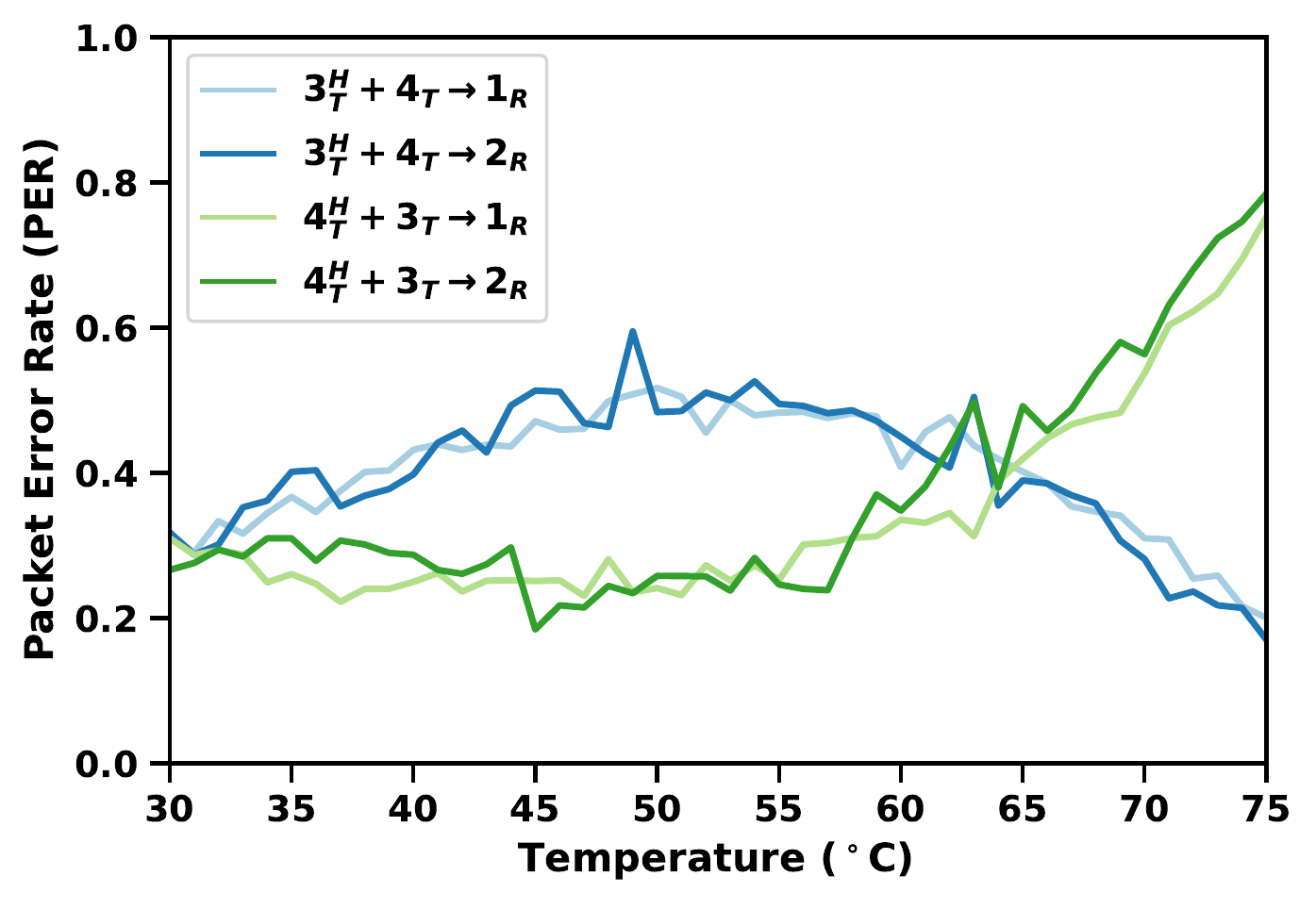}
		\vspace{-5.50mm}
		\caption{Transmission Pair Node 3 + Node 4}
		\label{fig:templab_n3n4_temp_per}
	\end{subfigure}
	\vspace{-3.50mm}
	\caption{\textbf{Effect of temperature on Packet Error Rate (PER) for node pairs in Fig.~\ref{fig:templab_all_freq_per}.} Beating frequency changes induced by temperature changes can be seen to directly affect reliability at the receiver. While PER is also governed by other factors such as noise floor, PER changes in response to temperature generally follow similar curve regardless of the receiver.}
	\label{fig:templab_temp_per}
	\vspace{-4.25mm}
\end{figure}

\begin{figure}[t]
	\centering
	\includegraphics[width=.92\columnwidth]{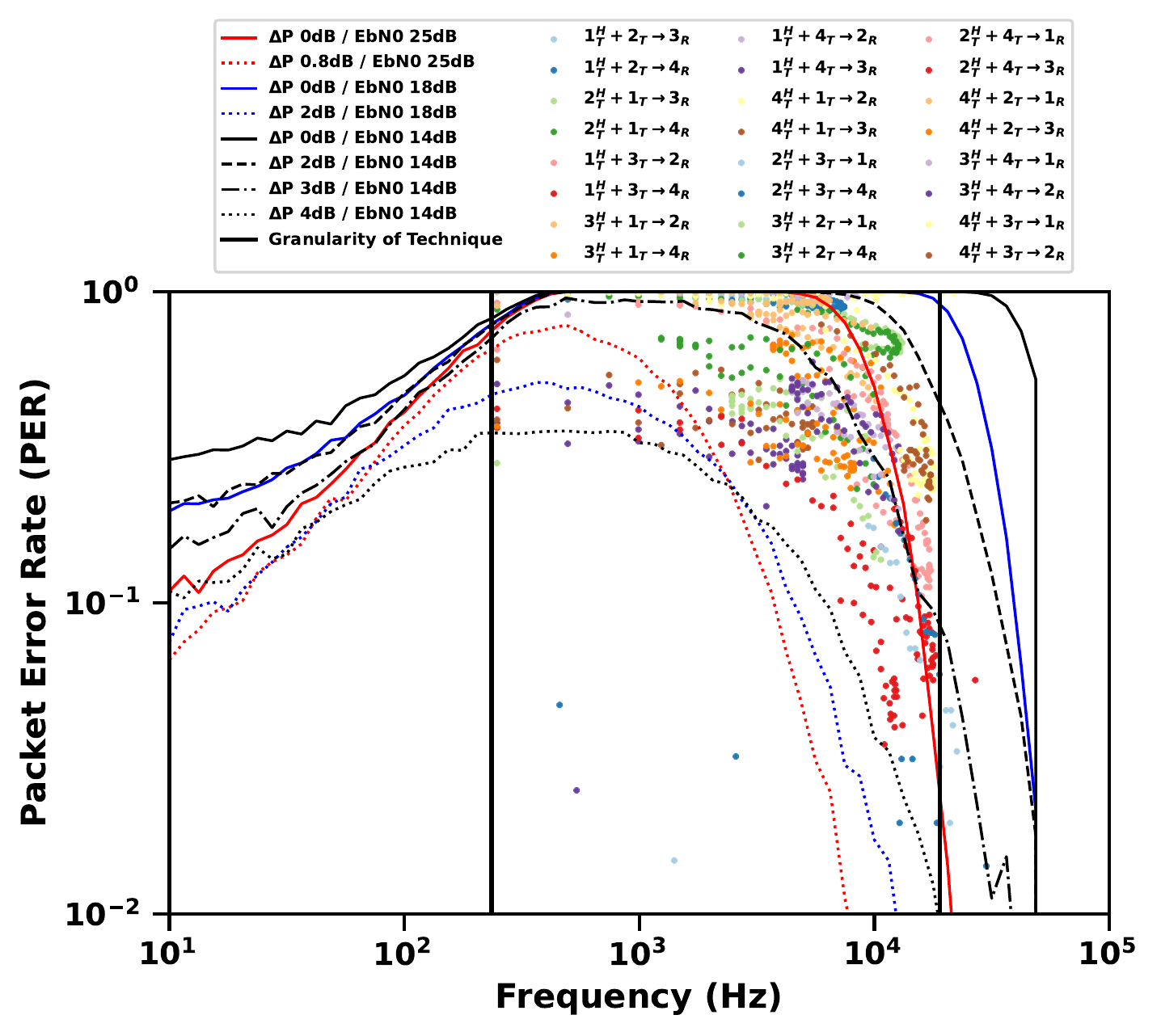}
	\vspace{-3.75mm}
	\caption{\textbf{Packet error rate vs. temperature induced beating frequency for all transmitting pair / destination configurations between a set of four nodes $\{1, 2, 3, 4\}$.} The PER response w.r.t. frequency is shown to be within the bounds expected by simulations depicting a 14\,dB $EbN0$ channel where transmission power delta is between 2\,dB to 4\,dB.}
	\vspace{-1.75mm}
	\label{fig:templab_all_freq_per}
\end{figure}

\begin{figure}[t]
	\vspace{-1.75mm}
	\centering
	\begin{subfigure}[t]{0.42\columnwidth}
		\centering
		\includegraphics[width=1\columnwidth]{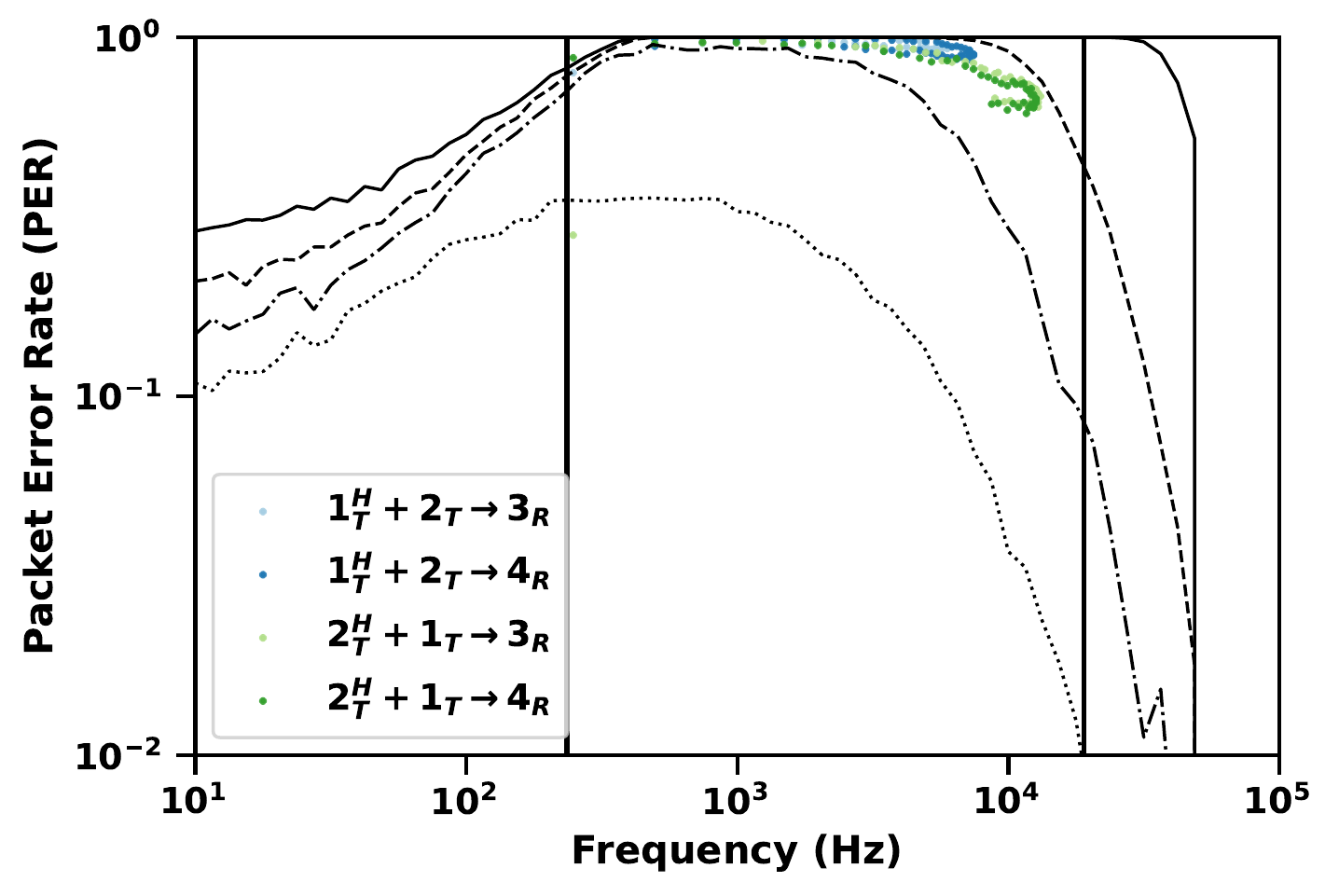}
		\vspace{-5.50mm}
		\caption{Transmission Pair Node 1 + Node 2}
		\label{fig:templab_n1n2_freq_per}
	\end{subfigure}%
	\begin{subfigure}[t]{0.42\columnwidth}
		\centering
		\includegraphics[width=1\columnwidth]{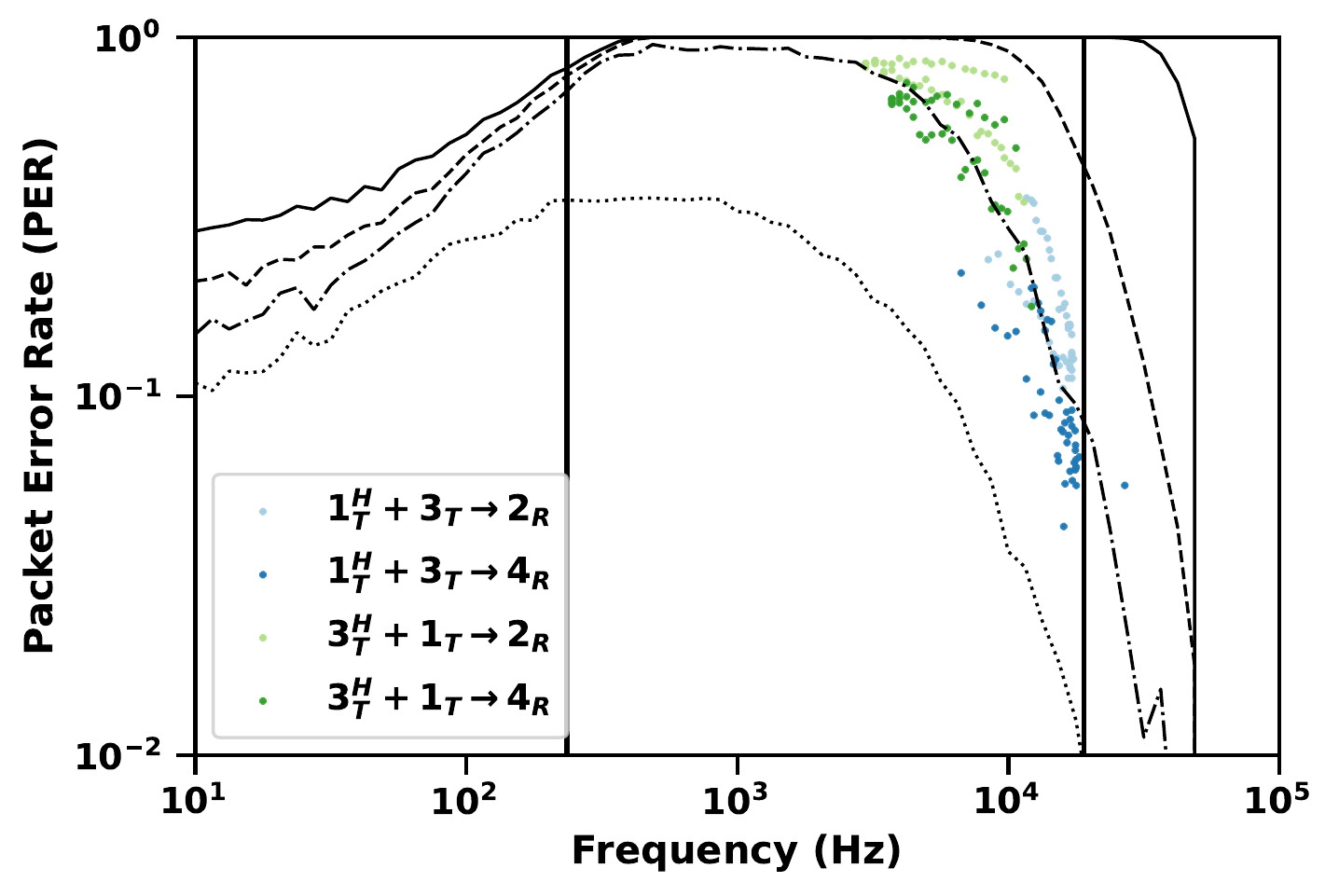}
		\vspace{-5.50mm}
		\caption{Transmission Pair Node 1 + Node 3}
		\label{fig:templab_n1n3_freq_per}
	\end{subfigure}
	\begin{subfigure}[t]{0.42\columnwidth}
		\centering
		\includegraphics[width=1\columnwidth]{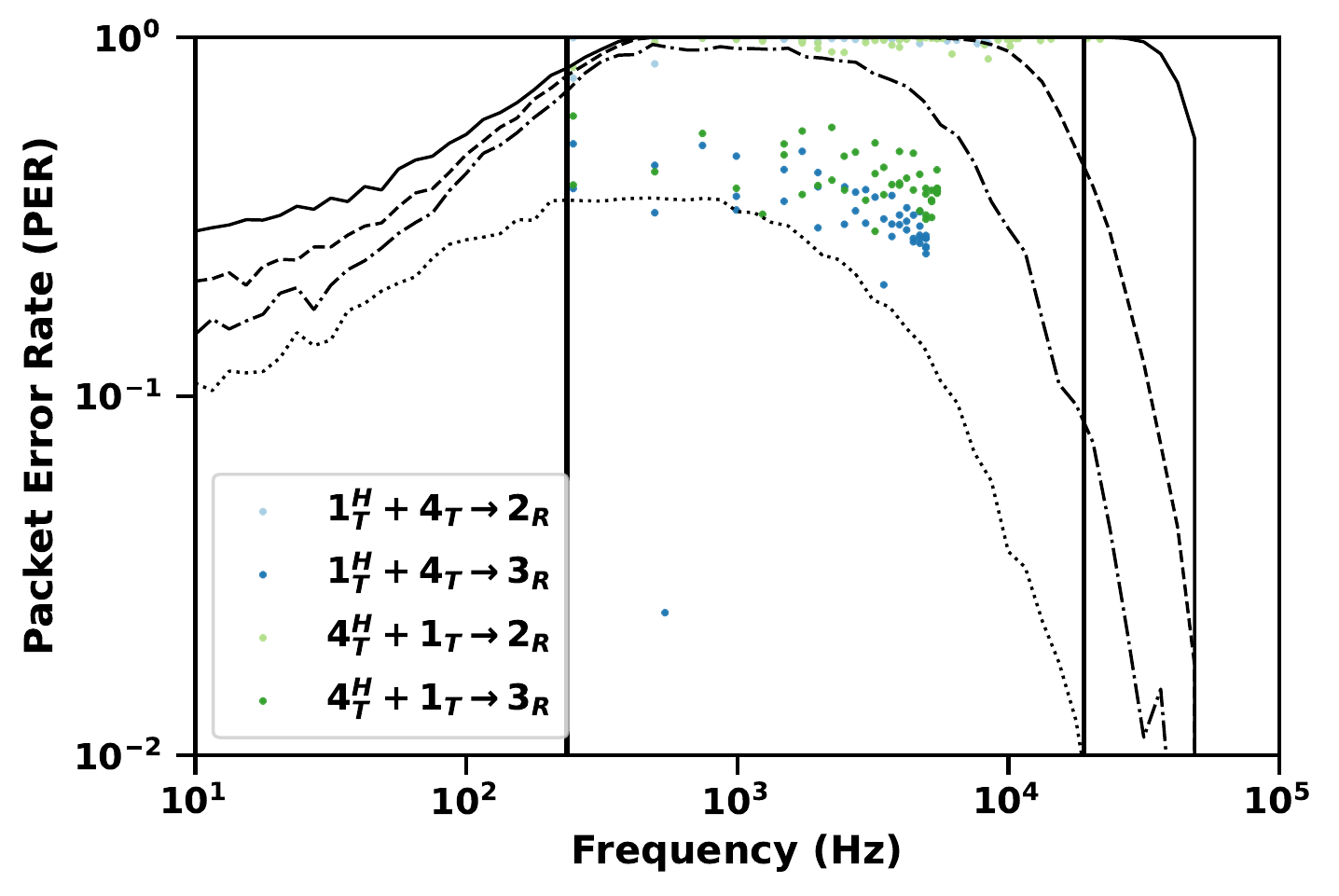}
		\vspace{-5.50mm}
		\caption{Transmission Pair Node 1 + Node 4}
		\label{fig:templab_n1n4_freq_per}
	\end{subfigure}
	\begin{subfigure}[t]{0.42\columnwidth}
		\centering
		\includegraphics[width=1\columnwidth]{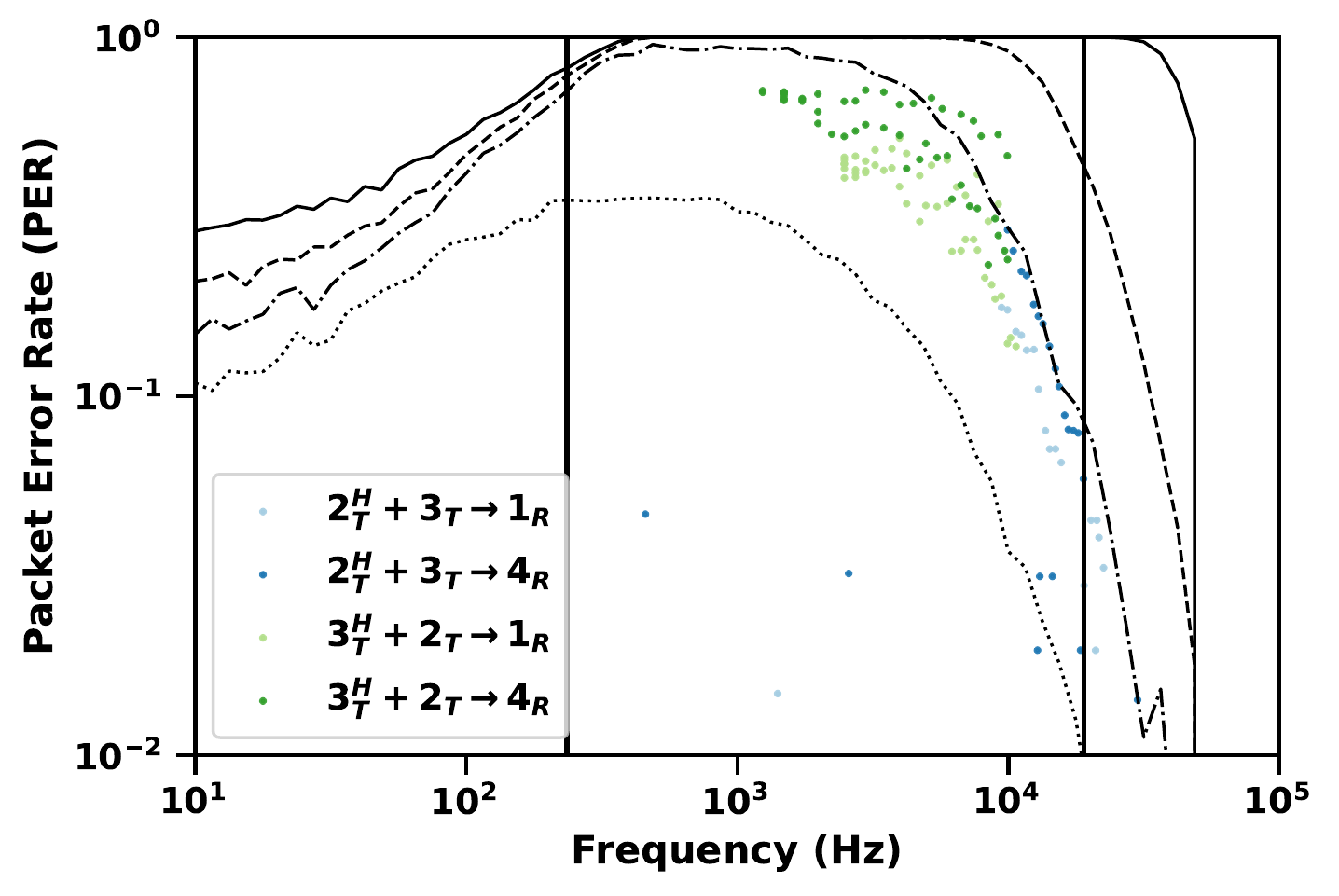}
		\vspace{-5.50mm}
		\caption{Transmission Pair Node 2 + Node 3}
		\label{fig:templab_n2n3_freq_per}
	\end{subfigure}
	\begin{subfigure}[t]{0.42\columnwidth}
		\centering
		\includegraphics[width=1\columnwidth]{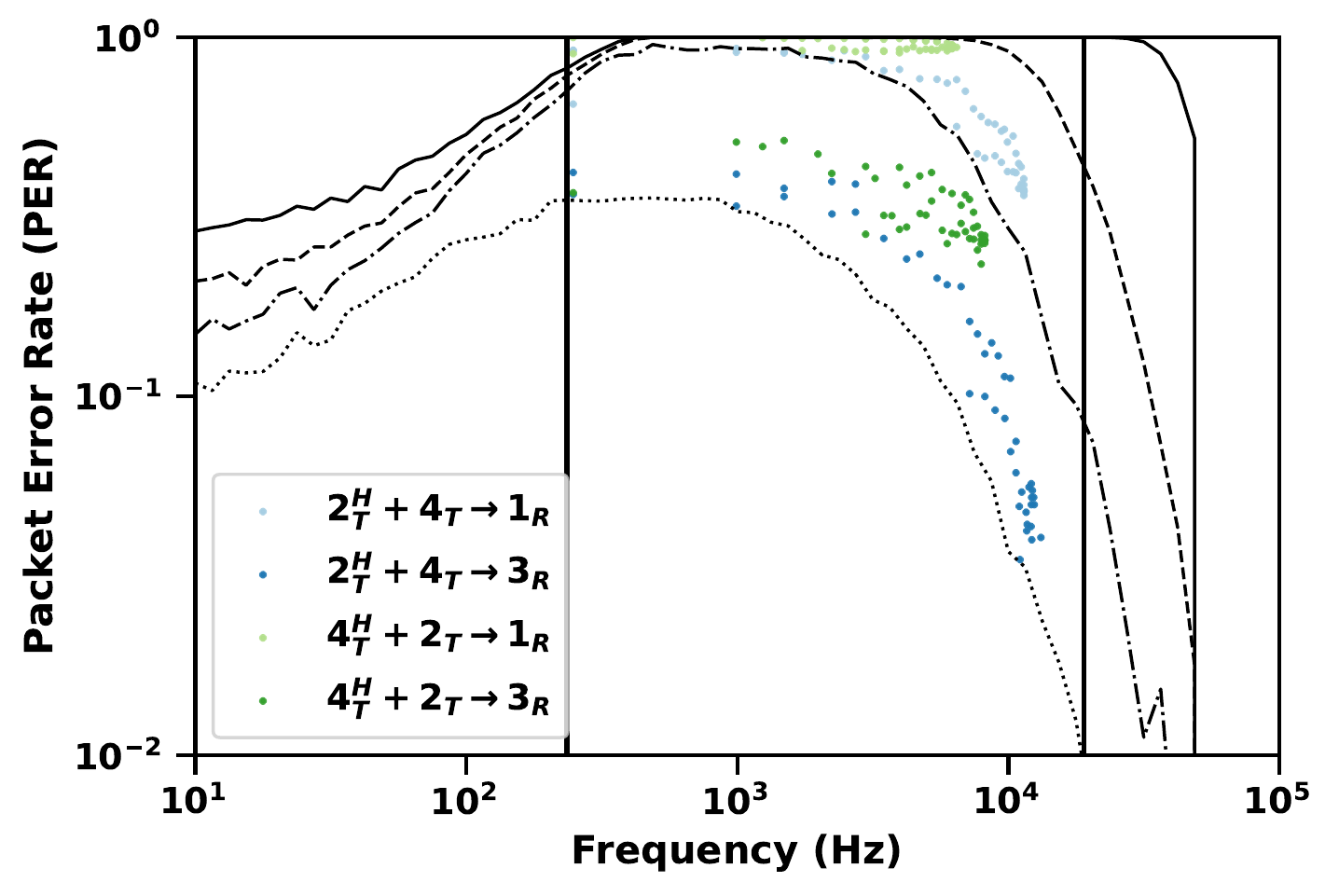}
		\vspace{-5.50mm}
		\caption{Transmission Pair Node 2 + Node 4}
		\label{fig:templab_n2n4_freq_per}
	\end{subfigure}
	\begin{subfigure}[t]{0.42\columnwidth}
		\centering
		\includegraphics[width=1\columnwidth]{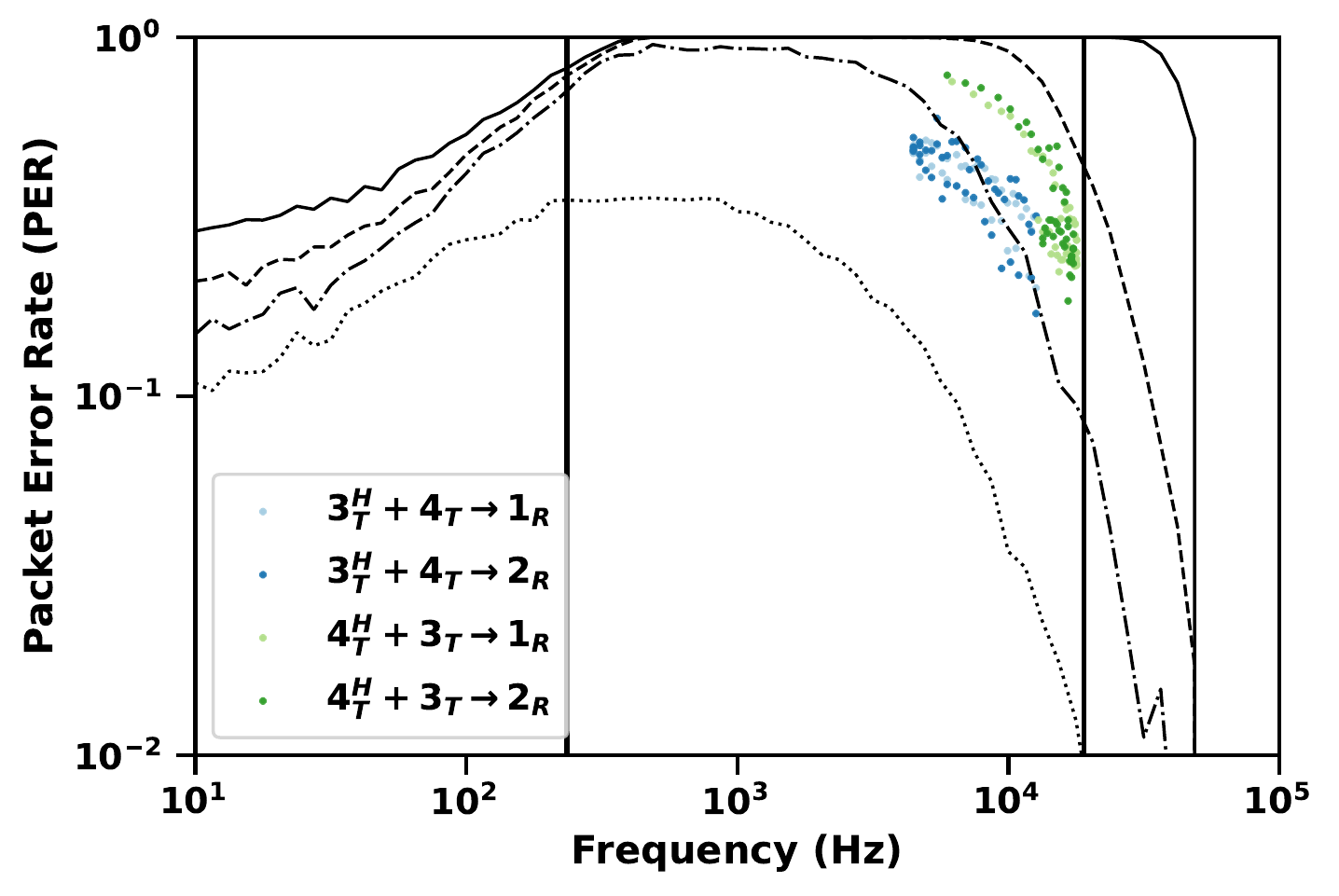}
		\vspace{-5.50mm}
		\caption{Transmission Pair Node 3 + Node 4}
		\label{fig:templab_n3n4_freq_per}
	\end{subfigure}
	\vspace{-3.00mm}
	\caption{\textbf{Beating frequency vs. Packet Error Rate (PER) for all node pairs in Fig.~\ref{fig:templab_all_freq_per}.} The black curves (solid, dashed, and dotted) indicate the $EbN0$=14\,dB simulation results for $\Delta$P=\{0,2,3,4\}\,dB, while the solid black vertical lines indicate the approx. beating frequency range detectable by our technique -- running an FFT over the bit error histogram.}
	\label{fig:templab_freq_per}
	\vspace{-2.50mm}
\end{figure}

\boldpar{Effect of temperature on beating}
\newtext{
\response{Since common off-the-shelf transceivers use non temperature compensated crystal oscillators}, as temperature increases, the frequency is expected to decrease~\cite{zhou2008frequency}. 
Since the CFO of a device is linked to this oscillator frequency, a change in temperature will result in a CFO change on a transmitting node. This subsequently affects the RFO between two concurrently transmitting nodes, and therefore the perceived beating pattern at the receiver. 
For example, if we have two concurrently transmitting nodes $TX1$ and $TX2$, where the carrier frequency of $TX1$ is greater than that of $TX2$ (i.e., $CF_{TX1} > CF_{TX2}$), \emph{increasing} the temperature at $TX1$ will \emph{decrease} its oscillator (carrier) frequency and therefore \emph{reduce the RFO between the two devices}.
As the temperature at $TX1$ continues to increase, the $CF_{TX1}$ may eventually drop below that of $CF_{TX2}$ (i.e., $CF_{TX1} < CF_{TX2}$). At such a point the RFO will no longer decrease, but will instead increase. Figs.~\ref{fig:templab_temp_freq},~\ref{fig:templab_temp_per},~\ref{fig:templab_all_freq_per}, and~\ref{fig:templab_freq_per} examine the effect of temperature on pairs of concurrently transmitting nodes, where the temperature of one node is gradually increased \response{(between 30$^{\circ}$C to 75$^{\circ}$C at 1$^{\circ}$C intervals}\footnote{\response{As TempLab does not offer cooling capabilities allowing testing below 30$^{\circ}$C, we select a temperature range emulating the variations in both ambient and direct sunlight scenarios during summer. Note that what is relevant is the \textit{difference} in temperature and not the absolute temperature, i.e., the CFO changes induced by temperature are expected to also affect beating frequency when temperature goes below 30$^\circ C$.}}). \response{This is done for all possible permutations of two transmitting nodes to a single receiving node within in the set of four nodes $\{1, 2, 3, 4\}$. In Figs.~\ref{fig:templab_temp_freq}--\ref{fig:templab_freq_per}, $X_T^H + Y_T - Z_R$ indicates a transmitting node pair $X_T + Y_T$ (where the subscript `$T$' denotes transmission and node $X$ is heated and denoted by the superscript `$H$') and the CT are recorded at the receiving node $Z$ (where the subscript `$R$' denotes reception)}. For these particular experiments, coaxial cabling was used to reduce the multipath effects that could potentially distort the beating histogram.
}


\newtext{
In Fig.~\ref{fig:templab_temp_freq} we observe the beating frequency curves follow similar trends as a function of temperature.
In each case, we heat one of the $T_x$ nodes while keeping the other fixed and send to two different destinations (i.e., the remaining two nodes in the set $\{1, 2, 3, 4\}$). We then repeat the experiment while heating the other $T_x$ node.
As the crystal oscillator of each node has unique imperfections of the manufacturing process, each response is slightly different. Figs.~\ref{fig:templab_n1n2_temp_freq}~to~\ref{fig:templab_n3n4_temp_freq} show each of the 6 possible transmitting node pairs in the set of nodes $\{1, 2, 3, 4\}$. We further observe that the response of the beating frequency remains consistent regardless of the destination, and conclude that the beating frequency at a receiving node depends \emph{solely} on the CFO of the transmitting nodes. 
}

\newtext{
Fig.~\ref{fig:templab_temp_per} shows the Packet Error Rate (PER) at each receiver during those experiments. It can be seen that, as with beating frequency in Fig.~\ref{fig:templab_temp_freq}, the PER response follows a curve w.r.t. temperature. However, while the rate of change of this response is agnostic to the receiving node (i.e., the shape of the curve remains consistent, regardless of which node receives), Figs.~\ref{fig:templab_n1n4_temp_per}~and~\ref{fig:templab_n2n4_temp_per}, in particular, indicate that other factors such as the noise floor at the receiving node may introduce a constant offset in the error rate. In these cases, we see considerable difference in the PER (around 60\%) when transmitting to different receivers. Interestingly, Figs.~\ref{fig:templab_n1n4_temp_freq}~and~\ref{fig:templab_n2n3_temp_freq} exhibit large erroneous spikes at around 37-62 degrees in Fig.~\ref{fig:templab_n1n4_temp_freq} and 40-56 degrees in Fig.~\ref{fig:templab_n2n3_temp_freq}. These can be explained when comparing to their respective plots in Fig.~\ref{fig:templab_temp_per}. In Fig.~\ref{fig:templab_n1n4_temp_per} there is a high PER for the $1_T^H + 4_T$ node pair (regardless of receiving node), where a high noise floor has reduced the definition of the waveform in the histogram, reducing the ability of the FFT to quantify the beating frequency. In Fig.~\ref{fig:templab_n2n3_temp_per} node pair $2_T^H + 3_T$ (again regardless of receiving node) experiences relatively high reliability, reducing the number of samples available for the FFT.
}

\newtext{
Fig.~\ref{fig:templab_all_freq_per} compares the results from Fig.~\ref{fig:templab_temp_freq}~and~\ref{fig:templab_temp_per} to show the PER response w.r.t. frequency changes (induced by temperature), while Fig.~\ref{fig:templab_freq_per} breaks down this comparison into the node pairs used in those previous figures. We use the simulator from Sect.~\ref{sec:simulation} to plot the expected PER response to beating frequency for various power delta~($\Delta$P) and normalised SNR~($EbN0$) values. Specifically, we simulate an SNR of 25\,dB, 18\,dB, and 14\,dB as a reference for high to low SNR values. Furthermore, while Nordic \texttt{nRF52840} radios may deviate from their set transmit power by $+/-$4\,dB across a Gaussian distribution (meaning that any two transmitting nodes may potentially have a power delta of up to 8\,dB), under normal conditions we assume this will generally be $+/-$2\,dB~\cite{nrf52840_productsheet}. We observe that the trend when plotting frequency against the PER generally matches that of the simulations. In particular, the curves most accurately follow the 14\,dB $EbN0$ simulations between 2\,dB to 4\,dB power delta. We therefore include these as reference in the breakdown of results into node pairs in Fig.~\ref{fig:templab_freq_per}. Furthermore the vertical black lines on the figures indicate the approximate granularity of our beating estimation characterization (generating a histogram of bit errors and running an FFT over the resulting waveform) -- i.e., our technique is able to provide adequate resolution to discern beating frequencies approximately $< 250$\,Hz and $> 20000$\,Hz in \blefive500K over 250 bytes packets, which results in 2000 data points sampled at $500$\,kHz. 
}





\subsection{Key Observations}
Recent literature has theorized that the beating effect should have a significant impact on CT performance~\cite{alnahas2020blueflood,schaper2019truth}.  
The experimental results presented in this section have shown that beating is present in both coded and uncoded \blefive PHYs, as well as in the DSSS-based \ieee PHY. We summarize these results by outlining a number of key observations.

\boldpar{Beating frequencies are device-specific} As shown by Fig.~\ref{fig:tosh_layoutcomp}, beating frequencies depend on the RFO between device pairs, and one cannot directly extrapolate results from a specific pair.

\boldpar{Preambles are sensitive to beating} While the start of a preamble can randomly coincide with a beating valley or peak, these results are relative to \emph{received} packets (i.e., those for which the preamble was successfully detected) and hence have bias towards a certain initial phase relationship.
This bias is further increased by calibration performed by a receiver during the preamble's reception~\cite{ble5specs}. As the packet is being received, the beating changes the signal properties and this calibration is no longer optimal; 
however, the sinusoidal beating frequency will periodically return the signal to that initial calibration.
This also explains why beating is visible through an error distribution analysis, and that with no bias (i.e., no preamble) it would present as a flat error distribution.

\boldpar{High data rate PHYs benefit from packet repetition} As shown by Fig.~\ref{fig:tosh_phycomp}, packet transmissions in high data rate PHYs span fewer beating valleys (potentially zero if the packet period is shorter than the beating period). Since the position of peaks and valleys is random, after several repetitions, i.e., with a higher \texttt{TX\_N}, it is likely that a packet will not experience a valley during the preamble and will be correctly received. Note that \texttt{TX\_N} is a core component of many CT-based protocols.

\boldpar{Low data rate PHYs benefit from the coding gain} Fig.~\ref{fig:tosh_phycomp_ratio} shows a significant difference in reliability between the \blefive500K and 125K\,PHYs. Indeed, \blefive500K exhibits the worst performance of any of the PHYs compared in this section. It is likely that the convolutional coding employed by \blefive is sensitive to beating errors, while the gain seen using the 125K PHY stems from the additional pattern mapper redundancy over the payload (as mentioned in Sect.~\ref{sec:simulation}). Similarly, while not achieving the same gains of \blefive 125K PHY, the DSSS used in \ieee halves the PER in comparison to the \blefive500K PHY. It is worth noting that results in this section do not consider significant external noise or interference, which may particularly penalize the very long packets of the \blefive125K PHY, as seen in the following section, and the use of the 500K PHY may again constitute an effective trade-off in harsh environments, particularly to survive intermittent jamming.

\boldpar{RFO between devices can be quantified by observation of the beating waveforms} \newtext{Fig.~\ref{fig:tii_beating_hist} shows how, by applying an FFT over the bit error histogram observed at a receiver, it is possible to calculate the beating frequency, and that this frequency is unique to each CT pair. 
Fig.~\ref{fig:templab_temp_freq} subsequently shows that the frequency curve w.r.t. temperature is similar regardless of the receiving (observing) device. This has important security and privacy implications. Given the increasing performance capabilities of modern low-power IoT radio chipsets, this technique could potentially be used to identify individual devices, either for authentication, or as an attack vector.}

\newtext{
\boldpar{Complex beating patterns indicate the number of transmitters}
Fig.~\ref{fig:tii_complex} demonstrates that it is possible to apply an FFT even over complex beating waveforms, and discern the \emph{number} of devices within a transmission. Not only does this again have important security implications in that nodes are no longer `hidden' within a flood, but it could potentially be used to improve CT performance. \response{Specifically, as per Fig.~\ref{fig:tosh_results_ct_comp}, it is notable that the performance of CT can potentially be severely affected by a high degree of nodes within a single transmission. As such, knowledge of the number of participating nodes could allow the design of new protocols that reduce the number of active nodes within a local area, depending on the underlying PHY, thereby improving performance}.
}

\boldpar{Beating frequencies are highly sensitive to temperature} \newtext{The sensitivity of crystal oscillators frequencies is well documented~\cite{zhou2008frequency}. As beating frequencies are ultimately linked to the relative carrier offset between devices, temperature variations also affect CT-induced beating frequencies. Fig.~\ref{fig:templab_temp_freq} and Fig.~\ref{fig:templab_temp_per} explore how temperature changes at individual transmitters can increase or decrease the beating frequency, and subsequently the overall PER of concurrent transmissions. Importantly, Figs.~\ref{fig:templab_all_freq_per}~and~\ref{fig:templab_freq_per} ultimately demonstrate that this impact is captured within our simulation and modeling from Sect.~\ref{sec:simulation}, and that the temperature sensitivity is predictable within known channel bounds. While chipsets such as the \texttt{nRF52840} don't permit fine-grained tuning of the carrier frequency, alternative platforms (such as an SDR) could potentially correct the CFO to improve error rates.}

\section{CT Performance over different PHY\lowercase{s}:\\ Experimental Evaluation with RF Interference}
\label{sec:interference}

\begin{figure}[!h]
	\centering
	\vspace{-1.25mm}
	\includegraphics[width=0.65\columnwidth]{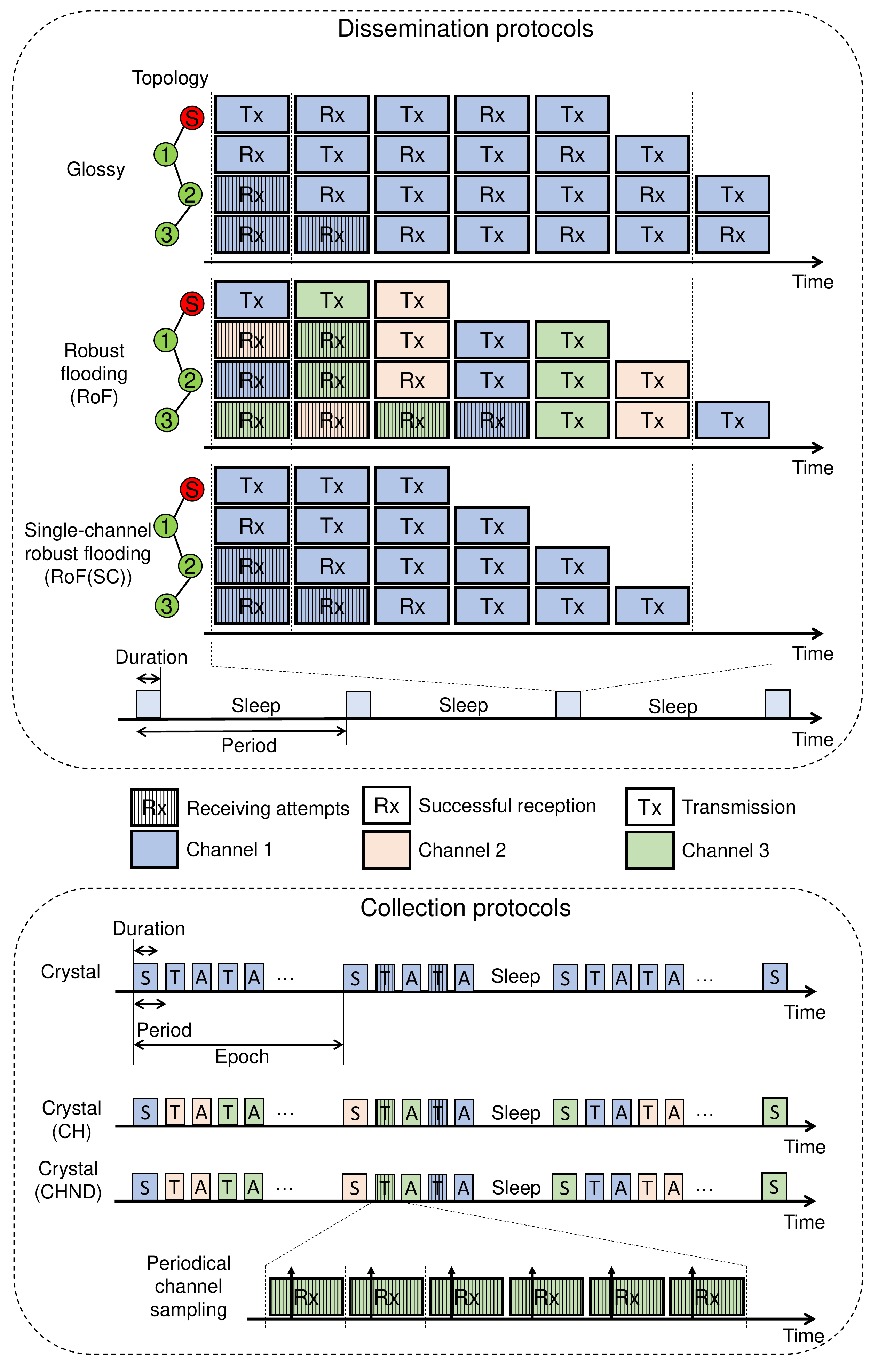}
	\vspace{-1.25mm}
	\caption{\textbf{High-level operations of the analysed CT-based protocols}. We consider three data dissemination protocols (Glossy, RoF, and RoF (SC)) performing one-to-many communication. In this illustration, the maximum numbers of transmissions (\texttt{TX\_N}) of these protocols is set to three. We then consider three variants of Crystal as data collection protocol (Crystal, $\rm{Crystal^{CH}}$, and $\rm{Crystal^{CH}_{ND}}$). Crystal works by exploiting Glossy as a flooding primitive and by splitting floods into transmission (T) and acknowledgment (A) pairs. $\rm{Crystal^{CH}}$ exploits multiple channels, whereas $\rm{Crystal^{CH}_{ND}}$ uses a noise detection mechanism in which a periodical channel sampling is enabled during the T period when no data is received.
	}
	\vspace{-2.25mm}
	\label{fig:protocols}
\end{figure}

This section presents an experimental study about the impact of different PHYs on the performance of CT-based protocols in the presence of RF interference.
Specifically, we evaluate three CT-based data dissemination protocols (i.e., Glossy~\cite{ferrari2011efficient}, Robust~Flooding~(RoF)~\cite{lim2017competition}, and RoF operating on a single channel~(RoF~(SC))), 
\newtext{
as well as three CT-based data collection protocols (i.e., Crystal~\cite{istomin2016data}, multi-channel Crystal ($\rm{Crystal^{CH}}$)~\cite{istomin18crystal}, and multi-channel Crystal with noise detection ($\rm{Crystal^{CH}_{ND}}$)~\cite{istomin18crystal}) 
}
on a large multi-hop network.
The operations of these CT-based protocols are depicted in Fig.~\ref{fig:protocols}; we refer the reader to~\cite{zimmerling20synchronous} for a detailed survey on CT-based protocols. 

The original Glossy~\cite{ferrari2011efficient} approach triggers transmissions after successful receptions, thereby alternating \emph{Rx} and \emph{Tx} slots at each hop.
However, not only does the Glossy approach operate on a single channel (meaning it is susceptible to RF interference at that frequency), but this reception-triggered \emph{Rx-Tx} technique means that \emph{Rx} failures will result in a missed transmission opportunity~\cite{lim2017competition, ma2018competition}. Using this technique, it is difficult to resume a CT flood if the latter is interrupted by RF interference.
An alternative approach was taken by the authors of~\cite{lim2017competition, raza2017competition}, introducing \emph{back-to-back} transmission slots (i.e.,~\emph{Rx-Tx-Tx}) alongside robust frequency diversity through per-slot channel hopping. In this Robust Flooding (RoF) approach, the first transmission of a node is still triggered by a correct reception, but further transmissions are time-triggered, with nodes synchronously hopping frequency at each slot according to a known offset on a global channel list. 
We compare these two approaches (Glossy and RoF), as they are commonly used as primitives to construct more complex CT-based protocols, and are hence representative of wider literature.  Furthermore, we introduce a variant of RoF operating on a single channel (RoF (SC)) to observe how this protocol performs with respect to the single-channel environment used by Glossy.

\newtext{
By using Glossy as a fundamental primitive, one can easily build many-to-one data collection protocols. 
Crystal~\cite{istomin2016data} is an example of such CT-based data collection protocol. 
In Crystal, the sink node initiates a synchronization flood to all the nodes in the network. 
Followed by the synchronization period, there are TA pairs (one transmission period and one acknowledgment period) as shown in Fig.\ref{fig:protocols} (bottom). 
A node can flood its message in a T period.
Relying on the capture effect, one of the messages can be probably received by the sink node successfully when multiple nodes have messages to flood in a T period.
In the following A period, the sink node floods its acknowledgment to indicate whose message has been just received.
This way, in the next T period, the received nodes would work as a relay node.
The synchronization period and TA pairs make up an epoch.
If there is no nodes sending messages in consecutive T periods (i.e., silent T periods), all the nodes would turn off the transceivers and no communication happens until the next epoch, thereby saving energy.
In order to work under interference, $\rm{Crystal^{CH}}$ and $\rm{Crystal^{CH}_{ND}}$~\cite{istomin18crystal} enable Crystal to work on multiple channels.
Rather than the slot-by-slot channel hopping mechanism utilized in RoF, where nodes rapidly hop frequency \emph{during} the flood, these protocols simply hop to a known frequency at the start of each flooding period, maintaining the same channel throughout the flood.}
\newtext{Table~\ref{table:protocol_params} summarizes the protocol parameters used.}
\newtext{
As mentioned before, a node would speculate that there is no new message if there are consecutive silent T \newtext{periods}.
However, it is difficult for a node to know whether the silent T period is ``real" or it is caused by interference.
Therefore, $\rm{Crystal^{CH}_{ND}}$ introduces the noise detection (ND) mechanism, where nodes detect the interference level by using RSS values if there is nothing received in current CT slot, and enter sleep mode once harsh interference is detected (i.e., the RSS values are greater than a specific threshold).
}

\captionsetup[subfloat]{labelformat=empty}
\begin{table*}%
	\centering
	\renewcommand{\arraystretch}{1.10}
	\footnotesize
	\begin{tabular}{ c l c c c c c c }
		\toprule
		\bf{Protocol type} & \bf{Protocol} & \bf{TX\_N} & \bf{Duration (ms)} & \bf{Period (ms)} & \bf{Epoch (ms)} & \bf{Channel} & \bf{Trigger mode}\\
		\midrule
		\multirow{3}{*}{Dissemination}
		& Glossy   & 6 & 100 & 200 & - & 2.480 GHz & Rx-Tx\\
		& RoF (SC) & 6 & 100 & 200 & - & 2.480 GHz & Rx-Tx-Tx\\
		& RoF      & 6 & 100 & 200 & - & 2.402, 2.426, 2.480 GHz & Rx-Tx-Tx\\
		\hline
		\multirow{3}{*}{Collection}
		& Crystal             & 6 & 20, 45, 15 & 30, 57, 27 & 1000 & 2.480 GHz & Rx-Tx\\
		& Crystal$^{CH}$      & 6 & 20, 45, 15 & 30, 57, 27 & 1000 & 2.402, 2.426, 2.480 GHz & Rx-Tx\\
		& Crystal$^{CH}_{ND}$ & 6 & 20, 45, 15 & 30, 57, 27 & 1000 & 2.402, 2.426, 2.480 GHz & Rx-Tx\\
		\bottomrule
	\end{tabular} 
	\vspace{+0.50mm}
	\caption{\newtext{Summary of protocol parameters used to evaluate CT-based flooding performance on D-Cube. Collection protocols based on Crystal employ three distinct \emph{sync} (S), \emph{transmit} (T), and \emph{acknowledge} (A) flooding phases with permitted maximum durations of 20, 45, and 15ms respectively, repeating every 1s (i.e., an epoch). For each phase, the periods (i.e., period of time in which flooding is allowed) are set to 30, 57, and 27 ms respectively, as depicted in Fig.\,\ref{fig:protocols}. Dissemination protocols employ a single flooding phase with a maximum duration of 100ms, periodically repeating every 200ms.}}%
	\label{table:protocol_params}
	\vspace{-0.50mm}
\end{table*}

\captionsetup[subfloat]{labelformat=empty}
\begin{table*}%
	\centering
	\renewcommand{\arraystretch}{1.10}
	\small
	\subfloat[][]{
		\begin{tabular}{ p{1cm}<{\centering} p{4cm}<{\centering} p{9cm} }
			\toprule
			\multicolumn{1}{c}{\textbf{No.}} & 
			\multicolumn{1}{c}{\textbf{Data Dissemination Layout}} & 
			\multicolumn{1}{c}{\textbf{Description}} \\ \midrule
			(d-i) & \dcubee \texttt{nRF52840 data dissemination}, layout \#\,1 & Messages sent from a sink (node \texttt{206}) to 5 destinations. 
			\\
			(d-ii) & \dcubee \texttt{nRF52840 data dissemination}, layout \#\,4 & Messages sent from a sink (node \texttt{200}) to 31 destinations. 
			\\ 
			\toprule
			\multicolumn{1}{c}{\textbf{No.}} & 
			\multicolumn{1}{c}{\textbf{Data Collection Layout}} & 
			\multicolumn{1}{c}{\textbf{Description}} \\ 
			\midrule
			(c-i) & \dcubee \texttt{nRF52840 data collection}, layout \#\,1 &  Messages sent from 5 source nodes to a sink (node \texttt{202}). 
			\\
			(c-ii) & \dcubee \texttt{nRF52840 data collection}, layout \#\,3 & Messages sent from 47 source nodes to a sink (node \texttt{227}). 
			\\ 
			(c-iii) & \dcubee \texttt{nRF52840 data collection}, layout \#\,4 & Messages sent from 10 source nodes to a sink (node \texttt{219}).
			\\
			\bottomrule
		\end{tabular} 
	}
	\vspace{+1.75mm}
	\caption{\textbf{Summary of layouts used to evaluate the performance of CT-based protocols in \dcubee.} The layout numbers refer to the official scenarios supported by the \dcubee benchmarking infrastructure, which involve nodes located in regions with both low and high node densities.}%
	\centering
	\label{table:layouts}
	\vspace{2.75mm}
\end{table*}
  
\subsection{Experimental Setup}
We use the \dcubee testbed~\cite{schuss20dcube} and all its $50$ nodes\footnote{\response{\url{https://iti-testbed.tugraz.at/wiki/index.php/Topology_of_Nodes}}} to evaluate in hardware three key metrics: \emph{end-to-end reliability}, \emph{latency}, and \emph{\newtext{total} energy consumption}. 
For each of the protocols (Glossy, RoF, RoF (SC), Crystal, $\rm{Crystal^{CH}}$, and $\rm{Crystal^{CH}_{ND}}$), we consider three scenarios characterized by the absence or presence of interference, denoted as \textit{no}, \textit{mild}, and \textit{strong} interference, respectively.

\dcubee's controllable RF interference is generated by observer nodes (Raspberry\,Pi\,3, directly attached to the nRF52840 boards) using \jamlabng.
\emph{Mild interference} (\dcubee level 2) uses a power of 30\,mW, generating intermittent (i.e., not continuous) interference for approximately 5\,ms every 13\,ms period.
\emph{Strong interference} (\dcubee level 3) emulates the transmissions of multiple \wifi devices across all the 2.4\,GHz band. Each \newtext{observer node} chooses a different channel, generating interference of 200\,mW for $\approx$\,8\,ms every 13\,ms. \response{Further details about the configuration of the generated interference are available in the original \jamlabng paper~\cite{schuss19jamlabng}.} 

Each dissemination protocol is run on \dcubee's data dissemination scenarios, i.e., those sending data from a single source to multiple destinations over a multi-hop network. 
\newtext{Each data collection protocol is run on \dcubee's data collection scenarios, i.e., those where a sink node collects data from multiple (or even all) nodes.
We further configure \dcubee to generate aperiodic messages with short (8B) and long (64B) payloads.
\response{To ensure that protocols span the minimum hop distance of the D-Cube testbed when using \blefive2M (known through prior experimentation)}, the maximum number of transmission attempts per node during a flooding period (defined as \texttt{TX\_N}) is set to six.
\newtext{The duration of a flooding and the periodicity in dissemination protocols are both set to 100 and 200\,ms respectively.}
The durations of the synchronization, T (transmission), and A (acknowledgment) periods are set to 20\,ms, 45\,ms and 15\,ms, respectively. 
Furthermore, there are 12 TA pairs in each epoch, which is set to one second. 
In single-channel protocols (Glossy, RoF (SC), and Crystal) the radio frequency is set to 2.480~GHz (which corresponds to channel 39 in BLE and channel 26 in \ieee), whereas multi-channel protocols (RoF, $\rm{Crystal^{CH}}$, and $\rm{Crystal^{CH}_{ND}}$) hop among three frequencies (2.402, 2.426, and 2.480~GHz, which correspond to BLE's three \newtext{advertising} channels 37, 38, and 39, respectively).}
\newtext{
For $\rm{Crystal^{CH}_{ND}}$ we further select a noise detection of -70\,dBm: this is a good choice given that the nRF52840DK exhibits RSS values of about -90\,dBm when there is no activity over the air. 

All the protocols (data dissemination and collection) are run on multiple \dcubee layouts to avoid making our results setup-specific. The topology of \dcubee can be regarded as two adjacent regions, one of which is a \emph{dense} region with around 20 nodes, whilst the remaining nodes make up a \emph{sparse} region of the network. 
For the data dissemination protocols, we run two \dcubee layout configurations: a single source node to (i) 5, and (ii) 31 destinations. 
For the collection protocols we run three \dcubee layout configurations featuring 5, 47, and 10 source nodes to a single destination, respectively. 
Additionally, we set the transmission power to 0\,dBm for all nodes, which ensures a network diameter between 6 and 10 hops depending on the employed layout. 
Table\,\ref{table:layouts} provides a summary of the \dcubee layout configurations used in our experiments. 

}

All the results shown in this section utilize publicly available implementations of Glossy\footnote{\url{https://sourceforge.net/p/contikiprojects/code/HEAD/tree/ethz.ch/glossy/}}, RoF\footnote{\url{https://github.com/ETHZ-TEC/robust-flooding} - \emph{a03f38a}}, $\rm{Crystal^{CH}_{ND}}$\footnote{\url{https://github.com/d3s-trento/crystal} - \emph{11ef31e}}, which we subsequently ported to the nRF52840-DK platform supported by \dcubee.

\subsection{Results}

\begin{figure}[!t]
	\centering
	\begin{subfigure}{0.45\columnwidth}
		\centering
		\includegraphics[width=1\columnwidth]{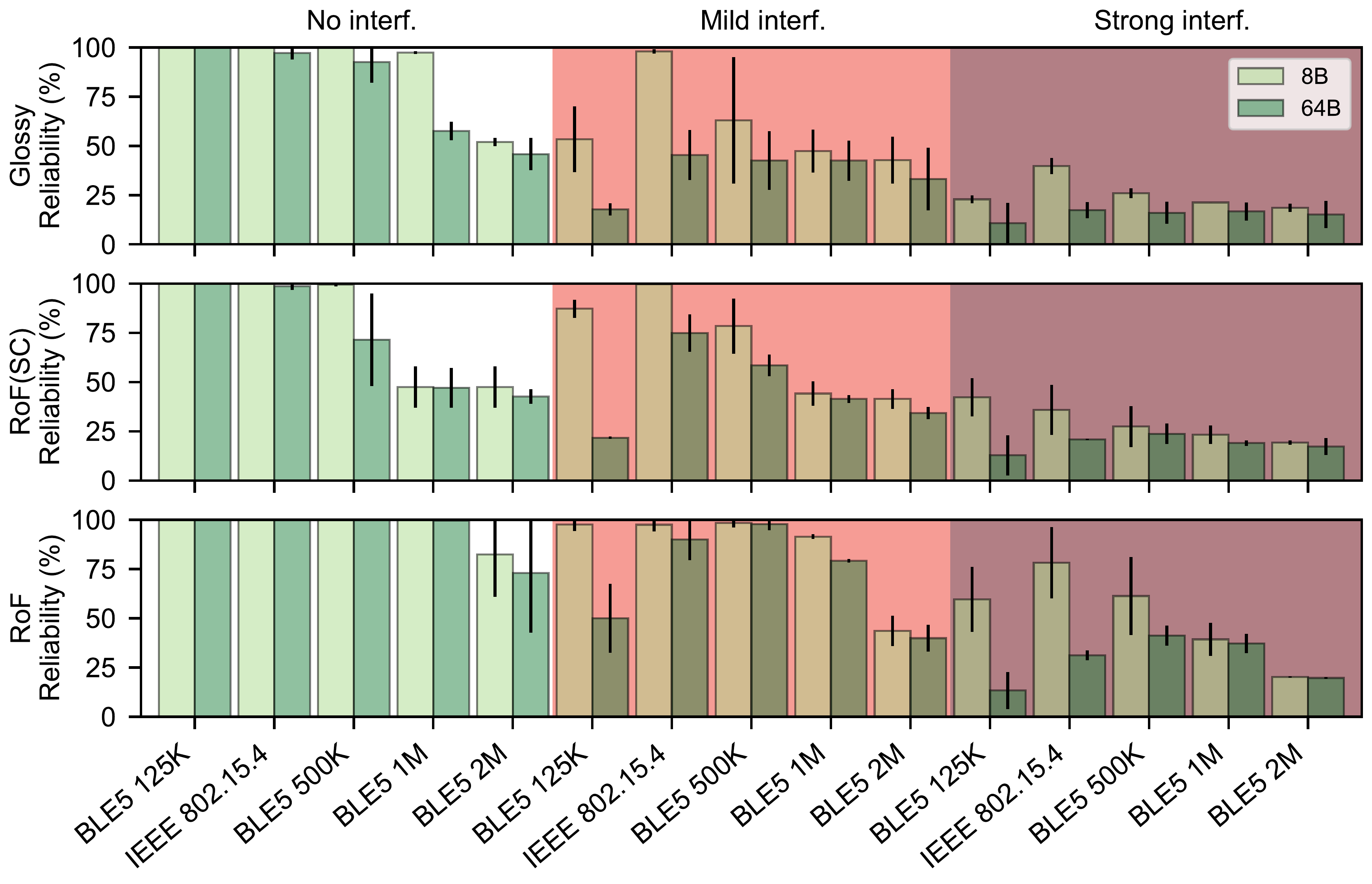}
		\vspace{-6.25mm}
		\caption{(a) Dissemination: average end-to-end reliability.}
		\label{fig:reliability_0dbm_dissemination}
		\vspace{+2.25mm}
	\end{subfigure}\hfill
	\begin{subfigure}{0.45\columnwidth}
		\centering
		\includegraphics[width=1\columnwidth]{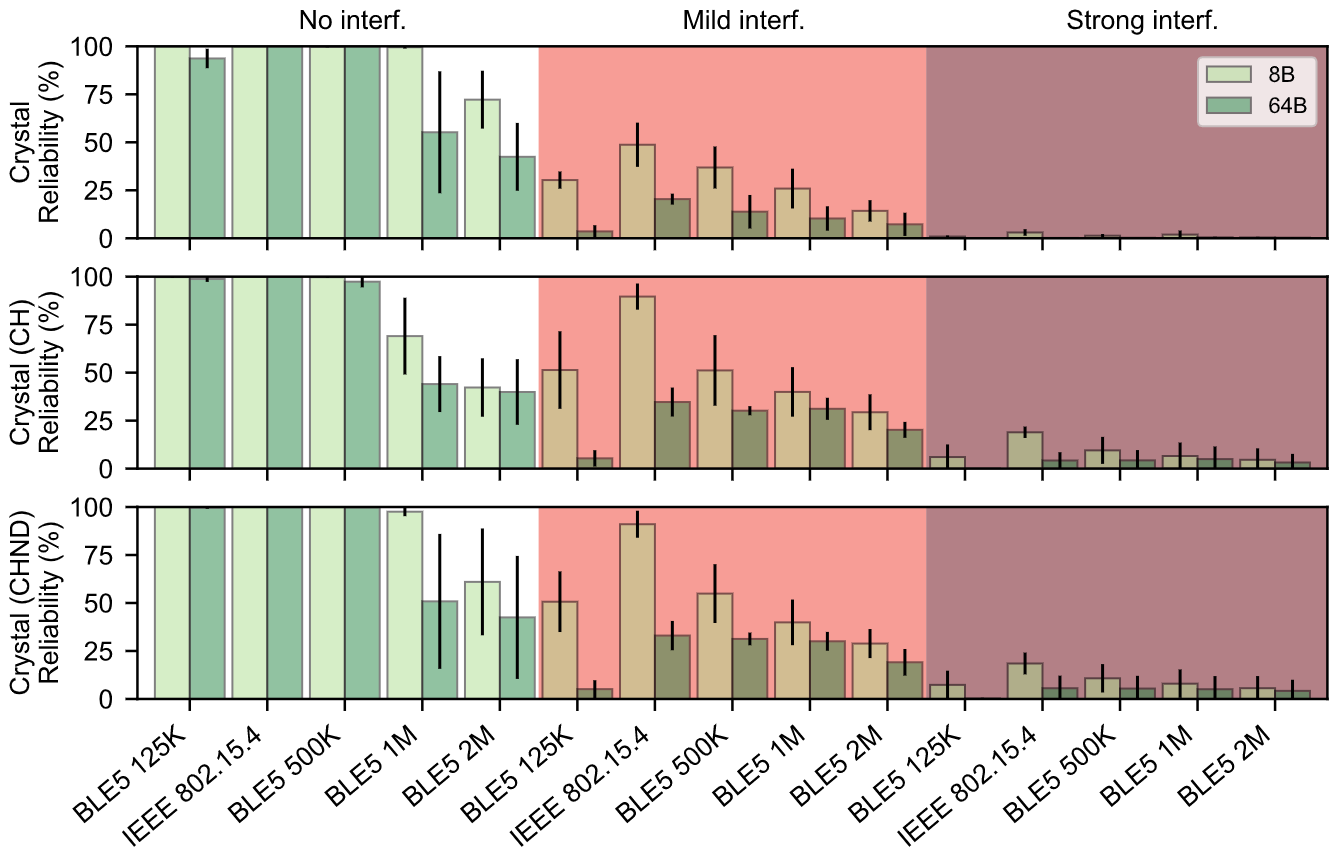}
		\vspace{-6.25mm}
		\caption{(b) Collection: average end-to-end reliability.}
		\label{fig:reliability_0dbm_collection}
		\vspace{+2.25mm}
	\end{subfigure}
	\begin{subfigure}{0.45\columnwidth}
		\centering
		\includegraphics[width=1\columnwidth]{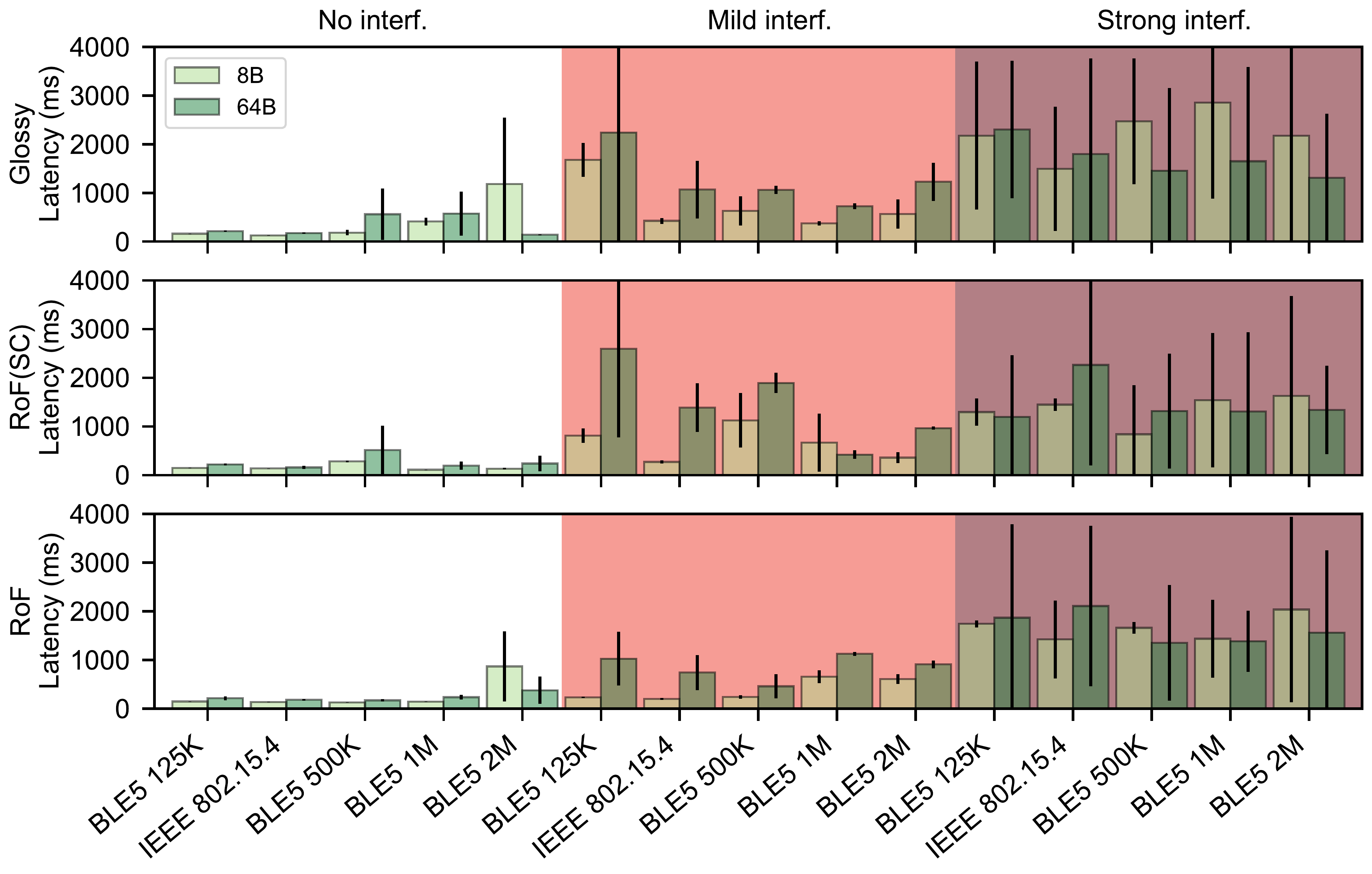}
		\vspace{-6.25mm}
		\caption{(c) Dissemination: average end-to-end latency.}
		\label{fig:latency_0dbm_dissemination}
		\vspace{+2.25mm}
	\end{subfigure}\hfill
	\begin{subfigure}{0.45\columnwidth}
		\centering
		\includegraphics[width=1\columnwidth]{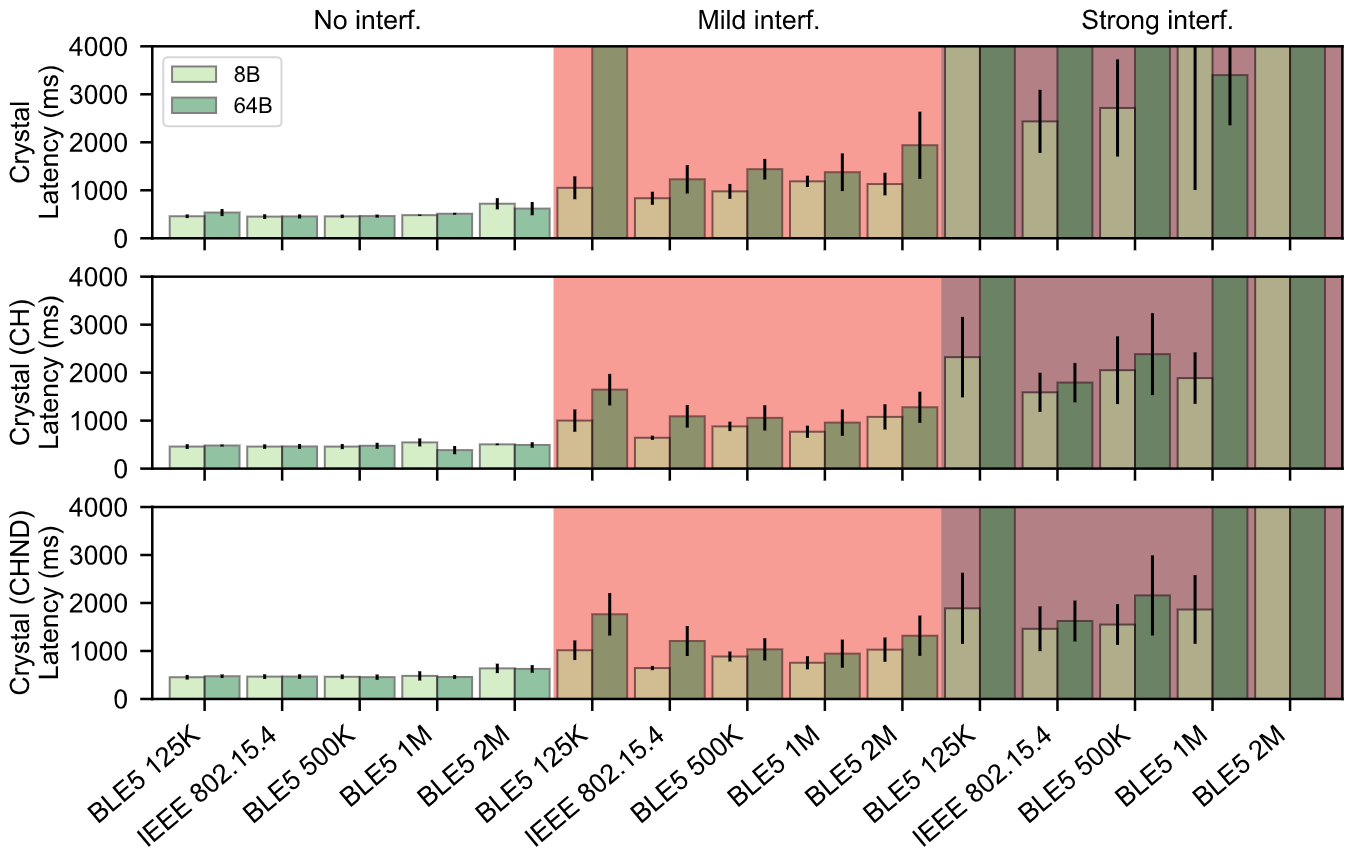}
		\vspace{-6.25mm}
		\caption{(d) Collection: average end-to-end latency.}
		\label{fig:latency_0dbm_collection}
		\vspace{+2.25mm}
	\end{subfigure}

	\begin{subfigure}{0.45\columnwidth}
		\centering
		\includegraphics[width=1\columnwidth]{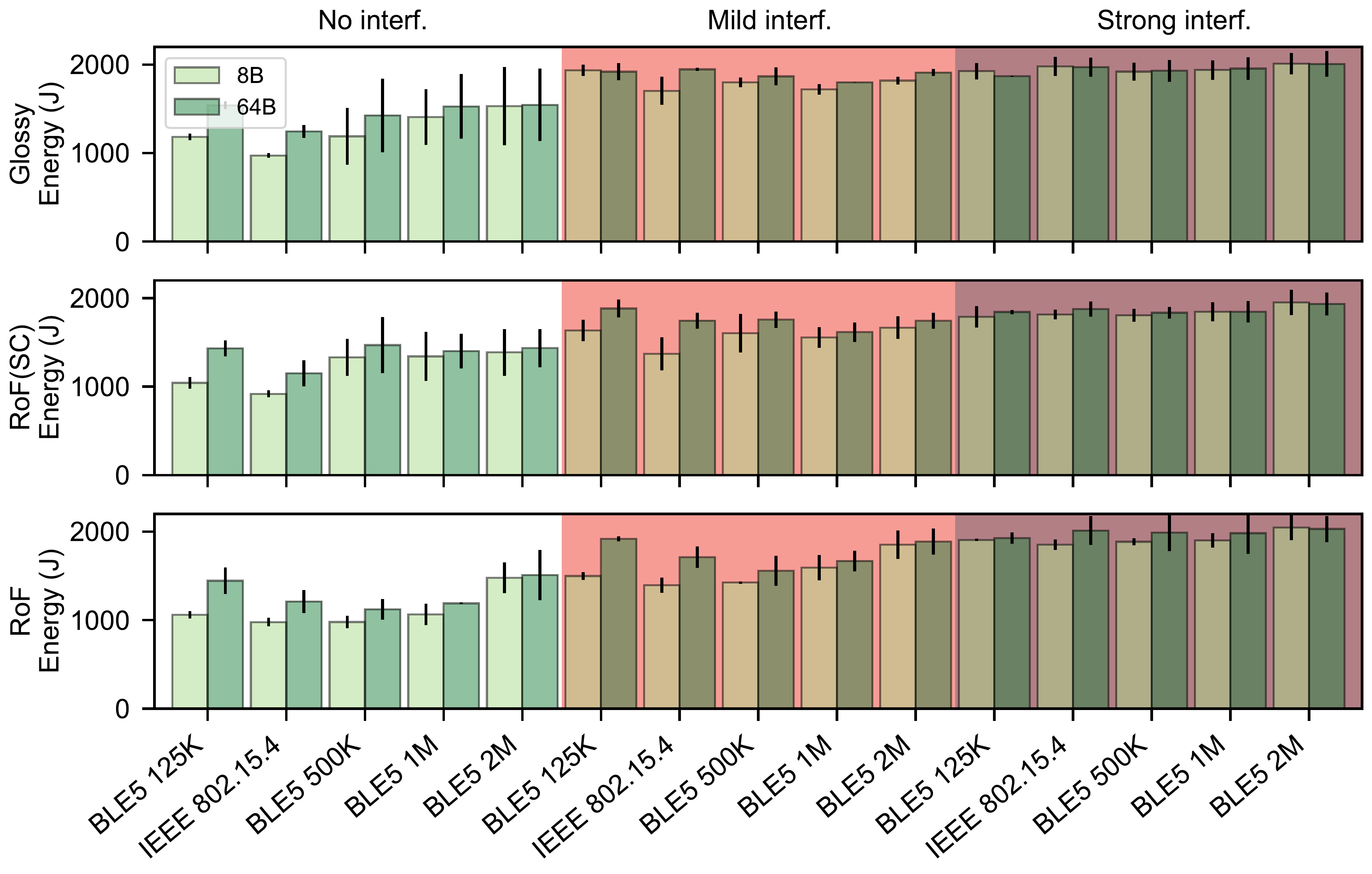}
		\vspace{-6.25mm}
		\caption{(e) Dissemination: total energy consumption.}
		\label{fig:energy_0dbm_dissemination}
	\end{subfigure}\hfill
	\begin{subfigure}{0.45\columnwidth}
		\centering
		\includegraphics[width=1\columnwidth]{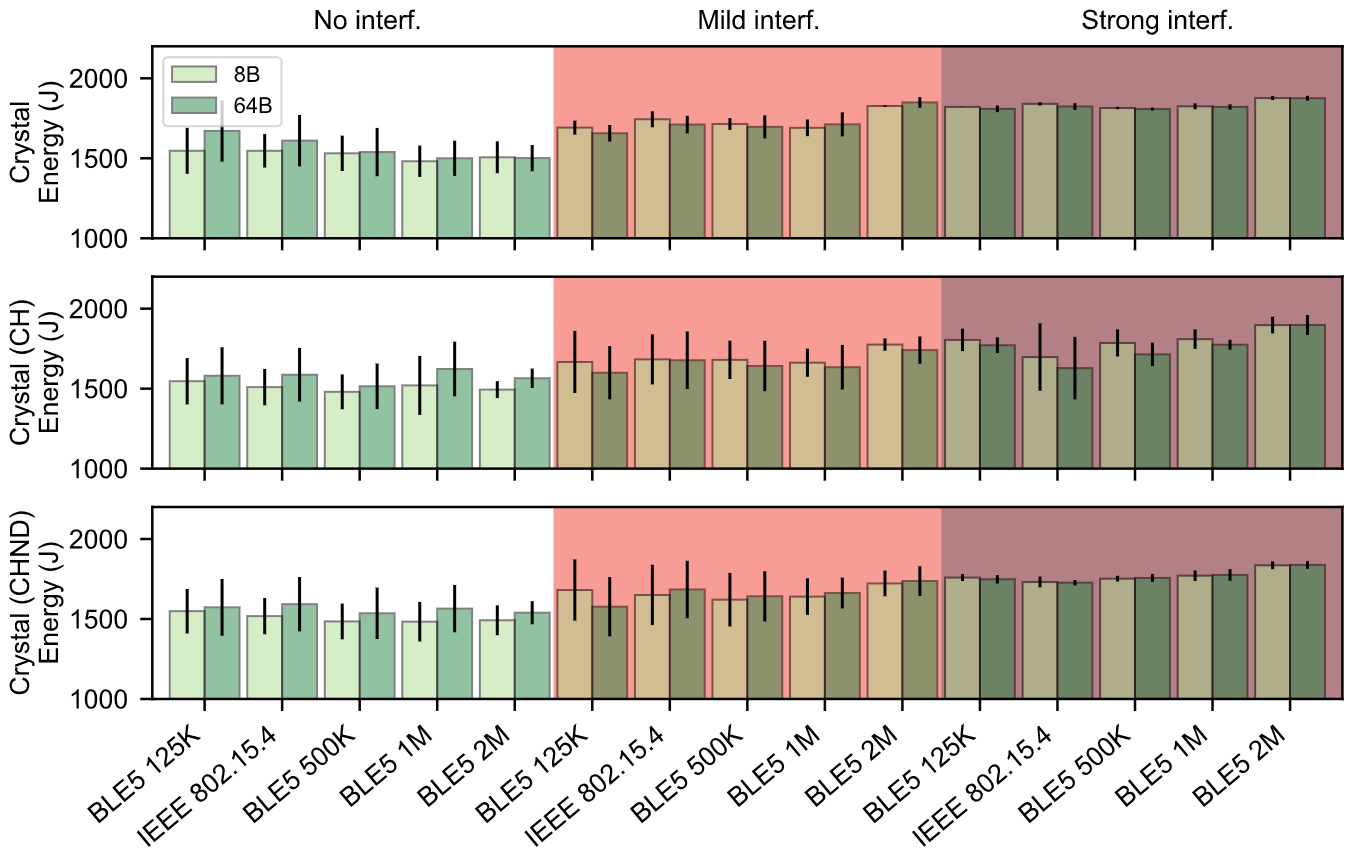}
		\vspace{-6.25mm}
		\caption{(f) Collection: total energy consumption.}
		\label{fig:energy_0dbm_collection}
	\end{subfigure}
	\vspace{-1.75mm}
	\caption{\textbf{Performance of CT-based data collection and dissemination protocols in the presence of RF interference.} These six plots show the average reliability, latency, and total energy consumption across the data dissemination and collection layouts of \dcubee listed in Table\,\ref{table:layouts}.}
	\label{fig:protocol_results}
	\vspace{-3.00mm}
\end{figure}

Fig.\,\ref{fig:protocol_results} summarizes the results of our experiments and provides an overview of the end-to-end reliability and latency, as well as of the total energy consumption of the analysed \newtext{data collection and} dissemination protocols in the absence and in the presence of RF interference.  
\boldpar{Reliability}
For the data dissemination protocols, Fig.~\ref{fig:reliability_0dbm_dissemination} shows that, under no interference, as data rate increases, the evaluated PHYs struggle to maintain reliability.
RoF exhibits the highest reliability, while the packet reception rate of RoF\,(SC) drops at higher data rates compared to that of Glossy.
Since the time-triggered transmission approach of RoF and RoF (SC) does not allow nodes to resynchronize at every \emph{Rx} slot (as in Glossy), nodes can be subject to synchronization errors due to drift. The high data rate PHYs are particularly sensitive to such errors: \blefive2M is only able to tolerate CT synchronization errors of up to 0.25\,$\mu$s~\cite{alnahas2019concurrentBLE5}.

In general, longer transmission times increase the probability that interference corrupts the packet. This is reflected in the reliability between \dcubee's long (64B) and short (8B) payloads under all three interference scenarios. Crucially, long transmission times in \blefive125K means that it struggles to escape interference, and results in surprisingly poor reliability across all three protocols. 
We also observe that \emph{back-to-back} repetition of packets in RoF and RoF~(SC) improves reliability over Glossy under mild and strong interference.
As expected, frequency diversity from RoF's channel hopping mechanism significantly improves performance of all PHYs -- except for the \blefive2M. This is likely due to the higher data rate, as interference generated by \jamlabng is \emph{periodic} across \emph{multiple} channels. High data rate floods are more likely to fall completely within the interference duration, while lower data rates may have transmission slots falling in-between interference periods. 
The results presented in Fig.~\ref{fig:reliability_0dbm_dissemination} also hint that the \ieee and the \blefive500K PHYs exhibit the highest reliability under strong interference when using short and long payloads, respectively. 

\newtext{
Data collection protocols exhibit similar trends to those of the data dissemination ones, i.e., shorter messages are delivered more reliably, and channel hopping mechanisms improve the reliability under interference.
However, data collection protocols cannot sustain a comparably high packet reception rate to that of data dissemination protocols when the interference becomes very strong, as visible in Fig.~\ref{fig:reliability_0dbm_collection}. 
The reason for this is that a message in a many-to-one network has less flooding chances than in a one-to-many network: in fact, multiple nodes are required to send their packets to the individual sink node and the throughput of the sink node is limited even if a channel hopping mechanism has been adopted. 
The noise detection feature improves the reliability by 15\% and 62\% for short and long payloads, respectively, under strong interference, which is coherent with the results presented in \cite{istomin18crystal}. 
}

\boldpar{Latency}
The end-to-end latency of CT-based protocols is inherently linked to reliability. Brute-force repetition means that packets \emph{may} successfully be received on poor channels, but much later in the flood.
\newtext{
Fig.~\ref{fig:latency_0dbm_dissemination} and~\ref{fig:latency_0dbm_collection} support this, and we observe significant latency jumps as interference increases.
Especially for data collection protocols (Fig.~\ref{fig:latency_0dbm_collection}), the latency can be much higher (i.e., as high as four seconds or more) under strong interference when only a single channel is used.}
As \dcubee's latency is only computed based on \emph{received} messages, it is conceivable that low latencies can be achieved even with poor reliability. This is evident under \textit{mild} interference, where the uncoded PHYs exhibit low latency despite poor reliability.
In general, coded PHYs (i.e., \blefive125K, \blefive500K, and \ieee) have better performance w.r.t. latency.

\boldpar{Energy efficiency}
Similarly, the energy consumption of nodes is intrinsically linked to reliability, as missed receptions increase radio-on time. Although, in principle, higher data rate PHYs should have a lower energy consumption, this relationship means that for \emph{short} payloads \blefive1M and 2M are less energy efficient than the coded PHYs.
Under interference, nodes take a longer time to attempt to overhear a packet. 
Therefore, interference also causes more energy consumption.
However, the underlying PHY rate is still a fundamental factor in energy consumption as shown in Fig.~\ref{fig:energy_0dbm_dissemination}. For \emph{long} payloads the lower data rates of the coded PHYs means that the radio can take considerable time to transmit, consuming far more energy. 
\newtext{
Coded PHYs offer a lower energy consumption also for data collection protocols (as shown in Fig.~\ref{fig:energy_0dbm_collection}), although the difference is less pronounced than in the dissemination protocols.
Interestingly, in the presence of RF interference, the energy consumption is often lower when transmitting longer payloads -- a behaviour that is opposite to that observed in dissemination scenarios. 
We believe that the reason behind lies in Crystal's design: after several consecutive silent T \newtext{periods} (in which nothing is received), the sink node lets all nodes enter sleep mode by sending a command in the next A period. 
}
\newtext{
In other words, packets with shorter payloads are more likely to survive mild interference and complete a Crystal round, which causes a higher energy consumption compared to the use of longer payloads (which abort the Crystal round earlier and hence consume less energy). 
In the presence of strong interference, also packets with short payloads have a lower chance to be received: the trend shown in Fig.~\ref{fig:energy_0dbm_collection} is hence flatter. 
}

\newtext{
We also note that the energy consumed by the nodes in $\rm{Crystal^{CH}_{ND}}$ is about 1\% higher than that in $\rm{Crystal^{CH}}$ due to the noise detection mechanism.
This is necessary to enable nodes to still overhear the network during silent T slots under interference.
This way, the sink node is more likely to receive the packets despite the presence of RF interference.
}

\subsection{Key Observations}
Based on these results, we make a number of observations on the network-wide performance of CT-based flooding protocols under RF interference as a function of the employed PHY.

\boldpar{At a network level, high data rate PHYs struggle even in absence of external RF interference}
Without the redundancy gains of coded PHYs, high data rate PHYs are sensitive to both de-synchronization and beating at greater CT densities (as per Fig.\,\ref{fig:tosh_results_ct_comp}). 
Even with the added benefit of frequency diversity, RoF's reliability drops when using the \blefive2M PHY.

\boldpar{\blefive125K is not necessarily the answer}
Although performing well when there is no interference, \blefive125K suffers from poor performance as soon as interference kicks in and packet size increases, taking a relatively large hit with respect to latency and reliability in comparison to the other PHYs.

\boldpar{\blefive500K and \ieee perform well under interference}
Compared to the other coded PHYs, \blefive500K achieves higher reliability and similar or lower latency when under interference. \ieee, on which the majority of CT literature is based, demonstrates similar high reliability under interference, but has a lower data rate.
If the level of interference is unknown, CT protocols benefit from transmission on either the \blefive500K or \ieee.

\boldpar{RoF's time-triggered transmission and channel hopping produce significant gains}
It is clear from Fig.~\ref{fig:reliability_0dbm_dissemination} that the combination of time-triggered transmissions and channel hopping in RoF gives significant gains under all scenarios.
However, RoF (SC) shows that without frequency diversity there is a chance that at higher data rates the interference duration may be longer than the flooding period. Increasing \texttt{TX\_N} could therefore give greater temporal redundancy and improve protocol reliability under interference.

\newtext{
\boldpar{More probability of collision (less probability of capture effect) when using faster data rates}
Crystal (and its multi-channel versions) significantly rely on the capture effect when multiple nodes have a (\emph{different}) message to send in a T period.
Specifically, in a T period, all the nodes would initiate a flood by sending their messages actively, and which message can be flooded to the sink node depends on the capture effect.
However, when high data rate PHYs with shorter preambles are adopted (Fig.~\ref{fig:reliability_0dbm_collection}), the capture effect is not triggered so frequently, and reliability decreases~\cite{escobar2019competition}.
}

\section{Related Work} \label{sec:related_work}

We discuss next related work and highlight how the contributions presented in this paper advance the state-of-the-art.

\boldpar{CT on different PHYs}
After the influential work by Dutta et al.~\cite{dutta10amac} and Ferrari et al.~\cite{ferrari2011efficient} were published in 2010/2011, a large number of researchers has started to study CT and develop CT-based protocols~\cite{zimmerling20synchronous, ferrari2012low, suzuki13choco, istomin2016data, doddavenkatappa2013splash, du17pando, landsiedel13chaos, chang18constructive, baddeley216pp, koenig16constructive, yuan13together, yuan14sparkle, mohammad18codecast, doddavenkatappa14p3, alnahas2017consensus, poirot19paxos, cao21coflood, vonzengen18dcoss, himmelmann20dcoss, he16arpeggio, debardashini20lcn, debardashini20mass}.
While most of the early works targeted exclusively \ieee devices using the 2.4\,GHz band, in the last years, a few studies have shown the feasibility of CT on other physical layers supported by \ieee, such as the UWB PHY~\cite{kempke16surepoint, corbalan19secon, lobba20concurrent}, as well as sub-GHz short-range~\cite{liao16toward, beutel19ipsn} and long-range technologies~\cite{liao17lora, ma20chirpbox}.

A number of works have recently focused on studying the feasibility of CT on BLE~\cite{alnahas2019concurrentBLE5, schaper2019truth, alnahas2020blueflood, roest15ble}.
Specifically, Al Nahas et al.~\cite{alnahas2019concurrentBLE5} verified the feasibility of a CT-based flooding protocol, named BlueFlood, on the different \blefive PHYs experimentally, and also reported its performance on \ieee~\cite{alnahas2020blueflood}.
Schaper~\cite{schaper2019truth} studied the conditions to make CT successful in these PHYs in an anechoic chamber.

Different from these studies, our current work does not aim to prove the \textit{feasibility} of CT on different radio technologies or PHYs, but instead to provide an in-depth \textit{characterization of the role of the physical layer} on the reliability and efficiency of CT-based solutions employing \ieee and \blefive.
To the best of our knowledge, we are the first do this in a systematic manner by demonstrating experimentally the impact of errors induced by de-synchronization and beating distortion in CT-based protocols as a function of the employed PHY.

\boldpar{CT performance under interference}
Several works have shown that CT-based data collection and dissemination protocols can outperform conventional routing-based solutions in terms of reliability, end-to-end, and energy consumption even in the presence of harsh radio interference~\cite{lim2017competition, istomin18crystal, ma20harmony, escobar2019competition}, as also highlighted in the context of the EWSN dependability competition series~\cite{boano17competition, schuss17competition, schuss18benchmark}.
To sustain a dependable performance under interference, CT-based solutions have been enriched, among others, with mechanisms such as local opportunistic retransmissions~\cite{ferrari2011efficient, sutton_zippy_2015, suzuki15ewsn}, channel-hopping~\cite{lim2017competition, istomin18crystal, sommer16channelhopping, escobar2016redfixhop, doddavenkatappa14p3}, network coding~\cite{doddavenkatappa2013splash, du17pando, yuan15ripple, mohammad18codecast, herrmann18mixer}, noise detection~\cite{istomin18crystal, ma20harmony}, stretched preambles~\cite{escobar2019competition}, data freezing~\cite{ma20harmony}, as well as an improved understanding of the network state~\cite{carlson13forwarder, brachmann16laneflood, sarkar16sleepingbeauty}.

However, most of these protocols have been implemented for and evaluated with \ieee technology only.
In this paper, we are the first to study the performance of CT-based data collection \newtext{and dissemination} protocols on a large scale under interference \textit{as a function of the employed PHY}.
We did this by evaluating the dependability of CT-based protocols on a large scale using modern multi-radio platforms supporting several PHYs, and by analysing the impact of other factors such as the length of the transmitted messages and the harshness of the interference.

\boldpar{Impact of beating effect on CT}
A few studies have tried to underpin the foundations of concurrent transmissions on a signal level.
Ferrari et al.~\cite{ferrari2011efficient} have simulated CT signals with Matlab and explained how accurately packets should be aligned in order to design reliable protocols.
Other works~\cite{rao_murphy_2016, wang_triggercast_2013, wang_disco:_2015} have analyzed CT signals theoretically and argued that it is difficult to generate ideal constructive interference, due to the timing errors caused by radio propagation and clock drift.

Liao et al. have been the first to argue that there exists a beating effect caused by innate CFO between device oscillators~\cite{liao2016revisiting}.
Specifically, in~\cite{liao2016revisiting}, the resultant signals are generated by Matlab and a TelosB node was used to observe how the DSSS modulation in \ieee saves CT signals from the beating effect.
More recent studies have demonstrated these beating effects generate periods of both constructive and destructive interference by observing the raw IQ samples of devices connected to an SDR using coaxial cables~\cite{alnahas2020blueflood}.

In this paper we are the first to demonstrate how PHY effects such as CFO-induced beating, and de-synchronization from hardware clock drift, directly affect the signal observed by a receiving node \textit{using over-the-air testbed experiments}. 
\newtext{Furthermore, we also characterize the impact of environmental effects such as temperature on beating.}


\captionsetup[subfloat]{labelformat=empty}
\begin{table*}%
  \centering
  \renewcommand{\arraystretch}{1.10}
  \subfloat[][]{
	\begin{tabular}{ p{7.60cm} p{7.45cm} }
		\toprule
		\multicolumn{1}{c}{\textbf{Observations on CT Performance}} & \multicolumn{1}{c}{\textbf{Recommendations}} \\ \midrule
		\amark The \ieee and \blefive500K PHYs are effective against external RF interference and beating. They stuggle under strong wide beating, which may cause a significant drop in reliability, but provide the best overall trade-off among all PHYs. & \amark 
		In the presence of strong external RF interference, use \ieee for shorter packets and \blefive500K for longer packets. They provide a good combination of resilience against beating and interference, and extended range.
		 \\
		\amark High data rate PHYs help escaping strong and wide beating, but exhibit poor performance in the presence of external RF interference. & \amark In absence of external RF interference and with a low network density, use \blefive2M (or~1M) and repetitions to exploit temporal redundancy$^{(*)}$.
		\\ 
		\amark The \blefive125K PHY features the longest range, but performs poorly when sending long packets under external RF interference and strong wide beating. & \amark 	Use \blefive125K only for very short packets or when the extended range is a must$^{(*)}$.\\ 
		\amark\newtext{Temperature changes at transmitting nodes can effectively \textit{narrow} or \textit{widen} beating frequencies and subsequently affect error rates (depending on PHY).} & \amark \newtext{An FFT can be periodically employed to quantify beating frequency and predictively switch PHY in response to changing temperature conditions.} \\
		\bottomrule
	\end{tabular} 
  }
  \vspace{+1.75mm}
  \caption{\textbf{Summary of observations on CT performance over different PHYs and corresponding recommendations.} $^{(*)}$The choice of which PHY to use to cope with beating should also be made based on the application's latency, energy efficiency, and RF range requirements.}%
  \centering
  \label{table:conclusions1}
   \vspace{-2.75mm}
\end{table*}

\captionsetup[subfloat]{labelformat=empty}
\begin{table*}%
  \centering
  \setlength{\tabcolsep}{2.50pt}
  \renewcommand{\arraystretch}{1.20}
  \subfloat[][]{
  \begin{tabular}{ c | c | c | c | c }
		\toprule
		\textbf{PHY} & \multicolumn{2}{c|}{\textbf{Beating Errors}} & \multicolumn{2}{c}{\textbf{External RF Interference}} \\ 
		& \textbf{short packet} & \textbf{long packet} & \textbf{short packet} & \textbf{long packet} \\ \hline
		\blefive125K & $\nearrow$ & $\searrow$ & $\nearrow$ & $\downarrow$ \\ \hline
		\blefive500K & $\uparrow$ & $\nearrow$ & $\nearrow$ & $\uparrow$ \\ \hline
		\blefive1M & $\nearrow$ & $\searrow$ & $\searrow$ & $\searrow$ \\ \hline
		\blefive2M & $\nearrow$ & $\nearrow$ & $\downarrow$ & $\downarrow$ \\ \hline
		IEEE\,802.15.4 & $\uparrow$ & $\nearrow$ & $\uparrow$ & $\nearrow$ \\ \bottomrule
  \end{tabular}  
  }
  \vspace{+0.75mm}
  \caption{\textbf{Suitability of the different PHYs to mitigate the presence of beating errors and external RF interference.} Up arrows indicate a higher suitability and better performance; down arrows suggest a lower suitability and poor performance.}%
  \centering
  \label{table:conclusions2}
  \vspace{-3.75mm}
\end{table*}

\section{Discussion and Future Work} \label{sec:conclusions}
\newtext{An increasingly large body of work has shown that, particularly in environments with high RF interference, CT-based protocols are a valuable tool allowing low-power mesh networks to mitigate the impact of harsh RF conditions and provide robust, low-latency communication.}
This paper provides the first systematic experimental evaluation about the impact of the physical layer on CT performance. While these results are specific to the \ieee and \blefive PHYs, they provide important insights on how the choice of the PHY can exacerbate or reduce errors due to beating, de-synchronization, and external RF interference.



Our main observations and recommendations are summarized in Table~\ref{table:conclusions1}. With modern chipsets now supporting real-time PHY switching~\cite{cc2652_productsheet,nrf52840_productsheet}, these insights can guide further research into CT protocols that take advantage of multi-PHY capabilities \newtext{\textit{at runtime}}. For example, future protocols may take advantage of repetitions in uncoded PHYs to combat beating, while mitigating RF interference through coded PHYs. 

Yet, while these findings are important for the design of CT protocols, there are a number of key areas that require additional research \newtext{and further investigation}. 
Foremost, analytical and mathematical modeling would help to substantiate the observations presented in Sect.~\ref{sec:simulation}~and~\ref{sec:beating}. Crucially, a greater understanding is needed around how beating errors affect a protocol's scalability on a network level, as real-world RF conditions make it difficult to determine how beating will manifest at each node. 

\newtext{
As shown in this paper, and succinctly summarized in Fig.~\ref{fig:templab_all_freq_per}, RFO may change over time due to oscillator temperature sensitivity -- resulting in changes to the beating frequency, and subsequently the effective error rate. By maintaining a histogram of bit errors and running an FFT over the resulting waveform, it is possible to quantify the effective RFO between devices. While we have demonstrated this technique statically, the prospect of continuous monitoring of the beating effect at the receiver is intriguing, and well within the capabilities of modern dual-MCU chipsets such as the~\texttt{nRF5340}~\cite{nrf5340_productsheet}.
} 

\newtext{
Furthermore, results from Sect.~\ref{sec:beating} argue that additional work is needed to not on only explore how \emph{complex} beating patterns affect CT performance (rather than the simple sine waves observed when receiving from two transmitters), but how these waveforms relate to the transmitting devices.
}
\newtext{	
Such capability would have important and highly impactful security repercussions. Not only could monitoring of the RFO feasibly be employed to fingerprint devices as a Physical Unclonable Function (PUF) while accounting for fluctuations due to ambient temperature changes, but filtering and analysis of beating waveforms would allow estimation of the \emph{number} of concurrently transmitting devices. Specifically, chipsets such as the~\texttt{nRF5340}~\cite{nrf5340_productsheet} and~\texttt{nRF2833}~\cite{nrf52833_productsheet} offer access to I/Q values within the Constant Tone Extension (CTE) included as part of Bluetooth\,5.2, which would remove the reliance on `failed' CRC packets to examine the bit errors while also allowing more fine-grained sampling for dealing with complex beating waveforms.
}

Techniques such as \emph{interleaving}, i.e., bit shuffling, (which improves the robustness of FEC with respect to burst errors \response{-- the primary culprit of beating-induced errors as discussed in Sect.~\ref{sec:simulation}}) and \emph{bit voting} are missing in the analyzed physical layers, and would be an effective addition to increase the reliability of CT.

\newtext{Finally, while this paper has mainly focused on the negative properties of beating, specifically packet errors due to destructive interference, Sect.~\ref{sec:background} alluded to the fact that beating isn't \emph{necessarily} bad. While beating valleys do indeed disrupt CT communication, there is an argument that further work is needed to explore positive beating effects. Specifically, long transmissions should avoid periods of destructive interference and something will eventually be received. Exploiting beating peaks is an interesting proposition that could significantly improve the performance of CT-based protocols.}

\section*{Acknowledgments}
This work was partially supported by the National Natural Science Foundation of China under Grant \texttt{61902188}, and in part by the China Postdoctoral Science Foundation under Grant \texttt{2020T130304}. 
This work has also been partly performed within the LEAD project ``Dependable Internet of Things in Adverse Environments'' funded by Graz University of Technology.
\bibliographystyle{ACM-Reference-Format}
\bibliography{references}

\end{document}